\documentclass[twoside,11pt,preprint]{article}

\usepackage{blindtext}

\usepackage{jmlr2e}

\usepackage{amsmath,amsfonts,amssymb}

\usepackage{hyperref}
\usepackage{graphicx}
\usepackage[ruled]{algorithm2e}
\RequirePackage{bm} 

\usepackage{color}

\usepackage{MnSymbol}

\newcommand{\e}{\mathbb{E}}
\newcommand{\1}{\mathbf{1}}
\newcommand{\lk}{\left[ }
\newcommand{\rk}{\right] }
\newcommand{\lc}{\left(}
\newcommand{\rc}{\right)}
\newcommand{\R}{\mathbb{R}}
\newcommand{\Db}{\mathbb{D}}

\newcommand{\Gb}{\mathbb{G}}
\newcommand{\Pb}{\mathbb{P}}

\newcommand{\Eb}{\mathbb{E}}
\newcommand{\Nb}{\mathbb{N}}
\newcommand{\N}{\mathbb{N}}

\newcommand{\bc}{\mathcal{B}}

\newcommand{\Dc}{\mathcal{D}}
\newcommand{\Lc}{\mathcal{L}}
\newcommand{\Hc}{\mathcal{H}}

\newcommand{\Nc}{\mathcal{N}}
\newcommand{\Sc}{\mathcal{S}}
\newcommand{\Uc}{\mathcal{U}}
\newcommand{\Fc}{\mathcal{F}}
\newcommand{\Gc}{\mathcal{G}}
\newcommand{\Zc}{\mathcal{Z}}
\newcommand{\Qc}{\mathcal{Q}}
\newcommand{\Tc}{\mathcal{T}}
\newcommand{\Ec}{\mathcal{E}}
\newcommand{\Vc}{\mathcal{V}}

\newcommand{\Mc}{\mathcal{M}}

\newcommand{\sumn}{\sum_{i=1}^n}

\newcommand{\rmd}{\mathrm{d}}
\newcommand{\sumijn}{\sum_{\substack{i,j=1 \\ i\not=j} }^n}

\newcommand{\as}{{\alpha_\star}}
\newcommand{\bs}{{\beta_\star}}
\newcommand{\ts}{{\theta_\star}}
\newcommand{\an}{{\alpha_n}}
\newcommand{\bn}{{\beta_n}}
\newcommand{\eps}{{\epsilon}}

\newcommand{\wcv}{\stackrel{\text{law}}{\longrightarrow}}

\newcommand{\hmmden}{\widehat{\mmd^2_{\epsilon_n}}}
\newcommand{\hmmdq}{\widehat{\mmd^2_q}}

\DeclareMathOperator{\argmin}{argmin}
\DeclareMathOperator{\mmd}{MMD}
\DeclareMathOperator{\var}{Var}
\DeclareMathOperator{\cov}{Cov}
\newcommand{\hmmd}{\widehat{\mmd^2}}
\newcommand{\mds}{\medskip}

\newcommand{\Xbm}{\bm{X}}
\newcommand{\Ybm}{\bm{Y}}
\newcommand{\Ubm}{\bm{U}}
\newcommand{\Vbm}{\bm{V}}

\newcommand{\xbm}{\bm{x}}
\newcommand{\ybm}{\bm{y}}

\newcommand{\ubm}{\bm{u}}
\newcommand{\vbm}{\bm{v}}

\newtheorem{thm}{Theorem}[section]
\newtheorem{lem}{Lemma}
\newtheorem{rem}{Remark}
\newtheorem{assumpt}{Assumption}

\newtheorem{defn}{Definition}

\usepackage{lastpage}
\jmlrheading{xx}{2024}{1-\pageref{LastPage}}{09/23; Revised \textcolor{black}{XX}/24}{xx/xx}{23-xxxx}{Florian Brück, Jean-David Fermanian and Aleksey Min}

\ShortHeadings{Distribution free MMD tests for model selection}{Brück, Fermanian and Min}
\firstpageno{1}

\begin{document}

\title{Distribution free tests for model selection based on maximum mean discrepancy  with estimated parameters}

\author{\name Florian Brück \email florian.brueck@unige.ch \\
       \addr Research Institute for Statistics and Information Science\\
       University of Geneva, Switzerland\\
       \AND
       \name Jean-David Fermanian \email jean-david.fermanian@ensae.fr \\
       \addr Ensae-CREST, France\\
       \AND
       \name Aleksey Min \email min@tum.de \\
       \addr Department of Mathematics\\
       Technical University of Munich, Germany\\
       }

\editor{Kenji Fukumizu}

\maketitle

\begin{abstract}
\textcolor{black}{There exist some testing procedures based on the maximum mean discrepancy (MMD) to address the challenge of model specification. 
However, they ignore the presence of estimated parameters in the case of composite null hypotheses.} In this paper, we first illustrate the effect of parameter estimation in model specification tests based on the MMD. Second, we propose simple model specification and model selection tests in the case of models with estimated parameters. All our tests are asymptotically standard normal under the null, even when the true underlying distribution belongs to the competing parametric families. 
A simulation study and a real data analysis illustrate the performance of our tests in terms of power and level.
\end{abstract}


\section{Introduction}\label{sec:intro}

Model selection is an ``umbrella term'' that refers to different important statistical problems. 
Any parametric or semiparametric model has to be theoretically validated, most often by statistical tests, 
inducing the  field of ``specification testing''.
In this work, we will consider models that specify the law of a Data Generating Process (DGP), i.e., we consider models of probability measures. 
Thus, a model, say $\Mc$, is associated with a family of probability measures 
$\Mc:=\{P_\alpha,\alpha \in \Theta_1\}$, where $\Theta_1$ denotes some set of parameters.
The true  underlying law of the DGP is denoted by $P$. 
The empirical law of a sample of $n$ i.i.d.\ observations from $P$ is denoted $\Pb_n$. 
To obtain consistent specification tests, it is necessary to measure the distance between a suggested model and the true underlying DGP. This is commonly done in terms of a semi-metric $\pi$ between probability measures.  
The distance between the model $\Mc$ and the law $P$ of the DGP is then defined as 
$$\pi(\Mc,P):=\inf_{\alpha\in \Theta_1} \pi(P_\alpha,P).$$
This definition allows to conduct a hypothesis test of
$$\Hc^{(\pi)}_{0,\Mc}: \pi(\Mc,P) =0,$$
which is a natural way of testing the hypothesis that the model for the DGP is correctly specified. 
Typically, the best-fitting probability measure(s) in the family $\Mc$ are defined by so-called pseudo-true parameter(s) (also called pseudo-true value(s)) $\alpha_\star$ that minimize the distance between the model $\Mc$ and the DGP, i.e., $\alpha_\star\in \arg\min_{\alpha\in \Theta_1} \pi(P_{\alpha},P).$
\textcolor{black}{In general, pseudo-true values are unknown and not unique. Assume that
one of them, still denoted $\as$, can be consistently estimated 
by a random sequence $(\an)_{n\geq 1}$. Thus $P_{\an}$ should be close to an optimal model in $\Mc$.}  
Further, $P$ is often also unknown and needs to be estimated via $\Pb_n$. Therefore, in practical applications, a consistent test of $\Hc^{(\pi)}_{0,\Mc}$ 
needs to be based on the asymptotic distribution of (an estimator of) $\pi( P_{\an},\Pb_n)$ under the null hypothesis $\Hc^{(\pi)}_{0,\Mc}$.

Since the seminal works of~\citet{durbin1973distribution,durbin1973weak}, numerous consistent specification tests have been proposed in the statistical literature, depending on the chosen distance $\pi$, but only a few of them manage composite assumptions. This is the case with the Cramer-von Mises and Kolmogorov distances \citep{pollard1980minimum}, the uniform distance between multivariate cdfs' \citep{beran1989stochastic}, the total variation distance or, equivalently, the $L_1$ distance between the underlying densities \citep{cao2005goodness} and a $L_2$-type distance between characteristic functions \citep{fan1997goodness}. Recently, the topic has regained attention, since testing composite assumptions based on distances that are popular in generative machine learning/statistics has become an important problem in applied machine learning/statistics, \textcolor{black}{notably through Wasserstein's distance \citep{hallin2021multivariate}, Kernel Stein Discrepancies \citep{liu2016kernelized,chwialkowski2016kernel} and the Maximum-Mean Discrepancy \citep{key2021composite}}.

When there exist several competing models, ``model selection'' rather means validating one of the competing model as ``the better one''. The aim is then to identify the model that is the closest to the DGP, a task that is also called ``model comparison''.
The seminal paper~\citet{Vuong1989} was the first to propose a general framework for this task. The author proposed to use the Kullback-Leibler (KL) divergence for model comparison: a
model $\Mc_1$ is preferred over a model $\Mc_2$ when its distance to the true model \textcolor{black}{is smaller}, i.e., when $\pi(\Mc_1,P)< \pi(\Mc_2,P)$ and $\pi$ is chosen as the Kullback-Leibler divergence.
In general, model selection between competing models is based on tests of the null hypothesis 
$$\Hc^{(\pi)}_{0,\Mc_1,\Mc_2}: \pi(\Mc_1,P)= \pi(\Mc_2,P). $$
In most circumstances, the competing models are parametric and may be misspecified, i.e., none of the models will satisfy $\pi(\Mc_i,P)=0$, $i\in \{1,2\}$. For instance, when $\Mc_1=\{P_{\alpha}, \alpha\in \Theta_1\}$ and $\Mc_2=\{Q_\beta, \beta\in \Theta_2\}$,
their pseudo-true values are defined as above by
\begin{equation*}
\as\in\argmin_{\alpha\in \Theta_1} \pi(P_{\alpha},P),\;\text{and}\; \bs\in\argmin_{\beta\in \Theta_2} \pi(Q_\beta,P)  .
\end{equation*}
Then, for a given couple $(\as,\bs)$ of pseudo-true values, $\Hc^{(\pi)}_{0,\Mc_1,\Mc_2}$ may be rewritten as 
$$\Hc^{(\pi)}_{0,\Mc_1,\Mc_2}: \pi(P_{\as},P)=\pi(Q_{\bs},P).$$
Once some estimated pseudo-true values $\an$ and $\bn$ and a sample from $P$ are available, 
a model selection test is typically based on the random quantity
$$\Tc_n:=\hat\pi(P_\an,\Pb_n)-\hat\pi(Q_\bn,\Pb_n),$$
where $\hat\pi (P_1,P_2)$ denotes an estimator of $\pi(P_1,P_2)$ for any couple of probability measures $(P_1,P_2)$.
Unfortunately, the asymptotic distribution of $\Tc_n$ is generally complex. 
Indeed, the randomness of the estimated parameters of both models matters to state the limiting law of $\Tc_n$ under $\Hc^{(\pi)}_{0,\Mc_1,\Mc_2}$. 
This is particularly the case for overlapping models, i.e.\ for models $\Mc_1$ and $\Mc_2$ for which $P_\as=Q_\bs$, a situation that often cannot be excluded. For example, this is the case for the KL divergence, where $\Tc_n$ is no longer distribution free~\citep[Theorem 3.3]{Vuong1989}. 
Therefore, the methodology of~\citet{Vuong1989} requires pre-testing to know whether the competing models are overlapping, a feature that is generally considered as a drawback. Some variations of the traditional test of \cite{Vuong1989} have been proposed in the literature to avoid pre-testing and to keep its asymptotic standard normal distribution. See, notably,~\citet{shi2015nondegenerate},~\cite{schennach2017simple} and the references therein.

The aim of this paper is to provide a solution for ``specification testing'' and ``model comparison'' when $\pi$ is the Maximum Mean Discrepancy (MMD). \
The MMD has become a very popular distance in machine learning and statistics since its introduction in \cite{smola2007hilbertembeddingdistr}. The MMD can be easily estimated even in high-dimensions and it has a nice interpretation in terms of embeddings of probability measures in Reproducing Kernel Hilbert Spaces (RKHS). See \cite{muandetreview2017} for a more recent review of the topic. Using the MMD as a discrepancy measure has been proven useful in many statistical applications including robust inference \citep{cherief2022finite,alquier2024universal}, change-point detection \citep{arlot2019kernel}, goodness-of-fit tests \citep{gretton2012kernel,bounliphone2016test} or copula estimation \citep{alquier2023estimation}. Further, is has been applied in generative machine learning ~\citep{dziugaite2015,li2015generative,lichangchengyangpoczos2017,sutherland2017,zhou2020} and in a variety of other domains including \textcolor{black}{ Transfer Learning \citep{longzhuwangjordan2017}, the computation of functions of random variables \citep{scholkopf2015}, Bayesian statistics \citep{fukumizusonggretton2013,parkjittikrumsejdinovic2016,cheriefalquier2020}, clustering \citep{jegelkagrettonscholkopf2009}, conditional independence testing \citep{zhangpetersjanzigscholkopf2011}, adaptive MCMC methods \citep{sejdinovicstrathman2014}, causal inference \citep{lopezpazmuandetscholkop2015}, dynamical systems \citep{songsmoafumumizu2009}  and in the construction of sampling algorithms \citep{hofert,bruck2024}, among many others}. \textcolor{black}{In general, the computational complexity of the MMD is quadratic in the number of data points, which can be restrictive with large datasets. Nonetheless, several recent research papers have provided computationally efficient procedures: see, e.g., the linear statistic in \citet{gretton2012kernel} or \citet{chatalicschreuder2022}.} 

The universal attractive feature of the MMD is that it can be easily defined \textcolor{black}{in general topological spaces - particularly high-dimensional spaces -} and that it can be easily estimated by sampling from the underlying probability measures. 
Let us recall the basics about the MMD, following the approach of~\citet{smola2007hilbertembeddingdistr}.
Consider some probability measures defined on some topological space $\Sc$ \textcolor{black}{equipped with its corresponding Borel sigma-algebra $\mathcal{A}$.}
Instead of comparing these probability measures directly in the space of probability measures, they may be mapped into an RKHS of real-valued functions $\Hc$ defined on $\Sc$. The latter space is associated with a symmetric and 
positive definite \textcolor{black}{function} $k: \Sc\times\Sc\to\R$, called kernel, which may be chosen by practitioners.
For many kernels $k$ (called ``characteristic''), these mappings are injective and the MMD distance between probability measures is then defined as the distance between their respective embeddings in the space of functions $\Hc$. More specifically, consider some random element $X$ in a topological space $(\Sc,\mathcal{A})$, whose law is $P$.
The embedding of the probability measure $P$ into $\Hc$ is given by the map $P \mapsto \e_{P}\lk k(\cdot,X) \rk=:\mu_{P}$. The latter map
implicitly depends on the kernel $k$, but this dependence is suppressed to lighten notations. 
When $k$ is characteristic, $P\mapsto\mu_P$ is injective and we have $P_1\not= P_2$ iff $\mu_{P_1}\not=\mu_{P_2}$.
Thus, the MMD defines a distance on the space of probability measures on $\Sc$ via $ \mmd (P_1,P_2):=  \Vert \mu_{P_1}-\mu_{P_2}\Vert_{\Hc} $.
One can deduce that
\begin{align*}
    \mmd^2 (P_1,P_2)=  \e_{X,X'\sim P_1}\Big[ \e_{Y,Y'\sim P_2} \big[ k(X,X')-2k(X,Y)+k(Y,Y') \big]\Big]    . 
\end{align*}
Therefore, the computation of $\mmd(P_1,P_2)$ does not involve any operations in the Hilbert space $\Hc$ but is solely dependent on expectations of known functionals w.r.t.\ $P_1$ \textcolor{black}{and} $P_2$. Moreover, given an i.i.d.\ sample $(X_1,X_2,\ldots,X_n)$ from $P_1$ and $(Y_1,Y_2,\ldots,Y_n)$ from $P_2$, $\mmd^2(P_1,P_2)$ can be empirically estimated by
\begin{align}
    \hmmd(P_1,P_2):= \frac{1}{n(n-1)} \sumijn \big\{ k(X_i,X_j)-2k(X_i,Y_j)+k(Y_i,Y_j) \big\}. \label{defhmmd}
\end{align}
The latter unbiased estimator of $\mmd^2(P_1,P_2)$ is a $U$-statistic of degree two associated with the symmetric \textcolor{black}{map} 
\begin{align}
h\big( (x_1,y_1), (x_2,y_2)\big) :=k(x_1,x_2)-k(x_1,y_2)-k(x_2,y_1)+k(y_1,y_2),
\label{defustatkernelmmd}
\end{align}
where $(x_j,y_j) \in \Sc^2, j\in \{1,2\}$.  
\textcolor{black}{Hereafter, such maps will be called ``$U$-statistic kernel'' (``$U$-kernel'', to be short), to be distinguished from the kernel $k$ itself.}
Clearly, $\hmmd(P_1,P_2)$ may be used to test whether or not $(X_1,X_2,\ldots,X_n)$ and $(Y_1,Y_2,\ldots,Y_n)$ follow the same underlying probability measure, 
i.e.\ to test the null hypothesis $\Hc_0: P_1=P_2$. A consistent test may be deduced from standard $U$-statistics theory \citep[Section 5]{serflingapproximationtheorems1980} by observing that, under $\Hc_0$,
\begin{equation*}
n\hmmd(P_1,P_2)\stackrel{\text{law}}{\longrightarrow} \sum_{i=1}^\infty \lambda_i\lc \mathcal{X}^2_i -2\rc, 
\label{complex_limiting_law}    
\end{equation*}
where $\mathcal{X}_i\sim \Nc(0,2)$ and the $\lambda_i$ denote the (possibly infinitely many) eigenvalues 
associated with the functional equation $\e\Big[ \big\{  k(X,y)-\mu_{P_1}(X)-\mu_{P_1}(y)+\e_Y [\mu_{P_1} (Y) ] \big\}\psi(X)\Big]=\lambda\psi(y)$ for every $y\in \Sc$ \citep{gretton2012kernel}. However, for almost every practical application, the computation/estimation of the eigenvalues $\lambda_i$ is very challenging or even impossible, which is a major limitation of a test for $\Hc_0$ based on $n\hmmd(P_1,P_2)$.

Our corresponding null hypothesis for model specification will be defined hereafter as
\begin{align}
\Hc_{0,\Mc}:\ \mmd(P_\as,P) =0,
    \label{H0_specific}
\end{align}
whereas the null hypothesis for model selection will be defined as
\begin{align}
\Hc_{0,\Mc_1,\Mc_2}:\ \mmd(P_\as,P)=\mmd(Q_\bs,P).
\label{H0_model_comp}
\end{align} 
\textcolor{black}{When some of our results are valid only under $\Hc_{0,\Mc}$ or $\Hc_{0,\Mc_1,\Mc_2}$, this will be explicitly specified.}
For notational convenience, the dependence of $\Hc_{0,\Mc}$ and $\Hc_{0,\Mc_1,\Mc_2}$ on $(\as,\bs)$ will remain implicit. \textcolor{black}{Moreover, from now on, and unless explicitly stated otherwise,} we allow that $\as$ and $\bs$ are some general unknown ``optimal'' parameters, not necessarily pseudo-true values, to keep the mathematical framework as general as possible.
\textcolor{black}{For example, $P_\as$ or $Q_\bs$ may be an ``optimal model'' because of some particular properties such as sparsity, ease of calibration, fairness, etc. In such circumstances, the unknown $\as$ (resp. $\bs$) are minimizing some loss function that may not be the MMD distance between $\Mc_{1}$ (resp. $\Mc_{2}$)  and $P$.}

The latter assumption $\Hc_{0,\Mc}$ may be seen as a standard zero assumption for two sample testing, as in~\citet{gretton2012kernel}. The main difficulty comes from the fact that $\as$ is unknown and has to be estimated.
In principle, \textcolor{black}{when a sequence $(\an)_{n\geq 1}$ tends to $\as$ in probability}, resorting to the asymptotic distribution of $\hmmd(P_\an,P)$ allows to conduct specification testing, i.e., testing the null hypothesis $\Hc_{0,\Mc}: \mmd(P_\as,P)=0$. 
\textcolor{black}{Unfortunately}, the limiting law of $\hmmd(P_\an,P)$ might be more complex than that of $\hmmd(P_\as,P)$. 
Moreover, it is usually impossible to generate independent samples from $P_\an$ without resorting to the sample from $P$, since $\an$ is usually estimated on the sample from $P$, adding another layer of complexity. 
Nonetheless,~\citet{key2021composite} recently showed that some \textcolor{black}{consistent} test of $\Hc_{0,\Mc}$ can be obtained 
by bootstrapping $\mmd^2(P_\an,\Pb_n)$.  

For model comparison, a similar phenomenon as for the test of~\citet{Vuong1989} occurs, even if $\as$ and $\bs$ were known: when $P_\as=P=Q_\bs$, the rate of convergence of $\hmmd(P_\as,P)-\hmmd(Q_\bs,P)$ is $n^{-1}$; but, otherwise, the latter rate of convergence 
is $n^{-1/2}$, as noticed in \citet{bounliphone2016test}. Thus, to built a consistent test of $\Hc_{0,\Mc_1,\Mc_2}$, one needs to deal with different rates of convergences, depending on whether or not $P_\as=P=Q_\bs$, a situation that is clearly unknown a priori. A two-step procedure with a pre-test of $P_\as=P=Q_\bs$ seems natural, but it is undesirable due to the difficulty of determining the critical values of $n\big\{\hmmd(P_\an,P)-\hmmd(Q_\bn,P)\big\}$ and multiple testing issues.

\mds 

In this paper, we provide a solution to the latter problem of model specification and model selection tests based on the MMD. The contributions of our paper can be summarized as follows. First, we investigate the influence of parameter estimation on the asymptotic distribution of the estimator $\hmmd(P_\an,P)$. We show that it is necessary to account for the influence of parameter estimation 
\textcolor{black}{to test $\Hc_{0,\Mc}$}
: see Section \ref{illustrative_example} for an illustrative example. Second, as we find that the asymptotic distribution of $n\hmmd(P_\an,P)$ under the null is rather complex to handle in practical applications, we provide new asymptotically distribution free approaches for specification testing and model comparison, as defined above (Equations~(\ref{H0_specific}) and~(\ref{H0_model_comp})). 
In particular, we show that our novel test statistics converge to a standard normal distribution under the null, allowing straightforward calculations of critical values.  Our ideas stem from a generalization of the sample splitting approach introduced in \citet{schennach2017simple}, which was used to obtain a standard normal distributed test statistic for Vuong's likelihood ratio test with estimated parameters \citep{Vuong1989}. Adapting their core ideas, this allows to propose test statistics that are relatively simple to compute and whose asymptotic standard normal law is not influenced by parameter estimation. 

The paper is organized as follows. Our test statistics for model selection are introduced in Section~\ref{test_stat_motivation}. In Section \ref{sec:asympdistteststat}, we formalize the mathematical framework and state the asymptotic distribution of our test statistic for model specification. A similar structure is followed in Section \ref{sec:modelselection}  to manage model comparison tests. Section \ref{sec:simstudy} contains a short simulation study to  illustrate the empirical  performance of the proposed tests.
\textcolor{black}{We apply our methodology to stock returns data in Section~\ref{sec:empirics}.}
Section~\ref{sec:conclusion} summarizes the results and sketches further extensions of the framework.
Most technical details, particularly the proofs of the main theorems, have been postponed to appendices.

\section{Non-degenerate MMD-based tests for model selection}
\label{test_stat_motivation}

At this stage, a reader may wonder why our model specification test will not be based solely on the test statistic $\hmmd(P_\an,P)$, which is an estimate of $\hmmd(P_\as,P)$ whose asymptotic law has been investigated in \cite{grettonkernelmethodtwosampleproblem}.
To illustrate the technical difficulties in working with $\hmmd(P_\an,P)$, assume that the estimator $\an$ of a value $\as$ is obtained from the sample drawn under $P$, which is the natural setup when $P_\an$ is a model for $P$. Then, any sample from $P_\an$ is inherently dependent due to the common $\an$  and it is not independent of the initial sample from $P$. Note that \cite{grettonkernelmethodtwosampleproblem} requires i.i.d.\ samples from $P_\an$ to compute $\hmmd(P_\an,P)$, which is problematic.
This issue has first been noticed, but not further investigated, by \cite{lloyd2015statistical}. However, since the dependence between the samples from $P_\an$ and $P$ deteriorates with increasing $n$, one might hope that $n\hmmd(P_\an,P)$ has the same asymptotic law as $n\hmmd(P_\as,P)$. Unfortunately, the following example illustrates that this is not the case. Furthermore, it shows that the asymptotic distribution of $n\hmmd(P_\an,P)$ is significantly more complicated than the asymptotic distribution of $n\hmmd(P_\as,P)$. Therefore, we later introduce novel distribution free test statistics when conducting model specification \textcolor{black}{or} comparison tests based on the MMD.

\subsection{An illustrative example}
\label{illustrative_example}

\textcolor{black}{On a probability space $(\Omega,\mathcal{B},\Pb)$, consider the random variable} $X\sim P= \mathcal{N} \left(0, 1\right)$. Let $(X_i)_{i=1,\ldots,n}$ be an i.i.d.\ sample drawn from $P$, the law of the DGP.
Define a parametric model for the law of $X$ by $\Mc:=\{\Nc(\alpha, 1),\ \alpha\in \R\}$.
For a given parameter $\alpha$, set 
$$
Y_{i}(\alpha):= Y_i+  \alpha
 \sim P_\alpha,\;\; \text{where } \,Y_i\overset{i.i.d.}{\sim} \mathcal{N} \left(0,1 \right),\;\; i \in\{1,\ldots,n\}.
$$
\textcolor{black}{Select $\as=0$} and the ``optimal model'' $\Nc(\as, 1)$ is identical to the law of the DGP. In practice, $\as$ is unknown and will be estimated by
$\alpha_n:=n^{-1}\sum_{i=1}^n X_{i}$, for example. Assume we want to test whether $P_{\as}=P$, or equivalently $\mmd(P_{\as},P)=0$, using the test statistic $n\hmmd(P_{\alpha_n},P)$. To this purpose, we choose the Gaussian kernel $k(x_1, x_2)= \exp\big(-( x_1-x_2)^2\big)$, which is characteristic. This is a convenient choice since any map $\alpha\mapsto h\big((X_i,Y_{i}(\alpha)), (X_j,Y_{j}(\alpha))\big)$, \textcolor{black}{as defined in~(\ref{defustatkernelmmd})}, is differentiable. Therefore, a second order Taylor expansions around $\as=0$ yields
\begin{eqnarray*}
    \lefteqn{n\hmmd(P_{\alpha_n},P) = n\hmmd(P_{\as},P)} \\
    & &+\,\sqrt{n}\,(\alpha_n-0)\frac{\sqrt{n}}{n(n-1)}\sumijn  \frac{\partial }{\partial \alpha} h\big((X_i,Y_i+\alpha), (X_j,Y_j+\alpha)\big)\vert_{\alpha=0}\\
    &&+\,\frac{n(\alpha_n-0)^2}{2}\frac{1}{n(n-1)}\sumijn \frac{\partial^2 }{\partial \alpha^2} h\big((X_i,Y_i+\alpha), (X_j,Y_j+\alpha)\big)\vert_{\alpha=0} +o_{\Pb}(n\alpha_n^2)\\
    &= &n\hmmd(P_{\as},P)+\sqrt{n}\alpha_n\sqrt{n}\tilde{U}_{1,n} +\frac{n\alpha_n^2}{2}\tilde{U}_{2,n} +o_{\Pb}(1),
\end{eqnarray*}
where $\Tilde{U}_{1,n}$ and $\Tilde{U}_{2,n}$  denote the $U$-statistics of degree two corresponding to the first and second order derivatives of $\alpha\mapsto h\big((X_i,Y_{i}+\alpha), (X_j,Y_{j}+\alpha)\big)$ at $\as$. 
Thus, $n\hmmd(P_{\alpha_n},P)$ can be decomposed into the sum of the ``usual'' test statistic for a known and fixed $\as$ plus the random term $E_n:=\sqrt{n}\alpha_n\sqrt{n}\widetilde{U}_{1,n} + n\alpha_n^2\widetilde{U}_{2,n}/2$ that can be attributed to the noise created by the estimation of $\as$. Obviously, $n\hmmd(P_\an,P)$ has the same limiting law as $n\hmmd(P_\as,P)$ if and only if $E_n=o_{\Pb}(1)$.
Since $\sqrt{n}\an\wcv \Nc(0,1)$, $E_n=o_{\Pb}(1)$ if and only if $\sqrt{n}\tilde{U}_{1,n}+\sqrt{n}\an\tilde{U}_{2,n}/2=o_{\Pb}(1)$. By standard results of $U$-statistics theory and simple calculations, $\sqrt{n}\widetilde{U}_{1,n}$ weakly tends to a  $\Nc(0,\sigma_1^2)$ random variable with $\sigma_1^2=8\sqrt{3}/(63\sqrt{7})$ and
$\Tilde{U}_{2,n}\to 16/(5\sqrt{5})$ a.s. This implies $\sqrt{n}\tilde{U}_{1,n}+\sqrt{n}\an\tilde{U}_{2,n}/2\not=o_{\Pb}(1)$.
Therefore, the limiting law of $n\hmmd(P_{\alpha_n},P)$ is not equal to the limiting law of $n\hmmd(P_{\alpha\star},P)$, due to the influence of parameter estimation. 


\subsection{New asymptotically distribution free test statistics}
Let us first introduce some notation.    For sequences of random elements $(X_1,\ldots, X_n)$ and $(Y_1,\ldots, Y_n)$, we denote $[\Xbm]_{i:j}:=\big( X_i, X_{i+1}, \ldots, X_j\big)$, $[\Ybm]_{i:j}:=\big( Y_i, Y_{i+1}, \ldots, Y_j\big)$ and $[\Xbm,\Ybm]_{i:j}:=\big( (X_i,Y_i), (X_{i+1},Y_{i+1}),$ $\ldots,$ $ (X_j,Y_j)\big)$ for $1\leq i<j \leq n$. For $x_1, \ldots, x_n$ and $y_1, \ldots, y_n$, we similarly denote $[\xbm]_{i:j}:=\big( x_i, x_{i+1}, \ldots, x_j\big)$, $[\ybm]_{i:j}:=\big( y_i, y_{i+1}, \ldots, y_j\big)$ and $[\xbm,\ybm]_{i:j}:=\big( (x_i,y_i), (x_{i+1},y_{i+1}),$ $\ldots,$ $(x_j,y_j)\big)$ for  $1\leq i<j \leq n$. In the same manner, $\Xbm_{i_1, i_2, \ldots, i_k}$, $\Ybm_{i_1, i_2, \ldots, i_k}$ and $[\Xbm, \Ybm]_{i_1, i_2, \ldots, i_k}$ are defined, where the index set indicates the components of  concatenated variables.

Our new test statistics for model specification and model comparison will be based on a weighted combination of two test statistics, both being a potential candidate for this task. For the sake of simplicity, let us start by comparing two fixed probability measures $P_1$ and $P_2$. The first ingredient of our new test statistics is $\hmmd(P_1,P_2)$ as introduced in (\ref{defhmmd}).
Besides $\hmmd (P_1,P_2)$, a test of $\Hc_0: P_1=P_2$ may also be based on the $U$-statistic
\begin{eqnarray}
 \lefteqn{   \hmmdq(P_1,P_2):=\frac{1}{n/2(n/2-1)} \sum_{\substack{i,j=1 \\ i\not=j}}^{n/2} \big\{
    k(X_{2i-1},X_{2j-1})-k(Y_{2j},X_{2i})    \nonumber }\\
    &  &-\,  k(Y_{2i},X_{2j}) + k(Y_{2i-1},Y_{2j-1}) \big\} 
     = \frac{1}{n/2(n/2-1)} \sum_{\substack{i,j=1 \\ i\not=j}}^{n/2} q\big([\Xbm, \Ybm]_{2i-1,2i}\, , [\Xbm, \Ybm]_{2j-1,2j} \big),
    \nonumber \label{defhmmdnondeg}
\end{eqnarray}
introducing the $U$-statistic kernel
\begin{align*}
  q\big([\xbm, \ybm]_{1:2}\, ,  [\xbm, \ybm]_{3:4}\big):=k(x_1,x_3)-k(x_4,y_2)-k(x_2,y_4)+k(y_1,y_3).   \label{defnondegustatkernel}
\end{align*}
Hereafter, we will assume that $n$ is even for simplicity. Note that $\hmmdq$ is a $U$-statistic based on the sample $\big(  [\Xbm,\Ybm]_{1:2}, [\Xbm,\Ybm]_{3:4},\ldots,[\Xbm,\Ybm]_{(n-1):n}\big)$ from $P_1\otimes P_2\otimes P_1\otimes P_2$ of size $n/2$. It is easy to check that $\hmmdq(P_1,P_2)$ is an unbiased estimator of
$\mmd^2 (P_1,P_2)$.
By standard $U$-statistic arguments, we have
$$\sqrt{n}\big\{\hmmdq(P_1,P_2) - \mmd^2(P_1,P_2)\big\}\wcv \Nc(0,\sigma_q^2),$$
with $\sigma^2_q>0$, essentially if and only if $P_1$ or $P_2$ is not a Dirac (see (\ref{calc_sigma_q}) below). Therefore, a test of $\Hc_0:P_1=P_2$ may always be based on $\hmmdq(P_1,P_2)$. Unfortunately, it may result in a power loss compared to a test based on $\hmmd(P_1,P_2)$ since $\hmmdq(P_1,P_2)$ does not use all pairs of observations which are available from a sample of size $n$
drawn from $P_1\otimes P_2$. It should be noted that a test of $\Hc_0$ based on $\hmmdq(P_1,P_2)$ is similar to the  idea of \cite{shekhar2022}, who have recently proposed a $\mmd$-based test statistic of $\Hc_0$ that is essentially a non-degenerate two-sample $U$-statistic. Similarly to the expected behavior for $\hmmdq(P_1,P_2)$, \cite{shekhar2022} observe a power loss of their test statistic in comparison to a test based on $\hmmd(P_1,P_2)$. Therefore, it is desirable to approximately keep the power of a test based on $\hmmd (P_1,P_2)$, while resorting to the critical values of a normal distribution under the null hypothesis.

To this purpose, we will consider a linear combination of $\hmmd(P_1,P_2)$ and
$\hmmdq(P_1,P_2)$ in the spirit of~\cite{schennach2017simple}: introduce some (possibly random and data dependent) weights $\epsilon_n>0$ and define the test statistic
\begin{equation*}
    \hmmden(P_1,P_2):=\hmmd(P_1,P_2)+\epsilon_n\hmmdq(P_1,P_2). \label{defweightedustat}
\end{equation*}

If $\epsilon_n:=\epsilon>0$ is a constant, it is obvious that $\sqrt{n}\,\hmmden (P_1,P_2)  \wcv\epsilon \, \Nc(0,\sigma_q^2)$ under $\Hc_0$, since $\sqrt{n}\,\hmmd(P_1,P_2)$ \textcolor{black}{tends to zero in probability} when $P_1=P_2$. However, the choice $\epsilon_n=\epsilon>0$ may lead to a power loss, similar to a test based on $\sqrt{n}\,\hmmdq (P_1,P_2)$. Therefore, we impose that $\epsilon_n$ tends to zero in probability hereafter.

When $P_1$ and $P_2$ are two known probability measures, we will prove that a test of $\Hc_0$ can be conducted via the test statistic
\begin{align}
    \Tc_n(P_1,P_2):=\frac{\sqrt{n} \, \hmmden(P_1,P_2)}{\hat{\sigma}_n},  \label{defteststat}
\end{align}
where $\hat{\sigma}_n^2:=\hat{\sigma}_n^2(\epsilon_n,P_1,P_2)$ denotes an estimator of the asymptotic variance of the numerator of $ \Tc_n(P_1,P_2)$. With  our choice of $\hat{\sigma}_n$, which is described in \textcolor{black}{Section \ref{sec:asymp_var_estim} and \ref{case_differentiable_F}}, and under $\Hc_0$,  the test statistic  $ \Tc_n(P_1,P_2)$ converges weakly to $\Nc(0,1)$. Imposing  $\epsilon_n\to 0$ allows to approximately keep the power of $n\hmmd(P_1,P_2)$ under the alternative and also to avoid the problem of computing the critical values of the asymptotic law of $n\hmmd(P_1, P_2)$.

Instead of considering a couple of fixed probability measures $(P_1,P_2)$ we now consider an underlying parametric model $\Mc:=\{P_\alpha,\; \alpha \in \Theta_1\}$ for the DGP, as in Section~\ref{sec:intro}.
With the same notations as above, a ``specification test'' of $\Hc_{0,\Mc}: \mmd(P_\as,P)=0$  can be based on the statistic
\begin{align}
    \Tc_n(\Mc,P):=\sqrt{n}\,\frac{\hmmden(P_\an,P)}{\hat{\sigma}_n},  \label{defteststat_specif}
\end{align}
for some sequence of parameters $(\an)_{n\geq 1}$ that weakly converges to $\as$ at rate $n^{-1/2}$.
Typically, the estimated parameters $\an$ are obtained from an i.i.d.\ sample $(X_1,X_2,\ldots,X_n)$ from $P$. To mimic the previous situation and to calculate $\Tc_n(\Mc,P)$, we have to draw an i.i.d.\ sample $\big(Y_1(\an),Y_2(\an),\ldots,Y_n(\an)\big)$ from $P_\an$.
However, such a sample cannot be i.i.d.\ due to the common dependence on $\an$. Moreover, it cannot be independent of $(X_1,X_2,\ldots,X_n)$ when $\an$ is deduced from the latter sample. Additionally, the denominator $\hat{\sigma}_n$ in~(\ref{defteststat_specif}) depends on $\eps_n$ and needs to be calculated from the sample $\big(X_i,Y_i(\an)\big)_{i=1,\ldots,n}$.
Nevertheless, it is shown in Section~\ref{sec:asympdistteststat}  that $\Tc_n(\Mc,P)$ still converges to a standard normal distribution under $\Hc_{0,\Mc}$, regardless of the technical problems induced by $\an$ and $\hat{\sigma}_n$.

In the case of two competing models $\Mc_1$ and $\Mc_2$ for $P$, one may conduct a model comparison based  on the difference $\mmd^2(P_\as,P)-\mmd^2(Q_\bs,P)$, for two pseudo-true values $\as$ and $\bs$, to judge which family of probability measures is closest to the true underlying model $P$.
The idea of testing the null hypothesis $\mmd(P_\as,P)=\mmd(Q_\bs,P)$ based on $\sqrt{n}\big\{ \hmmd(P_\as,P)-\hmmd(Q_\bs,P) \big\}$ was first proposed by \cite{bounliphone2016test}. However, the latter authors only considered fixed competing probability measures and excluded the ``degenerate'' situation $P_\as=P=Q_\bs$.
In the case of parametric models for $P$, some estimates $\an$ and $\bn$ of $\as$ and $\bs$ are usually obtained from $(X_1,X_2,\ldots,X_n)$. Again, the framework of  \cite{bounliphone2016test} requires access to independent i.i.d.\ samples from $P_\an$ and $Q_\bn$, which also need to be independent of the sample from $P$. Obviously, this is impossible when $\an$ and $\bn$ are evaluated on the sample $(X_i)_{i=1,\ldots,n}$. Note that estimation of $\an$ and $\bn$ on separate hold-out sets from another $P$-sample does not resolve the issue, since the samples from $P_\an$ and $Q_\bn$ are inherently dependent due to the common dependence w.r.t. the estimated parameters. Therefore, a direct application of the test statistic proposed in \cite{bounliphone2016test} to model comparison problems is not mathematically justified. Moreover, since it has to be excluded that $P=P_\as=Q_\bs$, their test is not applicable to every modeling problem as the assumption $P=P_\as=Q_\bs$ may be reasonable for some competing generative machine learning models such as MMD GANs \citep{li2015generative,dziugaite2015}.

In our paper, we rectify these shortcomings by introducing a test for $\Hc_{0,\Mc_1,\Mc_2}: \mmd(P_\as,P)=\mmd(Q_\bs,P)$, based on
\begin{align}
    \Tc_n(\Mc_1,\Mc_2,P):= \sqrt{n}\frac{\hmmden(P_\an,P)-\hmmden(Q_\bn,P)}{\hat{\tau}_n},  \label{defteststatcomp}
\end{align}
where $\hat{\tau}_n^2=\hat{\tau}_n^2(\epsilon_n,P_\an,Q_\bn,P)$ denotes a natural estimator of the asymptotic variance of $\sqrt{n} \big\{ \hmmden(P_\an,P)-\hmmden(Q_\bn,P)\big\}$, which is specified in Section \ref{Notations_assumptions_comparison}. \textcolor{black}{Moreover, $(\an)_{n\geq 1}$ (resp.\ $(\bn)_{n\geq 1}$) denotes a sequences of parameters which weakly converges to $\as$ (resp.\ $\bs)$ at rate $n^{-1/2}$.} We prove that $\Tc_n(\Mc_1,\Mc_2,P) \wcv \Nc(0,1)$
under $\Hc_{0,\Mc_1,\Mc_2}$, even if $P=P_\as=Q_\bs$ and regardless of the dependence induced by $\an$ and $\bn$.

\section{Asymptotic behavior of MMD-based specification tests}
\label{sec:asympdistteststat}

Here, we state the asymptotic normality of our test statistic for model specification $\Tc_n(\Mc,P)$.

\subsection{Fundamental regularity conditions on $\Mc$, $k$ and $\an$} 
\label{Notations_assumptions}

Let us formalize our mathematical framework. Hereafter, we will assume that the sample space $\Sc$ is some topological space equipped with its Borel sigma-algebra. Due to \textcolor{black}{the Moore-Aronszajn Theorem \citep{aronszajn1950theory}}, there exists a unique RKHS $\Hc$ of real-valued functions on $\Sc$ that is associated with our kernel $k:\Sc\times \Sc \to\R$. It can be proved that $\Hc=\overline{\text{span} \{ k(x,\cdot)\, | \, x\in \Sc\}}$ where the closure is taken w.r.t.\ the RKHS norm.
\textcolor{black}{For instance, when $\Sc=\R$, the RKHS associated with the popular Gaussian kernel $k(x,y)=\exp(-(x-y)^2/\sigma^2)$ is the space of functions $f:x\mapsto \exp(-x^2/\sigma^2) \sum_{j=0}^{+\infty} v_j x^j$, for some coefficients $(v_j)_{j\geq 0}$ that satisfy $\sum_{j\geq 0} j! \sigma^{2j} v^2_{j}/2^j<\infty$. The scalar product of two elements  
$f(x)=\exp(-x^2/\sigma^2) \sum_{j=0}^{+\infty} v_j x^j$ and 
$g(x)=\exp(-x^2/\sigma^2) \sum_{j=0}^{+\infty} w_j x^j$ in $\Hc$ is 
then $<f,g>=\sum_{j\geq 0} j! \sigma^{2j} v_j w_j/2^j$  \citep[Theorem 1]{minh2010some}. 
See other examples of RKHS $\Hc$ in \citet[Chapter 7]{berlinet2011reproducing}}.
Let $(X_1,X_2,\ldots)$ denote an i.i.d.\ sequence drawn from the law $P$ of the DGP.
To state our results, we will need several conditions of regularity.

\begin{assumpt}
\label{A_characteristic} The kernel $k$ is \textcolor{black}{a measurable and bounded map} from $\Sc\times \Sc$ to $\R$. Moreover, it is characteristic: the map $\mu\mapsto \int_{\Sc} k(\cdot,x)\, \textcolor{black}{\mathrm{d}\mu(x)}$ from the space of Borel probability measures on $\Sc$ to $\mathcal{H}$ is injective.
\end{assumpt}

\textcolor{black}{
Note that the integral $\int_{\Sc} k(\cdot,x)\, \mathrm{d}\mu(x)$ has to be interpreted as a Bochner integral; see, e.g., \citet[Chapter 1]{dinculeanu2000vector}.  
Moreover, the boundedness of $k$ implies that the mean embedding is well-defined for any probability measure: see Section 3.1 in \cite{muandetcondmoment2020}, for instance.
Thus, Assumption~\ref{A_characteristic}} is sufficient to ensure that $\mmd(P_1,P_2)$ is a valid distance between two probability measures. For example, the Gaussian and Laplace kernels on $\R^d$ satisfy Assumption \ref{A_characteristic}: see \cite{fukumizu2007kernel} and~\cite{sriperumbudur2010hilbert} for a thorough account on kernels satisfying Assumption \ref{A_characteristic}. \textcolor{black}{The characteristic property is key in many applications of the MMD and has been studied in depth in the literature. It is closely related to, but different from, the concept of ``universality'' \citep{sriperumbudur2011universality,simon2018kernel}. It is required for some kernel measures of conditional dependence \citep{fukumizu2007kernel,fukumizu2009kernel}. Notably, \cite{nishiyama2016characteristic} stated the characteristicity of kernels defined by pdfs' of symmetric infinitely divisible distributions. Recently, \cite{szabo2018characteristic} studied the characteristic and universal properties of product kernels. See \citet[Section 3.3.1]{muandetreview2017} and the references therein too.}

Next, consider a parametric family of probability measures $\Mc:=\big\{ P_{\alpha};\alpha\in \Theta_1\big\}$ on $\Sc$.
\begin{assumpt}
\label{regular_model}
The space $\Theta_1$ is a compact subset of $\R^{p_\alpha}$ and its interior is non-empty.
There exists a topological space $\Uc$ equipped with its Borel sigma-algebra, a random element $U\sim P_U$ in $\Uc$ and a measurable map $F:\Uc\times \Theta_1 \rightarrow \Sc$ such that the law of $F(U;\alpha)$ is $P_\alpha$ for
any $\alpha\in \Theta_1$.
For a given parameter $\as$ that belongs to the interior of $\Theta_1$,
the map $ \alpha\mapsto \mmd^2(P_\alpha,P)$ is twice-continuously differentiable in a neighborhood of $\as$. Further, the random variable $\int_{\textcolor{black}{\Uc}}
k\big( X, F(u;\as) \big)\,\mathrm{d}P_U(u)$ is not constant a.e.\ \textcolor{black}{and $\text{supp}(P)\cap \text{supp}(P_\as)\not=\emptyset$.}
\end{assumpt}

Essentially, this assumption ensures that the parametrization of $\Mc$ is sufficiently regular. First, $P_\alpha$ smoothly varies w.r.t.\ $\alpha$ for the $\mmd$. Second, there exists a convenient way of simulating from the model $P_\alpha$ for every $\alpha\in\Theta_1$. In particular, Assumption~\ref{regular_model} ensures that a single i.i.d.\ sequence of random elements $(U_1,U_2,\ldots)$, $U_i\sim P_U$, is sufficient to obtain an i.i.d.\ sequence
$\big(F(U_1;\alpha),F(U_2;\alpha),\ldots\big)$ from $P_\alpha$, for every $\alpha\in\Theta_1$. Thus, from now on, we will assume that we have an i.i.d.\ sequence $(U_1,U_2,\ldots)$ from $P_U$ at hand, which is also assumed to be independent of $(X_1,X_2,\ldots)$. Furthermore, as is standard in mathematical statistics, we assume that the two independent i.i.d.\ sequences are induced by a common abstract probability space $(\Omega,\mathcal{B},\Pb)$. If not indicated otherwise, expectations $\e\lk\cdot\rk$ are always taken w.r.t.\ $\Pb$. \textcolor{black}{Moreover, $W_n=o_\Pb(1)$ (resp. $W_n=O_\Pb(1)$) means that a sequence of random elements $(W_n)_{n\geq 1}$ tends to zero in probability (resp. is bounded in probability) in the probability space $(\Omega,\mathcal{B},\Pb)$.} The last condition of Assumption \ref{regular_model} is very mild and of purely technical nature. It ensures that we are not working with constant random variables later.

Every model $\Mc$ can be defined in terms of the couple $(F,U)$ by setting $\Mc:=\{P_\alpha:=\text{Law}\big(F(U;\alpha)\big), \alpha \in \Theta_1\}$. 
Defining a model by a class of probability measures in this way is common when working with generative (also called simulation-based) models, which are very popular in machine learning and which often also implicitly appear in classical statistics\textcolor{black}{: see, e.g.,\ \cite{bond2021} for a review of the topic and \citet{dziugaite2015,li2015generative,lichangchengyangpoczos2017,sutherland2017,zhou2020} for generative models based on the MMD}.
 The map $F$ will therefore be called a generating function of the model $\Mc$. 
 To illustrate the role of $U$ and $F$, assume that $P_\alpha$ is a parametric family of probability measures on the real line. Then $U$ could be chosen as a uniformly distributed random variable on $(0,1)$ and
$F(\cdot;\alpha)$ could denote the inverse-quantile function of $P_\alpha$. Nevertheless, there might exist several tuples $(F,U)$ to describe the same model $\Mc$ and not every representation of $\Mc$ might satisfy Assumption \ref{regular_model}.

At this stage, we have not specified what we mean by ``optimal'' concerning $\as$. Formally, $\as$ could be arbitrarily chosen,
even if, in practice, this value is most often the minimizer of some distance between the family $(P_\alpha)_{\alpha\in\Theta_1}$ and the true probability measure $P$. In the latter case, we called $\as$ a pseudo-true value (see Section~\ref{sec:intro}). For the moment, we keep the discussion as general as possible: we do not impose that $\as$ is a pseudo-true value, even if this is implicitly implied by $\Hc_{0,\Mc}$.
Note that we neither require to specify the statistical method of inference for $\as$ by $\an$ nor which data is used to calculate $\an$ (full sample, sample splitting or overlapping).

To simplify the notations in the following, define
the functions
 \begin{eqnarray}
 h\big([\xbm, \ubm]_{1:2};\alpha\big)  &:=&
 h\big((x_1,F(u_1;\alpha)), (x_2,F(u_2;\alpha))\big)  \nonumber  \\
  &=& k(x_1,x_2)-k\big(x_1,F(u_2;\alpha)\big)-k\big(x_2,F(u_1;\alpha)\big)\nonumber \\
  &  &+\, k\big(F(u_1;\alpha),F(u_2;\alpha)\big),\;\; \text{and}
  \label{def_fct_h}
 \end{eqnarray}
\begin{eqnarray}
q\big([\xbm,\ubm]_{1:4};\alpha\big) &:=& k\big(x_1,x_3\big)-k\big(x_4,F(u_2;\alpha)\big) - k\big(x_2,F(u_4;\alpha) \big)
 \nonumber \\
&& +\, k\big(F(u_1;\alpha),F(u_3;\alpha)\big), \label{def_fct_q}
\end{eqnarray}
where the arguments $x_j$ and $u_k$ belong to $\Sc$ and $\Uc$ respectively.
Furthermore, define the family of maps
\begin{align}
    \tilde{h}(x,y;\alpha):\ \alpha\mapsto \e\Big[ h\big( (x, y) ,(X, F(U;\alpha))\big) \Big] ,\label{def_h_tilde}
\end{align}
which is indexed by $(x,y)\in \Sc\times \Sc$.
The gradient of any map $\alpha\mapsto H(\alpha)$ at some $\bar{\alpha}\in {\mathcolor{red}{\Theta_1}}$ will be denoted by $\nabla_\alpha H(\bar{\alpha})$, \textcolor{black}{and $\nabla_{\alpha^\top} H(\bar{\alpha})$ denotes its transpose (row) vector.
Concerning second order derivatives, $\nabla_{\alpha,\alpha^\top} H(\bar{\alpha})$ denotes the Hessian matrix of $H$ evaluated at $\alpha=\bar\alpha$.}

\subsection{Asymptotic variance estimation}
\label{sec:asymp_var_estim}
A key ingredient to obtain the asymptotic normality of $\Tc_n(\Mc,P)$ will be the choice of a suitable estimator of the asymptotic variance of $\sqrt{n}\hmmden (P_\an,P)$. In this section, we propose some intuitive estimators of the asymptotic variances of $\sqrt{n}\hmmd(P_\an,P)$ and $\sqrt{n}\hmmdq(P_\an,P)$, which we later combine to obtain a suitable estimator of the asymptotic variance of $\sqrt{n}\hmmden (P_\an,P)$.  To this purpose, recall that, for any fixed parameter $\alpha$, it is well-known \citep[p.\,192]{serflingapproximationtheorems1980} that the asymptotic variance of $\sqrt{n}\big\{\hmmd(P_\alpha,P) - \mmd^2(P_\alpha,P)\big\}$ is
$$\sigma_\alpha^2:=\var \Big( 2 \tilde{h} \big(X,F(U;\alpha);\alpha\big)\Big).$$

The corresponding empirical counterpart of $\sigma_\alpha^2$  is
\begin{align*}
\tilde\sigma^2_{\alpha} &:=\frac{4}{n}\sum_{i=1}^n \bigg\{ \frac{1}{n-1} \sum_{\substack{j=1\\i\not=j}}^n h\Big( \big(X_i,F(U_i;\alpha)\big),\big(X_j,F(U_j;\alpha)\big)\Big)- \hmmd(P_\alpha,P)\bigg\}^2 .
\end{align*}
Since our goal is to estimate $\sigma^2_\as$ but $\as$ is unknown, we replace $\as$ with $\an$ and define an estimator of $\sigma_\as^2$ by
\begin{align*}
\tilde\sigma^2_{\an} &:=\frac{4}{n}\sum_{i=1}^n \bigg\{ \frac{1}{n-1} \sum_{\substack{j=1\\i\not=j}}^n h\Big( \big(X_i,F(U_i;\an)\big),\big(X_j,F(U_j;\an)\big)\Big)- \hmmd(P_\an,P)\bigg\}^2 . \label{sigma_an}
\end{align*}
Similarly, the asymptotic variance of $\sqrt{n}\big\{\hmmdq(P_\alpha,P) - \mmd(P_\alpha,P) \big\}$ for any $\alpha\in \Theta_1$ is
$$\sigma^2_{q,\alpha}:=\var \big( 2\sqrt{2}\e_{[\Xbm, \Ubm]_{3:4}}\big[  q\big([\Xbm, \Ubm]_{1:4};\alpha\big)  \big] \Big).
$$
Analogously, this allows to define an estimator of $\sigma_{q,\as}^2$ via
\begin{equation*}
\tilde\sigma^2_{q,\an} :=\frac{16}{n}\sum_{i=1}^{n/2} \bigg\{ \frac{1}{n/2-1} \sum_{\substack{j=1\\i\not=j}}^{n/2} q\Big( [\Xbm, \Ubm]_{2i-1,2i,2j-1,2j} ;\an \Big)- \hmmd(P_\an,P)\bigg\}^2. \label{sigma_an_q}
\end{equation*}

One should observe that both estimators are always non-negative and of computational complexity $O(n^2)$. Further, $\sigma^2_\as=0$ when $P_\as=P$, but $\sigma_{q,\as}^2$ is always strictly positive under Assumption~\ref{regular_model}, which will later be important in our theoretical derivations. To see this, note that, if $\sigma_{q,\as}=0$ then the random element
\begin{eqnarray}
\lefteqn{  
\e_{[\Xbm, \Ubm]_{3:4}}\big[ q\big( [\Xbm, \Ubm]_{1:4};\as\big)  \big] =
\e\big[k(X_1,X_3) \big| X_1\big]  -\, \e\Big[k\big(X_4,F(U_2; \as)\big) \big| U_2\Big]    \nonumber  }\\
&&   - \e\Big[k\big(X_2,F(U_4; \as)\big) \big| X_2\Big] 
+ \e\Big[k\big(F(U_1; \as),F(U_3; \as)\big) \big| U_1\Big] \hspace{3cm}\label{calc_sigma_q} 
\end{eqnarray}
is constant almost surely.
Therefore, the four random variables on the r.h.s. of~(\ref{calc_sigma_q}) are constant almost surely.
In particular, this would imply that $\e\big[k\big(X,F(U; \as)\big) | X\big]$ is constant, a situation that has been excluded by Assumption~\ref{regular_model}.

\subsection{Differentiable generating functions \textcolor{black}{(model specification)}}
\label{case_differentiable_F}
In this section, we \textcolor{black}{treat the case of} $\alpha\mapsto F(u; \alpha)$ being twice differentiable for every $u\in \text{supp}(U)$.
Even if this assumption may appear relatively demanding (see the discussion in the beginning of Section \ref{general_specif_test} below),
the proofs of our results are significantly simpler and intuitive in this case. This is why we choose to first provide our results under the assumption of a \textcolor{black}{smooth} generating function, \textcolor{black}{once it is composed with the kernel}. Later, we generalize our results to the case of possibly non-differentiable maps. 

\begin{assumpt}
The maps $\alpha\mapsto k\big(x, F(u; \alpha)\big)$ and $\alpha\mapsto k\big(F(u; \alpha), F(\tilde{u}; \alpha)\big)$ are twice differentiable for every $x\in \Sc$ and $u,\tilde{u}\in\Uc$.
\label{cond_diff_k}
\end{assumpt}
As a consequence, the maps $h$ and $q$, as defined by~(\ref{def_fct_h}) and~(\ref{def_fct_q}) respectively, are twice differentiable w.r.t.\ $\alpha$. Note that Assumption \ref{cond_diff_k} is mainly an assumption on the smoothness of $F(u;\alpha)$, since most of the commonly used kernels $k$ are smooth functions.
Further, we require some usual conditions of regularity, expressed in terms of moments. These conditions are not only imposed on $h$ and $q$ but also on the auxiliary function
\begin{eqnarray}
g\big([\xbm, \ubm]_{1:3};\alpha\big)
&:=&\frac{4}{3}\big\{ h([\xbm, \ubm]_{1,2};\alpha)   h([\xbm, \ubm]_{1,3};\alpha) +  h([\xbm, \ubm]_{2,1};\alpha)   h([\xbm, \ubm]_{2,3};\alpha)  \nonumber \\
&  & +\, h ([\xbm, \ubm]_{3,2};\alpha)   h ([\xbm, \ubm]_{3,1};\alpha) \big\},  \label{def_g_alpha}
\end{eqnarray}
which frequently appears in the proofs.
For any $\delta >0$, let $B_\delta(\as)$ be an open ball in $\Theta_1$ that is centered at $\as$ and whose radius is $\delta$. \textcolor{black}{We require the following conditions of regularity.}

\begin{assumpt}
There exists a $\delta>0$ s.t.
$ \Eb\Big[ \sup_{\alpha_1 \in B_\delta(\as)} \| \nabla^2_{\alpha,\alpha^\intercal} h \big([\Xbm,\Ubm]_{1:2};\alpha_1\big) \| \Big] <\infty,$ and
$$ \Eb\Big[ \sup_{\alpha_1 \in B_\delta(\as)} \| \nabla^2_{\alpha,\alpha^\intercal} q \big([\Xbm,\Ubm]_{1:4};\alpha_1\big) \| \Big] +  \Eb\Big[ \sup_{\alpha_1 \in B_\delta(\as)} \| \nabla^2_{\alpha,\alpha^\intercal} g \big([\Xbm,\Ubm]_{1:3};\alpha_1\big) \| \Big] <\infty.$$
Moreover,  $\e\big[ \nabla_\alpha h\big([\Xbm,\Ubm]_{1:2};\as\big) \big]=\nabla_\alpha \e\big[   h\big([\Xbm,\Ubm]_{1:2};\as\big)\big]$, $\e\big[ \nabla_\alpha g\big([\Xbm,\Ubm]_{1:3};\as\big) \big]=$\\ $\nabla_\alpha \e\big[  g\big([\Xbm,\Ubm]_{1:3};\as\big)\big]$ and 
$\e\big[ \nabla_\alpha  q\big([\Xbm,\Ubm]_{1:4};\as\big) \big]=\nabla_\alpha \e\big[   q\big([\Xbm,\Ubm]_{1:4};\as\big)\big]$.
\label{cond_residual_deriv}
\end{assumpt}

Under the latter assumptions, Lemma~\ref{lem:asympdistrmmd_simple} in the appendix shows that $\tilde{\sigma}^2_\an=O_\Pb(n^{-1})$ when $P_\as=P$, whereas $\tilde{\sigma}^2_{q,\an}\to\tilde{\sigma}^2_{q,\as}>0$ remains positive.
This justifies a model specification test based on a weighted combination of $\sqrt{n}\hmmd(P_\an,P)$ and $\sqrt{n}\hmmdq(P_\an,P)$, which automatically ``switches'' between the two statistics, depending on whether or not $P_\as=P$.

\begin{thm}
\label{thm:nondegtestonemodel}
Assume $\epsilon_n\to 0$, $\epsilon_n \sqrt{n}\to\infty$ in probability, \textcolor{black}{$\sqrt{n}(\an-\as)=O_{\Pb}(1)$} and that Assumptions~\ref{A_characteristic}-\ref{cond_residual_deriv} are satisfied.
\begin{enumerate}
    \item If $ P=P_\as$, \textcolor{black}{i.e. under $\Hc_{0,\Mc}$}, we have 
    $$
    \Tc_n(\Mc,P)=\sqrt{n}\frac{\hmmden(P_\an,P)}{\hat{\sigma}_n} \stackrel{\text{law}}{\longrightarrow}
    \Nc( 0,1), 
    $$
    \textcolor{black}{where $\hat{\sigma}_n=\tilde{\sigma}_{\an}+\epsilon_n\tilde{\sigma}_{q,\an}.$}
    \item If $P\not =P_\as$, \textcolor{black}{i.e. if $\Hc_{0,\Mc}$ is not true}, then $\Tc_n(\Mc,P)$ tends to infinity in probability.
\end{enumerate}
\end{thm}
The proof is postponed to Section~\ref{proof_thm:nondegtestonemodel}. As a consequence of Theorem \ref{thm:nondegtestonemodel}, a consistent distribution free test of $\Hc_{0,\Mc}$ can be conducted with the test statistic $\Tc_n(\Mc,P)$ \textcolor{black}{(see Algorithm~\ref{algmodelspecification} below)}.
\begin{rem}
\label{rem_simple_assumpt}
Note that Theorem~\ref{thm:nondegtestonemodel} obviously covers the case of simple zero assumptions, i.e., testing $\Hc_0:P_1=P$ for some given probability $P_1$ and with the test statistic (\ref{defteststat}).
Our theory directly applies by defining $\Theta_1$ as the singleton $\{\as\}$, and setting $P_\as=P_1$.
The technical assumptions \ref{regular_model}-\ref{cond_residual_deriv} are no longer required in this case, since the sequence $(\an)$ becomes constant and it is no longer necessary to differentiate the kernels $h$ and $q$.
\end{rem}

\begin{rem}
\label{rem_estimaion_covar}
In Theorem~\ref{thm:nondegtestonemodel}, it is possible to replace the denominator $\textcolor{black}{\hat{\sigma}_n}=\tilde{\sigma}_{\an}+\epsilon_n\tilde{\sigma}_{q,\an}$ by $\big(\tilde{\sigma}^2_{\an}+\epsilon^2_n\tilde{\sigma}^2_{q,\an}\big)^{1/2}$ since both quantities
are asymptotically equivalent under our assumptions.
Moreover, to lighten our theoretical developments, we have neglected the covariance between $\hmmd(P_\an,P)$ and $\hmmdq(P_\an,P)$. Indeed, this covariance multiplied by $\epsilon_n$ is always asymptotically negligible compared to $\tilde{\sigma}^2_{\an}+\epsilon^2_n\tilde{\sigma}^2_{q,\an}$ (invoke the Cauchy-Schwartz inequality in the degenerate case, \textcolor{black}{as $\tilde{\sigma}_{\an}=o_\Pb( \epsilon_n)$}).
\end{rem}

\subsection{Non-differentiable generating function \textcolor{black}{(model specification)}}
\label{general_specif_test}

The assumption of differentiability of $\alpha \mapsto F(U; \alpha)$ may be considered as relatively strong in many practical applications in statistics and machine learning. \textcolor{black}{For example, denote $F(\cdot;\alpha)$ a deep neural network with the ReLu activation function and a parameter vector $\alpha$. Let $U$ be uniformly distributed on $[0,1]$. Then, defining $P_\alpha\stackrel{\text{law}}{:=}F(U;\alpha)$ yields a universal approximator of any probability distribution $P$ in terms of the $\mmd$: see \citet[Theorem 2.8]{yangliwang2022}. Obviously, $\alpha\mapsto F(u;\alpha)$ is not differentiable for any $u$, and the results from Section \ref{case_differentiable_F} cannot be applied. In this section, we show that our test may still be applied even if the generating function is not differentiable, imposing some regularity conditions on $\e\big[ k\big(F(U;\alpha),\cdot\big) \big]$ which are detailed in Appendix~\ref{reg_assump_nondiff_generating_fct}.}

Since we cannot apply a Taylor expansion w.r.t. $\alpha$ to $\Tc_n(\Mc,P)$ when $\alpha \mapsto F(U; \alpha)$ is not differentiable, we need more sophisticated tools than in Section~\ref{case_differentiable_F} to derive the asymptotic normality of $\Tc_n(\Mc,P)$.
Here, we rely on the framework of empirical $U$-processes introduced by \cite{arconesgine1993} and~\cite{arcones1994bootstrap}. \textcolor{black}{See Appendix~\ref{reg_assump_nondiff_generating_fct} for technical details}.  
Under our proposed assumptions, we obtain the limiting law of the test statistic defined in~(\ref{defteststat_specif}) when dealing with non-differentiable generating functions, which can be used to test the null hypothesis $\Hc_{0,\Mc}$. To be short, we recover the results of Theorem~\ref{thm:nondegtestonemodel}.

\begin{thm}
\label{thm:nondegtestonemodel_general}
Let Assumptions~\ref{A_characteristic}-\ref{regular_model} and~\ref{ass_equicont}-\ref{ass_degenerateuprocess} \textcolor{black}{in Appendix~\ref{reg_assump_nondiff_generating_fct}} hold. If  $\epsilon_n\to 0$, $\epsilon_n \sqrt{n}\to\infty$ in probability \textcolor{black}{and $\sqrt{n}(\an-\as)=O_{\Pb}(1)$}, then the conclusions of Theorem~\ref{thm:nondegtestonemodel} apply.
\end{thm}

\textcolor{black}{
The proof can be found in Section~\ref{reg_assump_nondiff_generating_fct} in the appendix. Since Assumptions
\ref{ass_equicont}-\ref{ass_degenerateuprocess} are quite abstract, we further illustrate the practical relevance of the proposed framework by verifying that a ReLu-type generative neural network satisfies all imposed regularity conditions in Appendix~\ref{app:relu-type_example}.}
We summarize our proposed model specification test in Algorithm~\ref{algmodelspecification}.
\SetKwInput{KWParam}{Requirements}
\SetKwInput{KWResult}{Test result}
\begin{algorithm}
 \label{algmodelspecification}
\caption{$\mmd$-based test of $\Hc_{0,\Mc}:\mmd(P_\as,P)=0$}
\SetAlgoLined \LinesNumbered
\KWParam{I.i.d.\ sample $\lc X_i\rc_{1\leq i\leq n}$ from $P$, generative model $F(U;\alpha)\sim P_\alpha$, estimator $\an$ of $\as$, tuning parameter $\epsilon_n$ and confidence level $\gamma$.}
Sample $\big( F(U_i,\an)\big)_{1\leq i\leq n} $, where $(U_i)_{1\leq i\leq n}\overset{i.i.d.}{\sim}P_U$ \;
Compute $ \Tc_n(\Mc,P)=\sqrt{n}\hmmden(P_\an,P)/(\tilde{\sigma}_{\an}+\epsilon_n\tilde{\sigma}_{q,\an})  $ \;   
Reject $\mmd(P_\as,P)=0$ when $\big\vert \Tc_n(\Mc,P)\big\vert>\Phi^{-1}(1-\gamma/2)$; otherwise, accept.
\end{algorithm}

\section{Asymptotic behavior of MMD-based tests for model comparison}
\label{sec:modelselection}

Let us specify the mathematical framework that is required to prove the asymptotic normality of the test statistic for model comparison introduced in~(\ref{defteststatcomp}).  First we  introduce  additional notation. For some sequences of random elements $(X_1,X_2,\ldots )$,   $(U_1,U_2,\ldots)$ and $(V_1,V_2,\ldots )$, we denote $[\Xbm,\Ubm, \Vbm]_{i:j}:=\big( (X_i,U_i, V_i), (X_{i+1},U_{i+1}, V_{i+1}),$ $\ldots,$ $ (X_j,U_j, V_j)\big)$ for $1\leq i<j \leq n$. Similarly, we  denote     $[\xbm,\ubm, \vbm]_{i:j}:=\big( (x_i,u_i, v_i), (x_{i+1},u_{i+1}, v_{i+1}),$ $\ldots,$ $(x_j,u_j, v_j)\big)$. In the same manner,   $[\Xbm, \Ubm, \Vbm]_{i_1, i_2, \ldots, i_k}$ and $[\xbm,\ubm, \vbm]_{i_1, i_2, \ldots, i_k}$ are defined, where the index set indicates the components of the concatenated variables.

\subsection{Regularity assumptions and asymptotic variance estimation}
\label{Notations_assumptions_comparison}
Consider two competing parametric models $\Mc_1=\{P_\alpha;\alpha \in \Theta_1\}$ and
$\Mc_2=\{Q_\beta;\beta \in \Theta_2\}$ for the DGP. The goal is to evaluate whether or not one of the models is closer to the true law $P$ of the data than the other in terms of the MMD. Our null hypothesis is then written
$ \Hc_{0,\Mc_1,\Mc_2}: \mmd(P_\as,P)= \mmd (Q_\bs,P) $.

\textcolor{black}{For convenience, the asymptotic behavior of our test statistic $\Tc_n(\Mc_1,\Mc_2,P)$ will be stated 
under the assumption that the optimal parameters are pseudo-true values for the MMD, i.e.,
$\as:=\argmin_{\alpha\in\Theta_1}\mmd(P_\alpha,P)$ and $\bs:=\argmin_{\beta\in\Theta_2}\mmd(Q_\beta,P)$.}     
This constraint is due to the fact $\Hc_{0,\Mc_1,\Mc_2}$ can be satisfied even though $P_\as\not =P\not=Q_\bs$, which would then introduce additional randomness in our test statistic. In this case, some terms of the form $\nabla_\alpha \mmd(P_\alpha,P)\vert_{\alpha=\as}\not=0\not= \nabla_\beta \mmd(Q_\beta,P)\vert_{\beta=\bs}$ would appear in the asymptotic variance, \textcolor{black}{adding significant complications in the estimation procedure, which is why we have refrained from investigating this case}. 
\textcolor{black}{To estimate $\as$ (resp. $\bs$), we could set $\an\in\argmin_{\alpha\in\Theta_1}\hmmd(P_\alpha,P)$ (resp. $\bn\in\argmin_{\beta\in\Theta_2}\hmmd(Q_\beta,P)$), which would yield consistent and asymptotically normal estimators under some regularity conditions \citep{briol2019statistical}. Nonetheless, this is not mandatory. Thus, the choice of $\an$ and $\bn$ will remain unspecified hereafter.}

Similarly to the existence of a random variable $U$ and a generating function $F(\cdot; \cdot)$ for the model $\Mc_1$, we assume
the existence of a random variable $V$ in some topological space $\Vc$ and a generating function $G(\cdot; \cdot): \Vc\times\Theta_2\to \Sc$ of $\Mc_2$ such that $G(V; \beta)\sim Q_\beta$ for every
$\beta \in \Theta_2$. As in the previous section, we assume that we have access to an i.i.d.\ sequence $(V_1,V_2,\ldots)$ from $P_V$, which is also independent of $(X_1,X_2,\ldots)$. We also assume the random vectors $(X_i,U_i,V_i)$, $i\in \{1,\ldots,n\}$, are independently drawn and they are induced by the same abstract probability space $(\Omega,\bc,\Pb)$.

In terms of notations and to distinguish quantities that are related to either model $\Mc_1$ or $\Mc_2$, we will use the same notation as in Section \ref{sec:asympdistteststat} but an upper index $\cdot^{(1)}$ (resp. $\cdot^{(2)}$) will refer to a quantity related to $\Mc_1$ (resp. $\Mc_2$). For instance, the map $h$ introduced in~(\ref{def_fct_h}) will be denoted as
$ h^{(1)}\big( [\xbm, \ubm]_{1:2};\alpha\big)  :=h\big((x_1,F(u_1; \alpha)), (x_2,F(u_2; \alpha))\big)$
when referring to $\Mc_1$, whereas we will denote
$ h^{(2)}\big( [\xbm,\vbm]_{1:2};\beta\big)  :=h\big((x_1,G(v_1; \beta)), (x_2,G(v_2; \beta))\big)$
when referring to $\Mc_2$.  To distinguish between the two parametric models, the letter $u$ (resp.\ $v$) will be reserved for the first model (resp.\ second model).

In the following, we will mimic the ideas of Section~\ref{sec:asympdistteststat}.
Recall that in the definition of $\Tc_n(\Mc_1,\Mc_2,P)$ in~(\ref{defteststatcomp}) we have not yet specified the estimator $\hat{\tau}^2_n$ of the asymptotic variance of $\sqrt{n} \big\{ \hmmden(P_\an,P) -\hmmden(Q_\bn,P) \big\} $. Again, we will use an estimator of the form $\hat{\tau}_n=\hat{\tau}_1+\epsilon_n\hat{\tau}_2$, where $\hat{\tau}^2_1$ (resp. $\hat{\tau}^2_2$) denotes an estimator of the asymptotic variance of $\sqrt{n}\big\{ \hmmd(P_\an,P)-\hmmd(Q_\bn,P) \big\}$ (resp. of $\sqrt{n}\big\{ \hmmdq(P_\an,P)-\hmmdq(Q_\bn,P) \big\}$).
To this aim and in accordance with the definitions of $\sigma_\alpha^2$ and $\sigma_{q,\alpha}^2$ in Section~\ref{Notations_assumptions} define
    \begin{equation}
    \sigma^2_{\alpha,\beta}:=\var \Big( 2\,\e_{X_2,U_2,V_2}\big[  h^{(1)}\big( [\Xbm, \Ubm]_{1:2};\alpha\big)-h^{(2)}\big( [\Xbm, \Vbm]_{1:2};\beta\big)  \big] \Big),\; \;\text{and}
    \label{def_sigma_alpha_beta}
    \end{equation}
    \begin{equation}
        \sigma^2_{q,\alpha,\beta}:=\var \Big( 2\sqrt{2}\,\e_{[\Xbm, \Ubm, \Vbm]_{3:4}}\big[  q^{(1)}\big( [\Xbm, \Ubm]_{1:4};\alpha\big) -q^{(2)}\big( [\Xbm, \Vbm]_{1:4};\beta\big) \big] \Big).
    \nonumber 
    \end{equation}
Defining
\begin{eqnarray}
h\big([\xbm, \ubm, \vbm]_{1:2};\alpha,\beta\big) &:= & h^{(1)}\big([\xbm, \ubm]_{1:2};\alpha\big)-h^{(2)}\big([\xbm, \vbm]_{1:2};\beta\big),\;\; \text{and} \nonumber   \\ 
q\big([\xbm, \ubm, \vbm]_{1:4};\alpha,\beta\big) &:=& q^{(1)}\big([\xbm, \ubm]_{1:4};\alpha\big)
- q^{(2)}\big([\xbm, \vbm]_{1:4}; \beta\big) \label{def_q_alpha_beta}
\end{eqnarray}
we can introduce their corresponding estimators, for a given tuple $(\alpha,\beta)$, via
\begin{align*}
\tilde\sigma^2_{\alpha,\beta} &:=\frac{4}{n}\sum_{i=1}^n \bigg\{ \frac{1}{n-1} \sum_{\substack{j=1\\i\not=j}}^n h\Big( [\Xbm, \Ubm, \Vbm]_{i:j};\alpha,\beta \Big)- \lc \hmmd(P_\alpha,P)-\hmmd(Q_\beta,P)\rc\bigg\}^2,\;\text{and} 
\label{def_sigma_tilde_alpha_beta}
\end{align*}
\begin{align*}\tilde{\sigma}^2_{q,\alpha,\beta}&:=\frac{8}{n}\sum_{i=1}^n \bigg\{ \frac{1}{n-1} \sum_{\substack{j=1\\i\not=j}}^n q\Big( [\Xbm, \Ubm, \Vbm]_{2i-1,2i,2j-1,2j} ;\alpha,\beta\Big)-\lc \hmmd(P_\alpha,P)-\hmmd(Q_\beta,P)\rc\bigg\}^2 .
\end{align*}

Again, $\tilde\sigma^2_{\an,\bn}$ and $\tilde\sigma^2_{q,\an,\bn}$ are used as estimators of $\sigma^2_{\as,\bs}$ and $\sigma^2_{q,\as,\bs}$ respectively.

\subsection{Differentiable generating functions \textcolor{black}{(model comparison)}}
\label{case_differentiable_FG}

We are able to derive a test for model comparison that is never degenerate when $(U_i)_{i\geq 1}$ and $(V_i)_{i\geq 1}$ are independent. We need a slight extension of Assumption \ref{cond_residual_deriv} to prove the results.

\begin{assumpt}
    \label{cond_resid_model_comp}
    We have $\text{supp}(P_\as)\cap\text{supp}(P)\cap\text{supp}(Q_\bs)\not=\emptyset$ and there exists a $\delta>0$ s.t.
    $$ \Eb\bigg[ \sup_{(\alpha_1,\beta_1) \in B_\delta((\as,\bs))} \| \nabla^2_{(\alpha,\beta),(\alpha,\beta)^\intercal} g \big([\Xbm,\Ubm,\Vbm]_{1:3};(\alpha_1,\beta_1)\big) \| \bigg] <\infty $$
    $$ \text{ and } \nabla_{(\alpha,\beta)} \e\big[  g\big([\Xbm,\Ubm,\Vbm]_{1:3};(\as,\bs)\big)\big]=\e\big[ \nabla_{(\alpha,\beta)}   g\big([\Xbm,\Ubm,\Vbm]_{1:3};(\as,\bs)\big)\big]. $$    
\end{assumpt}

\begin{thm}
\label{thm:nondegtestmodelcomp_diff}
Assume that $\epsilon_n\to 0$, $\epsilon_n\sqrt{n}\to\infty$ in probability, $\as\in\argmin_{\alpha\in\Theta_1} \mmd^2(P_\alpha,P)$ and  $\bs\in\argmin_{\beta\in\Theta_2} \mmd^2(Q_\beta,P)$, 
\textcolor{black}{$\sqrt{n}(\an-\as)=O_{\Pb}(1)$ and $\sqrt{n}(\bn-\bs)=O_{\Pb}(1)$},
the samples $(U_i)_{i\geq 1}$ and $(V_i)_{i\geq 1}$ are independent and that Assumptions~\ref{A_characteristic}-\ref{cond_resid_model_comp} are satisfied by the competing models $\Mc_1$ and $\Mc_2$.
\begin{enumerate}
    \item Under $\Hc_{0,\Mc_1,\Mc_2}:\mmd(P_\as,P)=\mmd(Q_\bs,P)$, we have
    \begin{align}
        \Tc_n(\Mc_1,\Mc_2,P)=\sqrt{n}\frac{\hmmden(P_\an,P)-\hmmden(Q_\bn,P)}{\hat{\tau}_n}\stackrel{\text{law}}{\longrightarrow} \Nc(0,1) , \nonumber 
    \end{align}
    \textcolor{black}{where $\hat{\tau}_n=\tilde{\sigma}_{\an,\bn}+\epsilon_n\tilde{\sigma}_{q,\an,\bn}$.}
    \item If $\mmd(P_\as,P)>\mmd(Q_\bs,P)$, then
    $$\sqrt{n}\frac{\hmmden(P_\an,P)-\hmmden(Q_\bn,P)}{\hat{\tau}_n}\to +\infty \;\;\text{in probability}. $$
    \item If $\mmd(P_\as,P)<\mmd(Q_\bs,P)$, then $$\sqrt{n}\frac{\hmmden(P_\an,P)-\hmmden(Q_\bn,P)}{\hat{\tau}_n}\to -\infty \;\;\text{in probability}.  $$
\end{enumerate}
\end{thm}
\textcolor{black}{The proof can be found in Appendix~\ref{techn_proofs_model_comp}}. As in Remark~\ref{rem_simple_assumpt}, the latter theorem also covers the case of known parameters $\as$ and $\bs$, i.e.\ the case when $\Theta_1$ and $\Theta_2$ are singletons, solely requiring Assumption \ref{A_characteristic}.

\subsection{Non-differentiable generating functions \textcolor{black}{(model comparison)}}
\label{general_comparison_test}

When $\alpha \mapsto F(u; \alpha)$ or $\beta\mapsto G(v; \beta)$ is not \textcolor{black}{twice} differentiable, we rely on similar techniques as in Section~\ref{general_specif_test} to deduce the asymptotic normality of $\Tc_n(\Mc_1,\Mc_2,P)$.
Some technical conditions related to empirical $U$-processes are required: see Section~\ref{proofs_model_comp_nonsmooth} in the appendix.

\begin{thm}
\label{thm:nondegtestmodelcomp}
Assume $\epsilon_n\to 0$, $\epsilon_n \sqrt{n}\to\infty$ in probability, $(U_i)_{i\geq 1}$ and $(V_i)_{i\geq 1}$ are independent, $\as\in\argmin_{\alpha\in\Theta_1} \mmd(P_\alpha,P)$ and $\bs\in\argmin_{\beta\in\Theta_2} \mmd(Q_\beta,P)$,
\textcolor{black}{$\sqrt{n}(\an-\as)=O_{\Pb}(1)$ and $\sqrt{n}(\bn-\bs)=O_{\Pb}(1)$},
and that Assumptions~\ref{A_characteristic}-\ref{regular_model} and~\ref{regularity_tildeh_modelcomp}-\ref{ass_degenerateuprocess_modelcompp} in Appendix \ref{proofs_model_comp_nonsmooth} are satisfied for the two competing models $\Mc_1$ and $\Mc_2$.
Then the conclusions of Theorem~\ref{thm:nondegtestmodelcomp_diff} apply.
\end{thm}

\textcolor{black}{We summarize our proposed model selection test in Algorithm~\ref{algmodelselection}.}

\SetKwInput{KWParam}{Requirements}
\begin{algorithm}
 \label{algmodelselection}
\caption{$\mmd$ based test of $\Hc_{0,\Mc_1,\Mc_2}:\mmd(P_\as,P)=\mmd(Q_\bs,P)$}
\SetAlgoLined \LinesNumbered
\KWParam{I.i.d.\ sample $\lc X_i\rc_{1\leq i\leq n}$ from $P$, generative models $F(U;\alpha)\sim P_\alpha$ and $G(V,\beta)\sim Q_\beta$, estimator $\an$ of $\argmin_{\alpha \in\Theta_1}\mmd(P_\alpha,P)$, estimator $\bn$ of $\argmin_{\beta \in\Theta_2}\mmd(Q_\beta,P)$, tuning parameter $\epsilon_n$ and confidence level $\gamma$.}
Sample $\big( F(U_i,\an)\big)_{1\leq i\leq n} $ and $\big( G(V_i,\bn)\big)_{1\leq i\leq n} $, where $(U_i)_{1\leq i\leq n}\overset{i.i.d.}{\sim}P_U$ and $(V_i)_{1\leq i\leq n}\overset{i.i.d.}{\sim}P_V$ are independent\;
Compute $ \Tc_n(\Mc_1,\Mc_2,P)=\sqrt{n}\big\{ \hmmden(P_\an,P)-\hmmden(Q_\bn,P)\big\}/(\tilde{\sigma}_{\an,\bn}+\epsilon_n\tilde{\sigma}_{q,\an,\bn})  $ \;   
Reject $\mmd(P_\as,P)=\mmd(Q_\bs,P)$ when $\big\vert \Tc_n(\Mc_1,\Mc_2,P)\big\vert>\Phi^{-1}(1-\gamma/2)$; otherwise, accept.
\end{algorithm}

\section{Simulation study}
\label{sec:simstudy}

\subsection{\textcolor{black}{Monte Carlo study for model specification} }

To investigate the performance of the model specification test based on the test statistic $\Tc_n(\Mc,P)$, let us generalize  the example from Section \ref{illustrative_example} to an arbitrary dimension $p$, inspired by the toy example from \cite{gretton2012kernel}. Consider a $p-$dimensional random vector
$X\sim P= \mathcal{N_n} \left(0, I_p\right),$
where $P$ still denotes the law of the DGP and $I_p$ is the $p-$dimensional identity matrix.

As a first example, the model $\Mc$ for the law of $X$ is defined by
$$
Y(\alpha)= Y+  \alpha
 \sim P_\alpha,\; \text{with}\;\ Y \sim \mathcal{N} \big(0,
\sigma^2 I_p\big),
$$
for some known variance $\sigma^2$, and $\alpha:=(\alpha_1, \ldots, \alpha_p)$ is a $p$-dimensional vector to be estimated.
For every $\sigma^2$, the ``optimal'' parameter is $\as=0$. Moreover, $P=P_\as$ if $\sigma^2=1$. We will vary the standard deviation $\sigma$ of the competing models by setting $\sigma\in\{1.0,\, 1.1, \, 1.2,\,  1.3,\,  1.4\}$. Furthermore, consider a $p-$dimensional Gaussian kernel $k(X_1, X_2)=e^{-||X_1-X_2||_2^2/p}$. In the literature (see, e.g.,~\cite{gretton2012kernel}), the exponent of the Gaussian kernel is often normalized by an expression containing the empirical median. Since an influence of this estimation step on the asymptotic distribution should also be investigated in the future, we do not consider this type of normalization.

As in Section \ref{illustrative_example}, we estimate $\alpha$ by the empirical mean of $(X_1, \ldots, X_n)$, an i.i.d.\ sample from $P$, i.e., $\alpha_n=n^{-1}\sum_{i=1}^n X_i$. We generate by simulation an i.i.d. sample $(Y_{1}, \ldots,Y_{n})$ from $\Nc(0,\sigma^2 I_p)$ to build the sample $\big(Y_1(\an),\ldots,Y_n(\an)\big)$ from $P_\an$. \textcolor{black}{Note that, given $\an$, the quantities $Y_k(\an)$ are mutually independent, but not unconditionally}. In this simulation study, we set the number of Monte Carlo replications to $1000$, i.e., we generate $1000$ independent replications of  the test statistics  $\Tc_n(\Mc,P)$ and report the empirical level/power of a test of $\Hc_{0,\Mc}$. As a comparison, we also provide  the empirical level/power of a test of $\Hc_{0,\Mc}$ which is solely based on $\sqrt{n}\hmmdq(P_\an,P)$.

\begin{figure}[!t]
    \centering
    \includegraphics[width=0.48\textwidth]{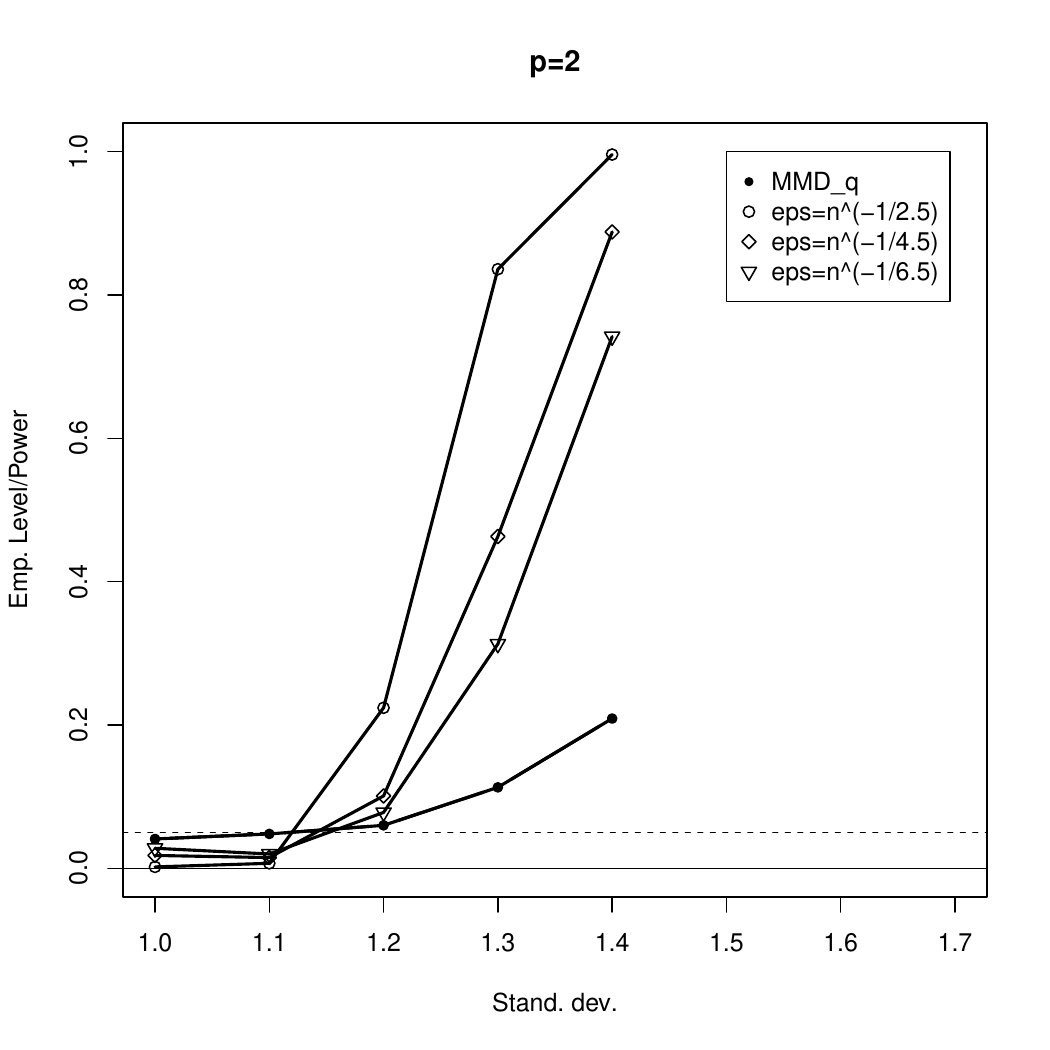}
    \includegraphics[width=0.48\textwidth]{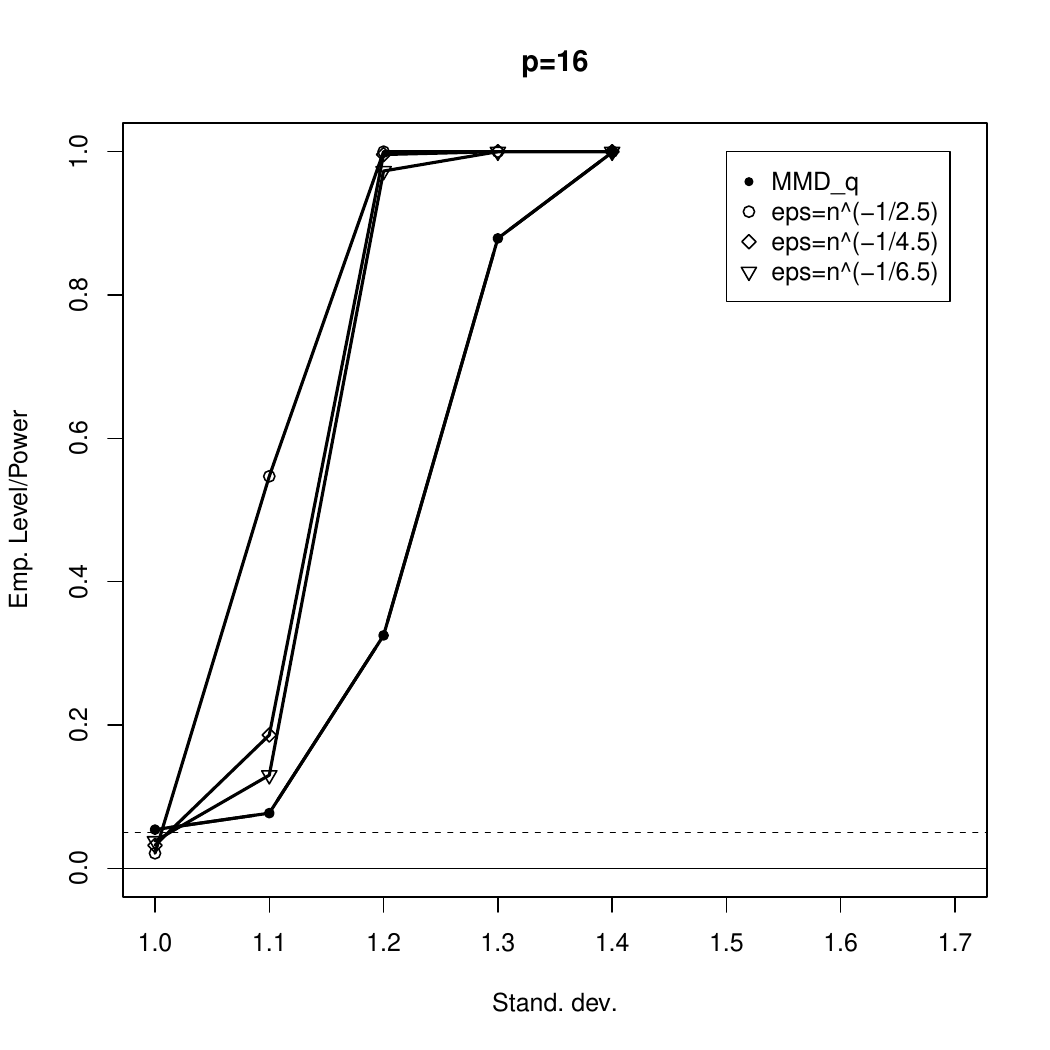}
    \caption{\footnotesize Empirical level and  power of the tests based on  $\Tc_n(\Mc,P)$  and $\sqrt{n}\hmmdq(P_\an,P)$ for dimensions $p=2$ (left) and  $p=16$ (right), a sample size $n=500$, as well as for different choices of $\epsilon_n$ (see the legend) and  varying {\bf \textit{standard deviation}}. The rejection probabilities are estimated using $1000$ replications of the tests based on samples of size $n$. The black dashed line indicates the significance level $0.05$.}
    \label{fig:pvalues-1}
\end{figure}

For the two tests based on the  statistics $\Tc_n(\Mc,P)$  and  $\sqrt{n}\hmmdq(P_\an,P)$ with a level $5\%$, Figure \ref{fig:pvalues-1} shows the  empirical proportion of rejections of the null hypothesis $\Hc_{0,\Mc}$ for dimensions $p \in\{2, 16\}$, sample size $n=500$   as well as for different choices of $\epsilon_n\in\{n^{-1/2.5}, n^{-1/4.5},n^{-1/6.5}\}$. In Figure \ref{fig:pvalues-1}, we restrict ourselves to this moderate sample size in order to empirically illustrate an influence of a choice of the weights $\epsilon_n$ on the performance of the proposed tests. 
Note that the  empirical proportions of rejections of $\Hc_{0,\Mc}$ when $\sigma=1.0$ are the empirical levels of these tests. When $\sigma > 1.0$, they are their empirical powers.

First of all, we observe in Figure \ref{fig:pvalues-1} that all tests keep their empirical level sufficiently well for all considered $\epsilon_n$. Furthermore, if the convergence rate of $\epsilon_n$ to zero is decreasing, then the empirical power of the test based on  $\Tc_n(\Mc,P)$ is also decreasing, confirming our intuition that $\epsilon_n$ has to tend to zero with $n$. However, the power is always higher than the empirical power of the test solely based on $\sqrt{n}\hmmdq(P_\an,P)$. For any considered tuning parameter $\epsilon_n$, the test based on  $\Tc_n(\Mc,P)$  always outperforms the test based on $\sqrt{n}\hmmdq(P_\an,P)$ for the considered sample size $n=500$ and all dimensions, \textcolor{black}{confirming the relevance of $\Tc_n(\Mc,P)$}. \textcolor{black}{In all our experiments, we have observed that the empirical power of all tests is increasing with increasing sample sizes and also for increasing dimension.} 
\textcolor{black}{In the sequel, we fix $\epsilon_n=n^{-1/2.5}$ since this choice empirically yields the highest powers of the proposed MMD specification test.}

As a second example, define the family of competing models for the law of $X$ as
$$
Y(\sigma)= \alpha_0 {\bf 1} + diag(\sigma_1, \ldots, \sigma_p) Y
 \sim P_\sigma,;\ Y\sim \mathcal{N} \left(0,
I_p\right)
$$
for some pre-specified marginal mean $\alpha_0\in\R$, where the marginal standard deviations are $\sigma_1, \ldots, \sigma_p$. We set $\sigma:=(\sigma_1, \ldots, \sigma_p)$ and ${\bf 1}=(1, \ldots, 1)$. If we fix $\alpha_0=0$, then the ``optimal'' parameters are $\sigma_1^*=\ldots=\sigma_p^*=1$ and $P=P_{\sigma^*}$, where $\sigma^*=(\sigma_1^*, \ldots,\sigma_p^*)$. Now, we vary the mean  $\alpha_0$ of the competing model $Y(\sigma)$ by setting $\alpha_0\in\{0, \, 0.1, \, 0.2, \ldots, \, 0.6\}$. As in the previous example, we  consider the $p-$dimensional Gaussian kernel $k(X_1, X_2)=e^{-||X_1-X_2||_2^2/p}$ and  set the significance level at  $0.05$. Furthermore, we estimate $\sigma_j$ by the empirical standard deviation of the $j-$th marginal i.i.d. sample
from $P$, namely $\sigma_{j,n}^2=n^{-1}\sum_{i=1}^n (X_{ij}-\overline{X}_{ij})^2$, and set $\sigma_{n}:=(\sigma_{1,n},\ldots, \sigma_{p,n})$.  Thus, we use the two samples $(X_1, \ldots, X_n)$ and $\big(Y_{1}(\sigma_n), \ldots, Y_{n}(\sigma_n)\big)$ from $P$ and $P_{\sigma_n}$ to test the null hypothesis $\Tc_n(\Mc,P)$. In the simulation study, 
the number of Monte Carlo replications is again $1000$
and we report the empirical level/power of a test of $\Hc_{0,\Mc}$.

\begin{figure}[!t]
    \centering
    \includegraphics[width=0.48\textwidth]{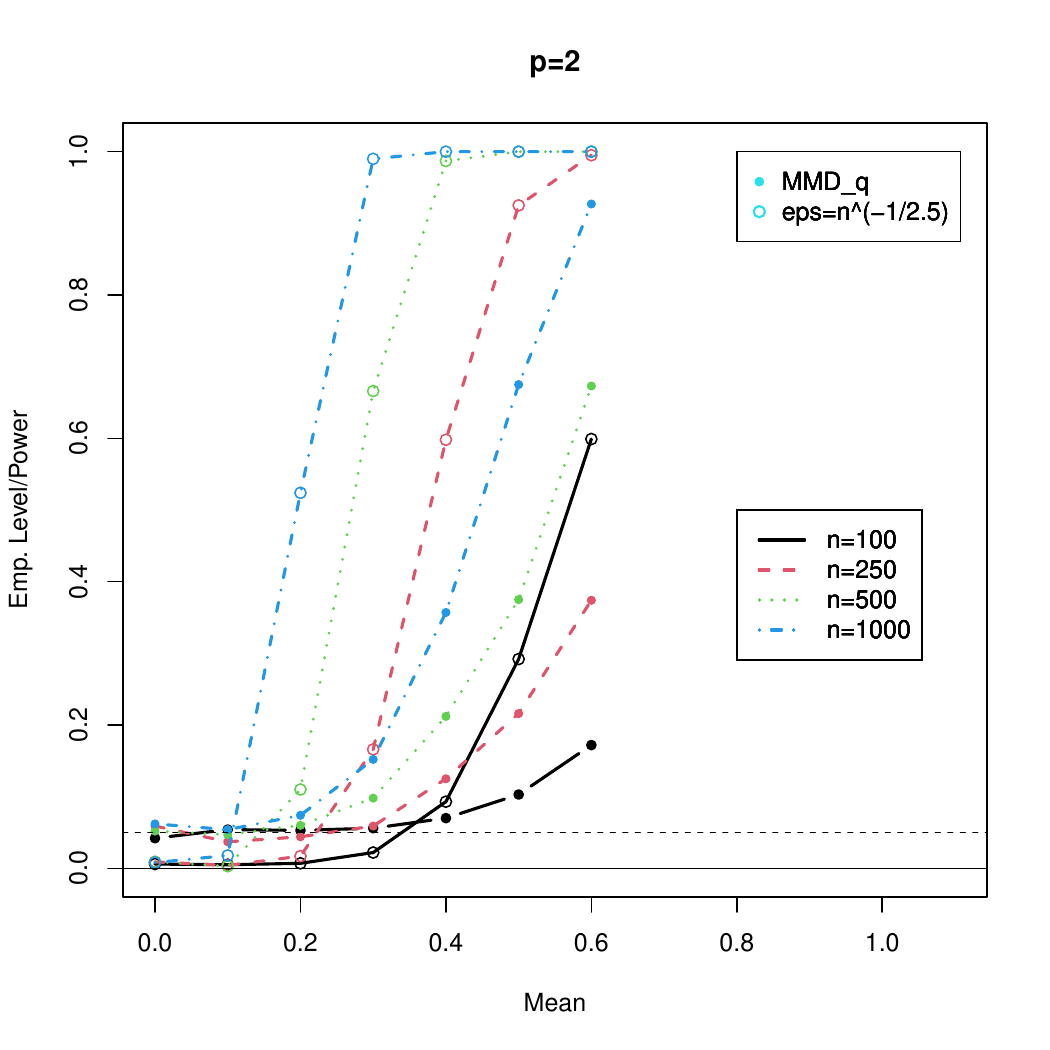}
    \includegraphics[width=0.48\textwidth]{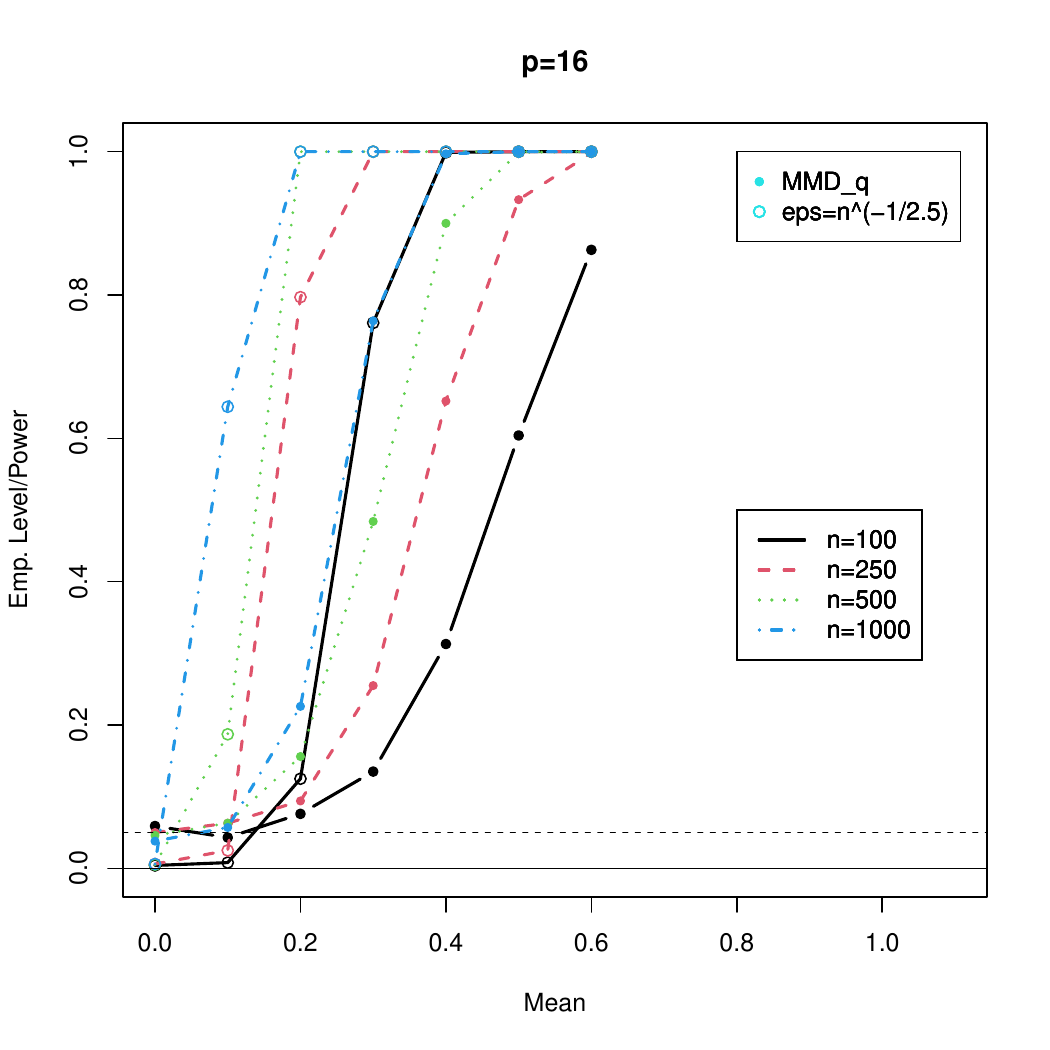}
    \caption{\footnotesize Empirical level and  power of the tests based on $\Tc_n(\Mc,P)$  and  $\sqrt{n}\hmmdq(P_{\sigma_n},P)$  for dimensions $p=2$ (left) and $p=16$ (right) as well as for sample sizes $n=100, 250, 500, 1000$  (see the legend), $\epsilon_n=n^{-1/2.5}$ and  varying {\bf \textit{mean}}. The rejection probabilities are estimated using 1000 samples of size $n$. The black dashed line indicates the significance level $0.05$.}
    \label{fig:pvalues-2}
\end{figure}

For the two tests based on $\Tc_n(\Mc,P)$ and  $\sqrt{n}\hmmdq(P_{\sigma_n},P)$, Figure \ref{fig:pvalues-2} shows the  empirical proportion of rejections of the null hypothesis $\Hc_{0,\Mc}$ for dimensions $p \in\{2,16\}$, sample sizes $n\in\{100, 250, 500, 1000\}$  and   $\epsilon_n=n^{-1/2.5}$. Note that the empirical proportions of rejections of $\Hc_{0,\Mc}$ for the case $\alpha_0=0$ are the empirical levels of the tests. When $\alpha_0> 0$, they are their empirical powers. All considered tests keep their empirical level reasonably well \textcolor{black}{and their power increases with an increasing sample size}. Again, the tests based on $\Tc_n(\Mc,P)$ are always more powerful than the test based on $\sqrt{n}\hmmdq(P_{\sigma_n},P)$. 
\textcolor{black}{In this framework, note  the poor power of our MMD specification tests for a small sample size $(n=100)$ and a small dimension $(d=2)$.}
%

\subsection{\textcolor{black}{Monte Carlo study for model comparison} }

\begin{figure}[!t]
    \centering
    \includegraphics[width=0.48\textwidth]{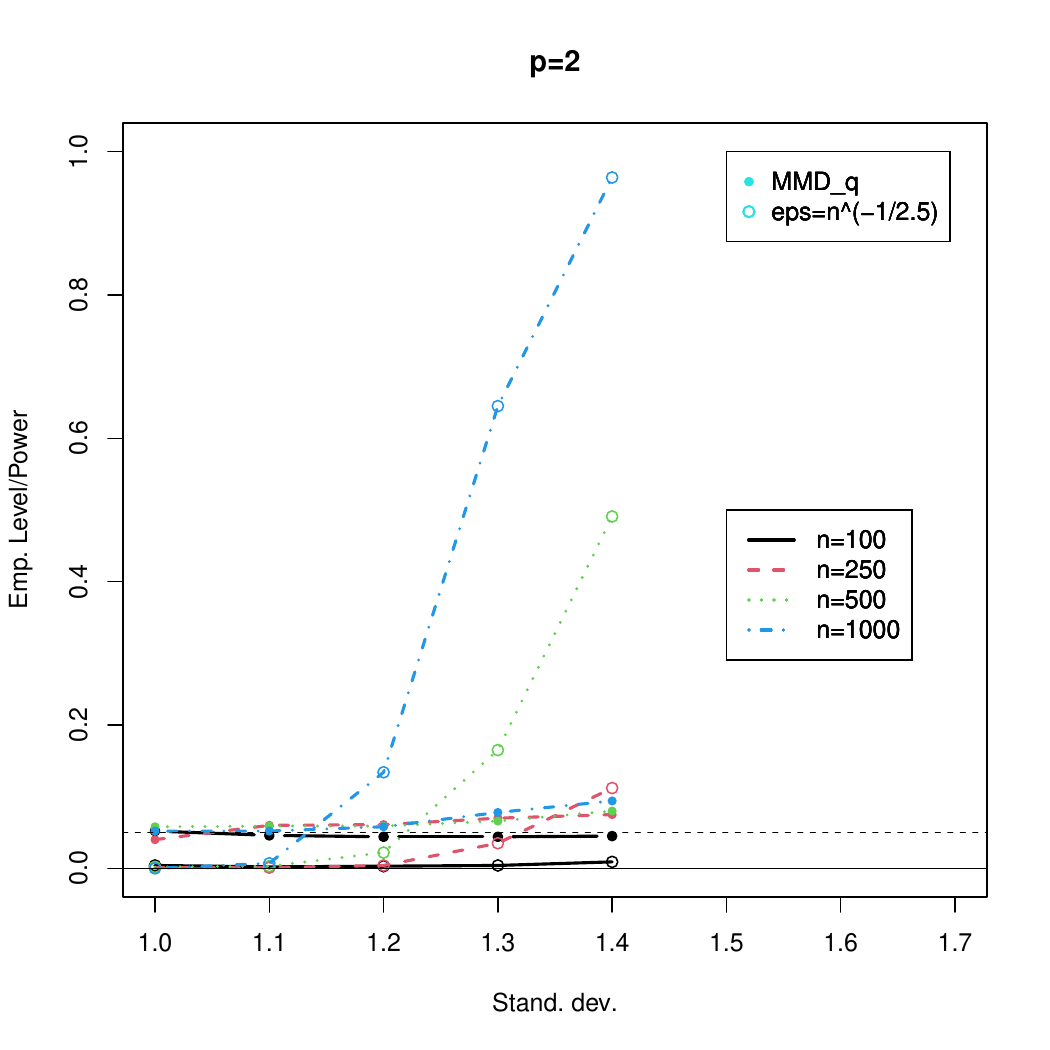}
    \includegraphics[width=0.48\textwidth]{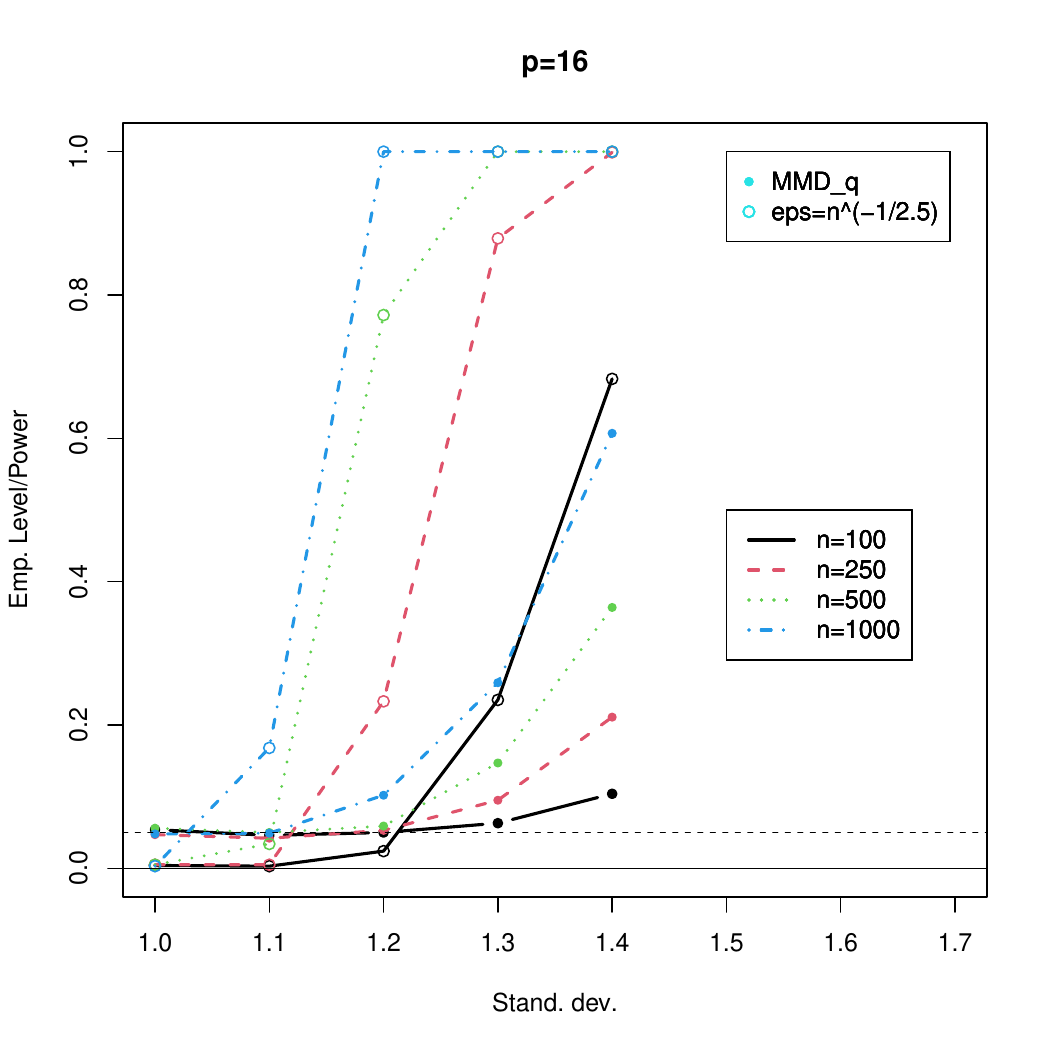}
    \caption{\footnotesize {\bf Degenerate case} for comparison of two models: Empirical level and  power of the tests based on  $\Tc_n(\Mc_1, \Mc_2, P)$  and $\sqrt{n}\big(\hmmdq(P_\an,P) - \hmmdq(Q_\bn,P) \big) $  for dimensions $p=2$ (left) and  $p=16$ (right) as well as for   sample sizes $n=100, 250, 500, 1000$ (see the legend),  $\epsilon_n=n^{-1/2.5}$ and  varying {\bf \textit{ standard deviation}} $\sigma$ in Model $\Mc_1$. Model $\Mc_2$ coincides with the true model ($\beta=0$). The rejection probabilities are estimated using $1000$ replications of the tests based on samples of size $n$. The black dashed line indicates the significance level $0.05$.}
    \label{fig:pvalues-3}
\end{figure}

In the third example, we focus on model comparison by considering two competing parametric models $\Mc_1$ and $\Mc_2$ for the law of $X$. 
Assume that the dimension $p$ is even. The first model $\Mc_1$ is defined by
$$
Y(\alpha)= Y+  \alpha
 \sim P_\alpha,;\ Y \sim \mathcal{N} \left(0,
diag(1,\ldots, 1, \sigma^2, \ldots, \sigma^2) \right),
$$
for some pre-specified variance $\sigma^2$, where $\alpha=(\alpha_1, \ldots, \alpha_p)$. Thus, the first $p/2$ margins of $Y(\alpha)$ have variance $1$ and the remaining $p/2$ margins have variance $\sigma^2$.  If $\sigma^2=1$, the model $\Mc_1$ coincides with the true model when $\alpha$ equals  the ``optimal'' parameter $\as=0$.  The second model $\Mc_2$ is defined by
$$
Z(\beta)= Z+  \beta
 \sim Q_\beta\, ;\ Z \sim \mathcal{N} \left(0, I_p\right)\, ,
$$
where $\beta=(\beta_1, \ldots, \beta_p)$. If $\beta=0$, the model $\Mc_2$ also coincides with the law of the DGP. Therefore, we may be in the degenerate situation, when the two competing models with optimal parameters coincide with the law of the DGP. As in the first example, we vary the standard deviation $\sigma$  by setting $\sigma\in\{1.0,\, 1.1, \, 1.2,\,  1.3,\,  1.4\}$. Further, we estimate $\alpha$ and $\beta$ by the empirical mean of the i.i.d.\ sample from $P$, $\alpha_n=\beta_n=n^{-1}\sum_{i=1}^n X_i$. Then, we independently generate the two samples $\big(Y_{1}(\alpha_n), \ldots, Y_{n}(\alpha_n)\big)$ and $\big(Z_{1}(\beta_n), \ldots, Z_{n}(\beta_n)\big)$ to test the null hypothesis $\Hc_{0,\Mc_1,\Mc_2}:\mmd(P_\as,P)=\mmd(Q_\bs,P)$.
In the simulation study, we set the number of Monte Carlo replications to $1000$, i.e., we generate $1000$ independent replications of  the test statistic $ \Tc_n(\Mc_1,\Mc_2,P)$ and report the empirical level/power of a test of $\Hc_{0,\Mc_1,\Mc_2}$.  As a comparison, we also provide the level/power of a test of $\Hc_{0,\Mc_1,\Mc_2}$ which is solely based on $\sqrt{n}\big( \hmmdq(P_\an,P)-\hmmdq(Q_\bn,P) \big)$.
For the two considered tests,
Figure \ref{fig:pvalues-3} shows the  empirical proportion of rejections of the null hypothesis $\Hc_{0,\Mc_1,\Mc_2}$  for dimensions $p \in\{2,16\}$, sample sizes $n\in\{100, 250, 500, 1000\}$   and   $\epsilon_n=n^{-1/2.5}$.
The  empirical proportions of rejection of $\Hc_{0,\Mc_1,\Mc_2}$ for the case $\sigma=1.0$ are the empirical levels of our tests.
The cases $\sigma > 1.0$ correspond to their empirical powers.

First of all, we observe in Figure \ref{fig:pvalues-3} that all tests keep their empirical level fairly well. \textcolor{black}{As previously, their power is relatively small for small a dimension ($d=2$) and a small sample size ($n=100, 250$)}. 
Note that the empirical power of our MMD test is always higher than that provided by the test solely based on $\sqrt{n}\big( \hmmdq(P_\an,P)-\hmmdq(Q_\bn,P) \big)$. For the considered tuning parameter $\epsilon_n=n^{-1/2.5}$, the test based on  $ \Tc_n(\Mc_1,\Mc_2,P)$  always outperforms the test based on the competitor test statistic  $\sqrt{n}\big( \hmmdq(P_\an,P)-\hmmdq(Q_\bn,P) \big)$ for the two considered dimensions \textcolor{black}{ and $n\in \{500, 1000\}$}. As expected, the empirical power of all tests is increasing with increasing sample sizes. As we have already observed, it is also increasing for increasing dimensions.

For the fourth example, we modify the third example to avoid the degenerate case. Now, the  models  $\Mc_1$ and  $\Mc_2$ are given by
$$
Y(\alpha)= Y+  \alpha
 \sim P_\alpha,;\ Y \sim \mathcal{N} \left(0,
diag(1.2^2,\ldots, 1.2^2, \sigma^2, \ldots, \sigma^2) \right),\; \text{and}
$$
$$
Z(\beta)= Z+  \beta
 \sim Q_\beta\, ;\ Z \sim \mathcal{N}\left(0,
1.2^2 I_p\right) \, ,
$$
respectively. Thus, both models cannot coincide with the DGP, reflecting the non-degenerate case.
However, for $\sigma=1.2$, they coincide and are therefore equally far away from the DGP.  We vary the standard deviation $\sigma$  by setting $\sigma\in\{1.2,\, 1.3, \, 1.4,\,  1.5,\,  1.6\}$.

\begin{figure}[!t]
    \centering
    \includegraphics[width=0.48\textwidth]{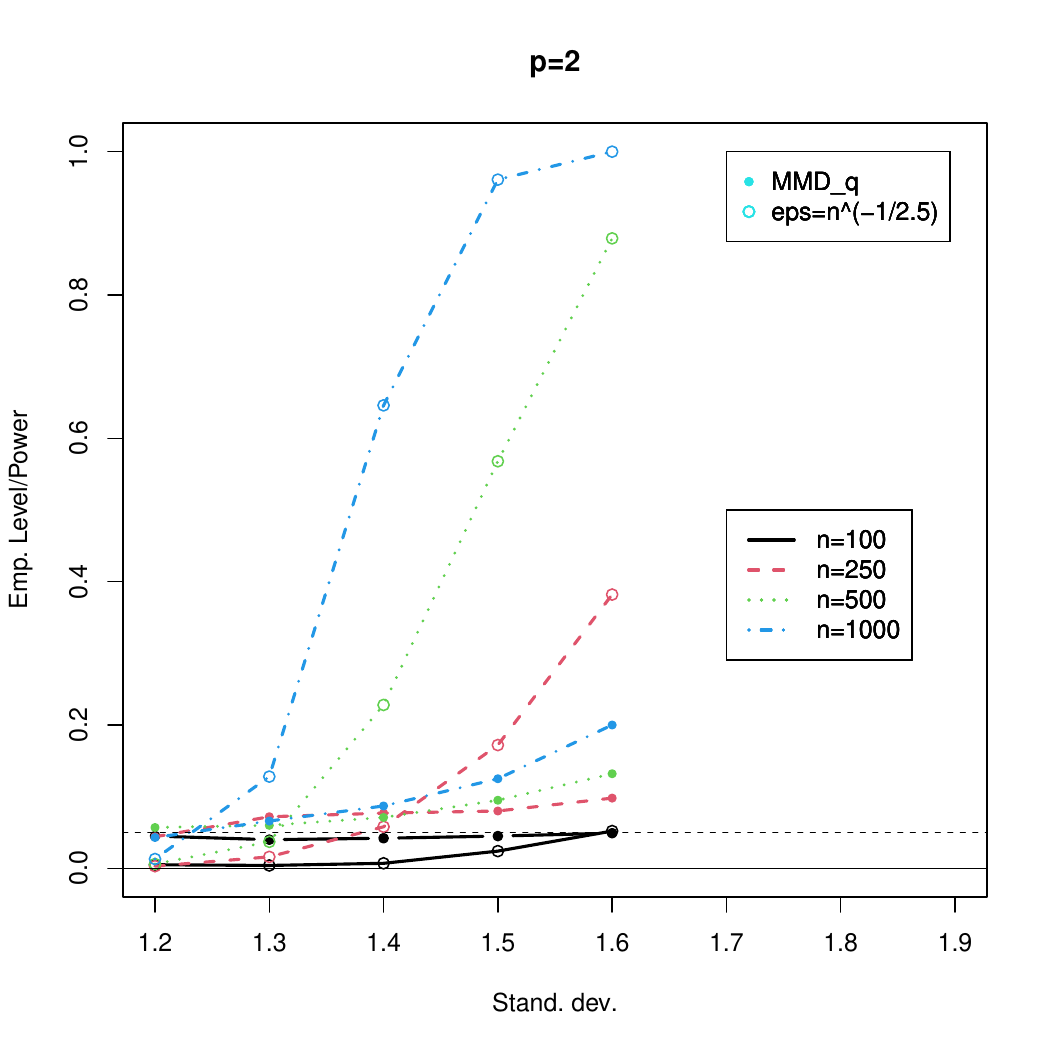}
    \includegraphics[width=0.48\textwidth]{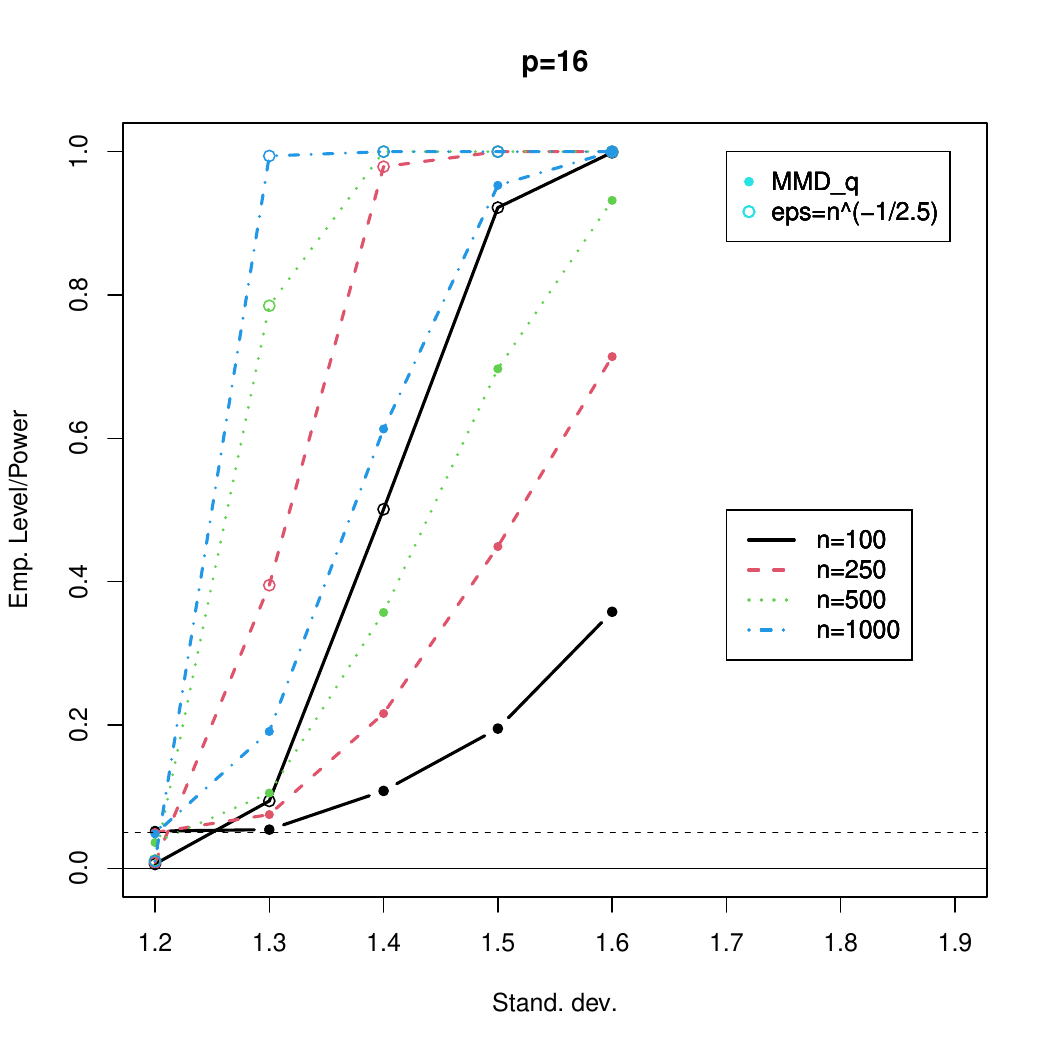}
    \caption{\footnotesize {\bf Non-degenerate case} for comparison of two models: Empirical level and  power of the tests based on  $\Tc_n(\Mc_1, \Mc_2, P)$  and $\sqrt{n}\big(\hmmdq(P_\an,P) - \hmmdq(Q_\bn,P) \big) $  for dimensions $p=2$ (left),  $p=16$ (right) as well as for sample sizes $n=100, 250, 500, 1000$ (see the legend),     $\epsilon_n=n^{-1/2.5}$ and  varying {\bf \textit{standard deviation}} in Model $\Mc_1$. Both models do not coincide with the true model. The rejection probabilities are estimated using $1000$ replications of the tests based on samples of size $n$. The black dashed line indicates the significance level $0.05$.}
    \label{fig:pvalues-4}
\end{figure}
For the two tests based on the statistics $ \Tc_n(\Mc_1,\Mc_2,P)$   and  $\sqrt{n}\big( \hmmdq(P_\an,P)-\hmmdq(Q_\bn,P) \big)$, Figure \ref{fig:pvalues-4} shows the empirical proportion of rejections of the null hypothesis $\Hc_{0,\Mc_1,\Mc_2}$  for dimensions $p \in\{2,16\}$, sample sizes $n\in\{100, 250, 500, 1000\}$  and $\epsilon_n=n^{-1/2.5}$. The empirical proportions of rejections of $\Hc_{0,\Mc_1,\Mc_2}$ for the case $\sigma=1.2$ (resp. $\sigma > 1.2$) are the empirical levels (resp. powers) of the tests. In Figure~\ref{fig:pvalues-4}, we clearly observe that the test based on  $ \Tc_n(\Mc_1,\Mc_2,P)$  is again more powerful than a test based on $\sqrt{n}\big( \hmmdq(P_\an,P)-\hmmdq(Q_\bn,P) \big)$.
\textcolor{black}{Further, similar conclusions as in the third example  can be drawn}.
%

\textcolor{black}{The code for the implemented tests that are used in the simulation study is accessible at \url{https://github.com/Flo771994/MMD_tests_for_model_selection}.}

\section{Empirical analysis}
\label{sec:empirics}
We illustrate the proposed MMD specification and model selection tests on historical stock returns, using the same dataset as~\citet{bfmclarke2022}. This dataset consists of the  daily log-returns of the MSCI indices for Austria, Germany, Ireland, Italy, the Netherlands, Singapore and Sweden from December 31, 1998 till March 12, 2018. The 5007 daily log-returns of seven countries constitute a  multivariate time series, for which the considered  MMD tests are not applicable. Therefore, we follow a standard econometrical approach by filtering every univariate time series of log-returns through ARMA(p,q)-GARCH(1,1) techniques (see Chapter 4 in \citet{mcneil-et-al-2005}, e.g.).  Using BIC, we find that the model ARMA(0,0)-GARCH(1,1) is the most suitable one for the considered returns.  Thus, we get so-called ``standardized'' residuals, that can reasonably be considered as i.i.d., with expectation zero and variance $1$. We then focus on the law of such residuals.

Using the MMD specification and model selection tests, we first answer the question of whether a scaled t-distribution or a mixture of two normal distributions are appropriate for fitting the univariate standardized residuals, two standard choices in financial econometrics. Obviously, all distributions in the two considered models are restricted to have expectation $0$ and variance $1$. Furthermore, as proposed in the previous section, we use the nuisance parameter $\epsilon_n=n^{-1/2.5}$ for $n=5007$. Table \ref{table-returns}  reports the p-values for the proposed MMD-based tests. Our conclusions are consistent with those obtained by \citet{bfmclarke2022} with the corrected Clarke test and the one-step Vuong test. Namely, none of the two considered univariate distributions for standardized residuals can be preferred 
with respect to another. This is not surprising because both parametric families are commonly used in applied financial econometrics.  
%
\begin{table}[!t]
\centering
\begin{tabular}{|r|c|c|c|}
  \hline
Country     & Specification test     &  Specification test        &   Model comparison          \\
             &  for the normal mixture     &   for the scaled t-distribution    &        test           \\ \hline
Austria      &0.3872                  & 0.8288                          &    0.6019                          \\
Germany      &0.7143                  & 0.4689                          &    0.6981                          \\
Ireland      &0.8301                  & 0.6783                          &    0.9001                          \\
Italy        &0.9638                  & 0.5020                          &    0.5140                          \\
Netherlands  &0.6206                  & 0.5335                          &    0.9051                          \\
Singapore    &0.6548                  & 0.7121                          &    0.9582                          \\
Sweden       &0.7638                  & 0.8206                          &    0.9426                          \\
   \hline
\end{tabular}
\caption{P-values for the model specification and model comparison tests with $\epsilon_n=n^{-1/2.5}$  for the standardized residuals of  Austria, Germany, Ireland, Italy, Netherlands, Singapore and Sweden (univariate models).}
\label{table-returns}
\end{table}

In the next step, we would like to find an appropriate multivariate model for the seven dimensional vector of standardized residuals. Since this problem is not an easy task, we split it into two sub-tasks: first, the choice of marginal models, and, second, the choice of a copula model, as in \citet[Chapter 5]{mcneil-et-al-2005} and many others. According to Table \ref{table-returns}, the  scaled t-distribution is appropriate for modeling the marginal standardized residuals. Thus, we select it as the univariate marginal model for each of the seven standardized residuals. In order to estimate our parametric copula models under consideration, 
we transform all univariate standardized residuals by their marginal empirical distributions to get pseudo-observations. 
The copula parameters are then estimated by semiparametric pseudo maximum likelihood \citep{genest-et-al-1995}. 
This standard technique is a robust approach to misspecification and yields asymptotically normal estimators.

Now, we can compare the normal copula and the t-copula \citep[Chapter 5]{mcneil-et-al-2005}, and perform also specification tests for them. 
The full multivariate distribution of the seven dimensional standardized residuals is finally 
specified by the marginal scaled t-distributions coupled with either the normal copula or the Student t copula. The first three p-values in the left part of  Table \ref{table-returns-mult} indicate that the specification test as well as the model selection test cannot reject the corresponding null hypotheses at the 5\% significance level. {It appears that the degree of freedom of the Student-t copula cannot fully capture all bivariate tail dependencies and, therefore, cannot be favored in this context, despite its common preference in applied financial econometrics.
\begin{table}[!t]
\centering
\begin{tabular}{|l|r|r|r|r|}
\hline
\textbf{Type of the test\quad / \quad Copula} & Normal & Student $t$ & Clayton & Gumbel \\ \hline
Model specification& 0.8444  & 0.7177    &   $1.9\cdot 10^{-29}$   & $9.79\cdot 10^{-25}$    \\ \hline
Model comparison      & \multicolumn{2}{|c|}{0.8555}      &\multicolumn{2}{|c|}{0.0006}       \\ \hline
\end{tabular}
\caption{P-values for the model specification and model comparison tests with $\epsilon_n=n^{-1/2.5}$ for seven-dimensional standardized residuals (multivariate models).}
\label{table-returns-mult}
\end{table}

Alternatively, we also consider the Clayton copula and the Gumbel copula. They are governed only by a single parameter, 
which is probably not enough to reasonably model a random vector of dimension seven.
When they are combined  with marginal scaled t-distributions, 
the p-values of the specifications tests for the Clayton copula and the Gumbel copula in the right part of Table \ref{table-returns-mult} are not surprisingly very small. Therefore, the null hypotheses of model specification can be rejected for both models, at any standard nominal levels. 
Moreover, the null hypothesis that these two models are equivalently well suited is rejected at the 5\% level. The test statistics is equal to $3.4271$, indicating 
that the multivariate model with the Gumbel copula is preferred to the one with the Clayton copula. 

\section{Conclusion and Outlook}
\label{sec:conclusion}
We have provided novel MMD-based model specification and model selection tests when the model parameters are estimated in a first-stage. In comparison with an approach solely based on $\hmmd(P_\an,P)$, these tests circumvent the major difficulty of computing the critical values of a complicated asymptotic distribution, but instead simply require to resort to the critical values of the standard normal distribution. Moreover, since our distribution free testing procedures are also valid when no parameter estimation is conducted, they yield valuable alternatives to the two sample test proposed by \cite{grettonkernelmethodtwosampleproblem}. The testing procedures are summarized in Algorithm \ref{algmodelspecification} and Algorithm \ref{algmodelselection}, respectively.

Both proposed test statistics depend on a tuning parameter, whose ``optimal'' choice is still an open problem that may be investigated in the future, \textcolor{black}{for example by a local power analysis in the same spirit as \citet{schennach2017simple}.} Moreover, it remains future work to derive certain properties of our testing procedures, such as the local power and uniformity of convergence over sets of DGPs. 
Moreover,  \cite{cherief2022finite} recently investigated parameter estimation based on the MMD-based on dependent input data. Since our test statistics are basically composed of $U$-statistics, future research might be concerned with relaxing the assumption of i.i.d.\ observations in the application of our MMD-based model specification and model selection tests by resorting to the vast literature on $U$-statistics of dependent data.



\appendix 
\section{Proofs}
\label{sec:proofs}

Hereafter, for any symmetric map $\psi$ and any i.i.d.\ sample $(Z_1,\ldots,Z_n)$, the associated $U$-statistic of degree $r$ 
is denoted as $\Uc_n \psi:=\textcolor{black}{\lc r!\binom{n}{r}\rc^{-1}} \sum_{(i_1,i_2,\ldots,i_r) \in I_{r,n}} \psi(Z_{i_1},\ldots,Z_{i_r})$, where 
$$I_{r,n}:=\{ (i_1,\ldots,i_r): i_j\in \Nb, 1\leq i_j \leq n; i_j\neq i_k \;\text{if}\; j\neq k  \}.$$
In our case, $Z_i$ will be
$(X_i,U_i)$ or the concatenation of several similar random vectors.
More specifically, consider a class $\mathcal{L}$ of symmetric real-valued functions on $\otimes_{i=1}^m (\Sc\otimes \Uc)$. 
With a slight abuse of notation, denote as $(\mathcal{U}^{(m)}_n \ell)_{\ell\in\mathcal{L}}$ the empirical $U$-process which acts on the sample
$$\big([\Xbm,\Ubm]_{1:m},[\Xbm,\Ubm]_{(m+1):2m}\ldots,[\Xbm,\Ubm]_{(n-m+1):n}\big),$$
and whose degree is determined by $\ell$.
The latter sample is drawn from $\otimes_{i=1}^m (P\otimes P_U)$ and is of size $ n/m$ (which is implicitly assumed to be an integer).
Note that $\mathcal{U}^{(1)}_n \ell$ is just $\mathcal{U}_n \ell$ and we drop the upper index in this case.

Let us illustrate our notation with the two classes of symmetric functions
\begin{equation}
\mathcal{F}:= \Big\{  [\xbm,\ubm]_{1:2} \mapsto h\big( [\xbm,\ubm]_{1:2} ;\alpha\big)  \big\vert \ \alpha\in B_\delta (\as) \Big\},\; \text{and}    
\label{def_Fc}
\end{equation}
\begin{equation}
\mathcal{F}_q:= \Big\{ [\xbm, \ubm]_{1:4} \mapsto q\big([\xbm, \ubm]_{1:4};\alpha\big)\big\vert \ \alpha\in B_\delta (\as) \Big\},
\label{def_Fcq}
\end{equation}
for some $\delta>0$, \textcolor{black}{where $h$ and $q$ are defined in \eqref{def_fct_h} and \eqref{def_fct_q}, respectively.}
The empirical $U$-process of degree one indexed by the set of functions $\Fc$ is defined as the stochastic process
$$\Uc_n:\ \mathcal{F}\mapsto\R;\ f\mapsto \Uc_n f:=\frac{1}{n(n-1)}\sum_{\substack{i,j=1\\i\not=j}}^{n} f \big( [\Xbm, \Ubm]_{i,j}\big).$$
Moreover, the empirical $U$-process of degree $2$ indexed by $\Fc_q$ is the stochastic process
$$\Uc^{(2)}_n:\ \mathcal{F}_q\mapsto\R;\ f\mapsto \Uc^{(2)}_n f:=\frac{1}{n/2(n/2-1)}\sum_{\substack{i,j=1\\i\not=j}}^{n/2} f \big( [\Xbm, \Ubm]_{2i-1,2i, 2j-1, 2j}\big)$$
This notation allows to rewrite the estimators of $\mmd^2(P_\as,P)$ in terms of empirical $U$-processes as
$$\hmmd(P_\an,P)=\Uc_n  h\lc \cdot; \an\rc \text{ and }\ \hmmdq(P_\an,P)=\Uc^{(2)}_n  q\lc \cdot ;\an\rc,$$
 Moreover, let the operator $\Pb \ell$ denote $\e\lk \ell(Z)\rk$, where the random vector $Z$ is in accordance with the arguments of $\ell$.
Thus, $\big((\Uc_n^{(m)}-\Pb)\ell\big)_{\ell\in\mathcal{L}}$ denotes the centered empirical $U$-processes whose asymptotic behavior was investigated in \cite{arconesgine1993},~\cite{arcones1994bootstrap}, among others.

\subsection{Theoretically convenient variance estimation}

Let us begin this section by proposing alternative estimators of $\sigma^2_\as$, $\sigma_{q,\as}^2$,$\sigma^2_{\as,\bs}$ and $\sigma_{q,\as,\bs}^2$, the asymptotic variances of $\hmmd(P_\as,P)$, $\hmmdq(P_\as,P)$, $\hmmd(P_\as,P)-\hmmd(Q_\bs,P)$ and $\hmmdq(P_\as,P)-\hmmdq(Q_\bs,P)$ respectively.
In Sections \ref{sec:asymp_var_estim} and \ref{Notations_assumptions_comparison}, we have defined empirical estimators $\tilde\sigma^2_\an$ and $\tilde\sigma_{q,\an}^2$ of $\sigma^2_\as$ and $\sigma_{q,\as}^2$. Unfortunately, these estimators are not $U$-statistics. Thus, an analysis of their asymptotic properties is not very convenient. In the following, we will introduce alternative $U$-statistics estimators and show that they are asymptotically equivalent - up to a term which tends to $0$ sufficiently fast - to the estimators defined in Sections \ref{sec:asymp_var_estim} and \ref{Notations_assumptions_comparison}. This will allow us to develop asymptotic theory for the simpler to analyze $U$-statistic estimators and to derive the corresponding asymptotic properties of the estimators introduced in Sections \ref{sec:asymp_var_estim} and \ref{Notations_assumptions_comparison} straightforwardly.

First, for any fixed $\alpha$, we symmetrize $\tilde\sigma^2_\alpha$ and discard the terms for which the indices $j$ and $k$ are equal, since their contribution is negligible. Doing so, we obtain a proper $U$-statistic, which is much more convenient to work with. This leads to the following estimator of $\sigma_\alpha^2$:
\begin{align*}
\hat{\sigma}^2_{\alpha}&:=\frac{4}{3n(n-1)(n-2)}  \sum_{\substack{i,j,k=1\\i\not=k, i\not=\textcolor{black}{j},k\not=j}}^n \big\{ h \big([\Xbm, \Ubm]_{i,j};\alpha\big)
   h\big([\Xbm, \Ubm]_{i,k};\alpha\big)  \nonumber \\
& \quad +\, h\big([\Xbm, \Ubm]_{j,i};\alpha\big)   h\big([\Xbm, \Ubm]_{j,k};\alpha\big)  + h\big([\Xbm, \Ubm]_{k,j};\alpha\big)   h\big([\Xbm, \Ubm]_{k,i};\alpha\big) \big\} \\
& \quad -\, 4\big\{\hmmd(P_\alpha,P)\big\}^2 =\Uc_n g(\cdot;\alpha)  -4\big\{\hmmd(P_\alpha,P)\big\}^2 .
\end{align*}
Thus, $\hat{\sigma}^2_{\alpha}$ can be decomposed into the weighted sum of the squared $\hmmd(P_\alpha,P)$ and a $U$-statistic of degree three with the symmetric $U$-kernel $g\big(\cdot;\alpha\big)$, as defined in (\ref{def_g_alpha}).
Obviously,  an estimator of $\sigma_\as^2$ can now be defined as
\begin{equation}
\hat\sigma^2_\an:= \Uc_n g(\cdot;\an) - 4\big\{\hmmd(P_\an,P)\big\}^2.
\label{hatsigma_an_V2}
\end{equation}


Similarly, we can define a $U$-statistic estimator of $\sigma^2_{q,\alpha}$ given by
\begin{eqnarray*}
\hat{\sigma}^2_{q,\alpha}&:=&\frac{2}{n(n/2-1)(n/2-2)}  \sum_{\substack{i,j,k=1\\i\not=k, i\not=\textcolor{black}{j},k\not=j}}^{n/2}  
\xi\big( [\Xbm, \Ubm]_{2i-1,2i, 2j-1,2j, 2k-1,2k };\alpha \big) \\
& & -\,  8\big\{\hmmd(P_\alpha,P)\big\}^2 =\Uc_{n}^{(2)} \xi(\cdot;\alpha) -  8\big\{\hmmd(P_\alpha,P)\big\}^2.
\end{eqnarray*}
where we define a symmetric (in $[\xbm, \ubm]_{1:2}$, $[\xbm, \ubm]_{3:4}$ and $[\xbm, \ubm]_{5:6}$) $U$-kernel as
\begin{eqnarray}
\lefteqn{\xi(\cdot;\alpha) : [\xbm, \ubm]_{1:6}
\mapsto  \frac{8}{3}\big\{  q\big([\xbm, \ubm]_{1,2,3,4};\alpha\big)  q\big([\xbm, \ubm]_{1,2,5,6};\alpha\big)} \nonumber \\
& &+\, q \big([\xbm, \ubm]_{3,4,1,2};\alpha\big) q\big([\xbm, \ubm]_{3,4,5,6};\alpha\big)    +  q\big([\xbm, \ubm]_{5,6,3,4};\alpha\big) q\big([\xbm, \ubm]_{5,6,1,2};\alpha\big) \big\} \label{def_xi_alpha}.
\end{eqnarray}
Note that the operator $\Uc_n^{(2)}\xi(\cdot,\alpha)$ is a usual $U$-statistic of degree three on the sample $[\Xbm, \Ubm]_{(2i-1):(2i)}$ for  $i=1,\ldots,n/2$. Now, an estimator of $\sigma_{q,\as}^2$ can be defined as $\hat{\sigma}^2_\an$.

In the model selection framework we can analogously define $U$-statistic estimators for the asymptotic variances $\sigma^2_{\alpha,\beta}$ and $\sigma^2_{q,\alpha,\beta}$. For a given tuple $(\alpha,\beta)$, we define
\begin{eqnarray}
\lefteqn{ \hat{\sigma}^2_{\alpha,\beta}:=\frac{1}{n(n-1)(n-2)}  \sum_{\substack{i,j,k=1\\i\not=j\not=k}}^n
g\big([\Xbm, \Ubm, \Vbm]_{i,j,k};\alpha,\beta\big) 
-  4\big\{\hmmd(P_\alpha,P)-\hmmd(Q_\beta,P)\big\}^2  \nonumber }\\
&& =\; \Uc_n g(\cdot;\alpha,\beta) -4\big\{\hmmd(P_\alpha,P)-\hmmd(Q_\beta,P)\big\}^2, \hspace{6cm}
\label{def_hat_sigma_alpha_beta}
\end{eqnarray}
as an estimator of $\sigma^2_{\alpha,\beta}$ where we define the symmetric (w.r.t.\ triplets $(\bm x,\bm u,\bm v)$ or concatenated pairs of triplets) $U$-statistic kernel
\begin{eqnarray}
\lefteqn{ g\big([\xbm, \ubm, \vbm]_{1:3};\alpha,\beta\big)
:=\frac{4}{3}\big\{ h\big([\xbm, \ubm, \vbm]_{1,2};\alpha,\beta\big)   h\big([\xbm, \ubm, \vbm]_{1,3};\alpha,\beta\big) \nonumber}\\
&&+\, h\big([\xbm, \ubm, \vbm]_{2,1};\alpha,\beta\big)  h\big([\xbm, \ubm, \vbm]_{2,3};\alpha,\beta\big) 
+\,  h\big([\xbm, \ubm, \vbm]_{3,2};\alpha,\beta\big)  h\big([\xbm, \ubm, \vbm]_{3,1};\alpha,\beta\big) \big\}. \hspace{1cm}\label{def_g_alpha_modelselection} 
\end{eqnarray}
Moreover, we can define an estimator of $\sigma^2_{q,\alpha,\beta}$ based on $\xi\big([\xbm, \ubm, \vbm]_{1:6};\alpha,\beta\big)$ by
\begin{eqnarray}
\hat{\sigma}^2_{q,\alpha,\beta}&:=&\frac{2}{n(n/2-1)(n/2-2)}
\sum_{\substack{i,j,k=1\\i\not=j\not=k}}^{n/2} \xi\big([\Xbm, \Ubm, \Vbm]_{2i-1,2i, 2j-1, 2j,2k-1, 2k};\alpha,\beta\big)  \nonumber \\
&  &-\, 8\big\{\hmmd(P_\alpha,P) - \hmmd(Q_\beta,P)\big\}^2  \nonumber \\
&=& \Uc_n^{(2)} \xi(\cdot;\alpha,\beta)  - 8\big\{\hmmd(P_\alpha,P) - \hmmd(Q_\beta,P)\big\}^2, \nonumber 
\end{eqnarray}
where we define
\begin{eqnarray}
\lefteqn{\xi\big([\xbm, \ubm, \vbm]_{1:6};\alpha,\beta\big)
:= \frac{8}{3} \big\{ q\big([\xbm, \ubm, \vbm]_{1,2,3,4};\alpha,\beta\big)  q\big([\xbm, \ubm, \vbm]_{1,2,5,6};\alpha,\beta\big) \label{def_xi_alpha_modelselection}}\\
&&+\, q\big([\xbm, \ubm, \vbm]_{3,4,1,2};\alpha,\beta\big)  q\big([\xbm, \ubm, \vbm]_{3,4,5,6};\alpha,\beta\big)    
+\, q\big([\xbm, \ubm, \vbm]_{5,6,3,4};\alpha,\beta\big)
q\big([\xbm, \ubm, \vbm]_{5,6,1,2};\alpha,\beta\big) \big\}. \nonumber
\end{eqnarray}
Therefore, $U$-statistic estimators for $\sigma_{\as,\bs}^2$ and $\sigma_{q,\as,\bs}^2$ are given by
\begin{align*}
    &\hat\sigma^2_{\an,\bn}:=\Uc_n g(\cdot;\an,\bn) -4\big\{\hmmd(P_\an,P)-\hmmd(Q_\bn,P)\big\}^2,\;\; \text{and} \\ 
&\hat\sigma^2_{q,\an,\bn}:= \Uc_n^{(2)} \xi(\cdot;\an,\bn)  - 8\big\{\hmmd(P_\an,P) - \hmmd(Q_\bn,P)\big\}^2,
\end{align*}
which are sums of parameter dependent $U$-statistics. 
The following Lemma shows that $\tilde{\sigma}^2_{\an}$, $\tilde{\sigma}^2_{q,\an}$, $\tilde\sigma^2_{\an,\bn}$ and $\tilde\sigma^2_{q,\an,\bn}$ are asymptotically equivalent to  $\hat{\sigma}^2_{\an}$, $\hat{\sigma}^2_{q,\an}$, $\hat{\sigma}^2_{\an,\bn}$ and  $\hat{\sigma}^2_{q,\an,\bn}$.

\begin{lem}
    \label{lemrelationvarest}
    \textcolor{black}{Under Assumptions~\ref{A_characteristic}-\ref{regular_model}}, $\tilde\sigma^2_{\an}= \hat{\sigma}^2_{\an}+ O_\Pb(n^{-1})$, $\tilde{\sigma}^2_{q,\an}=\hat{\sigma}^2_{q,\an}+O_\Pb(n^{-1})$, $\tilde\sigma^2_{\an,\bn}= \hat{\sigma}^2_{\an,\bn}+ O_\Pb(n^{-1})$ and $\tilde{\sigma}^2_{q,\an,\bn}=\hat{\sigma}^2_{q,\an,\bn}+O_\Pb(n^{-1})$. Additionally, $\sigma_\as>0$  whenever $P_\as \neq P$ and, under Assumption \ref{cond_resid_model_comp}, $\sigma_{\as,\bs}>0$ whenever $P_\as \neq P$ or $Q_\bs\not=P$.  
\end{lem}

\begin{proof}
First, a simple calculation shows that
   \begin{align*}
    \tilde\sigma^2_{\an}&= \hat{\sigma}^2_{\an} +\lc \frac{n-2}{n-1}-1\rc \Uc_n g(\cdot;\an) + \frac{4}{n-1} \Uc_n h(\cdot;\an)^2,  
\end{align*}
where $\Uc_n h(\cdot;\alpha)^2 $ is  a $U$-statistic of degree two with $U$-kernel $h(\cdot;\alpha)^2$ on the sample $(X_i,U_i)_{i=1,\ldots,n}$. The boundedness of $k$ implies that there exists a constant $C>0$ such that $\vert g(\cdot,\alpha)\vert\leq C$ and $\vert h^2(\cdot,\alpha)\vert \leq C $, which gives $\vert \Uc_n g(\cdot;\an)\vert \leq C$ and $ \vert\Uc_n h(\cdot;\an)^2\vert \leq C$. Thus,
$$\bigg\vert \lc \frac{n-2}{n-1}-1\rc \Uc_n g(\cdot;\an) + \frac{4}{n-1} \Uc_n h(\cdot;\an)^2\bigg\vert \leq  \frac{5}{n-1}C=O_\Pb(n^{-1}),$$
proving that $\tilde\sigma^2_{\an}= \hat{\sigma}^2_{\an}+ O_\Pb(n^{-1})$. Similar arguments show that $\tilde{\sigma}^2_{q,\an}=\hat{\sigma}^2_{q,\an}+O_\Pb(n^{-1})$, $\tilde\sigma^2_{\an,\bn}= \hat{\sigma}^2_{\an,\bn}+ O_\Pb(n^{-1})$ and $\tilde{\sigma}^2_{q,\an,\bn}=\hat{\sigma}^2_{q,\an,\bn}+O_\Pb(n^{-1})$. 

To prove that $\sigma_\as>0$ whenever $P_\as\neq P$, assume $\sigma_\as = 0$ and $\mmd(P_\as, P) >0$, seeking a contradiction. 
Note that $0=\sigma^2_\as=\var\lc \tilde{h}(X_1, U_1;\as)\rc$ implies $\tilde{h}(X_1, U_1;\as)=C$ a.s., and, w.l.o.g., assume $C\geq 0$. 
Since $\mathrm{MMD}(P_\as, P) =\Eb_{X,U}[\tilde{h}(X, U;\as)]$, $\tilde{h}(X_1, U_1;\as)=0$ a.s. implies 
$\mathrm{MMD}(P_\as, P) = 0$, which would be a contradiction. Thus $C>0$.
Moreover, observe that we can rewrite $\tilde{h}(X_1, U_1, \as) = T_1(X_1) + T_2(U_1)$, for some maps $T_1$ and $T_2$. Since $X_1$ and $U_1$ are independent, there exist two constants $C_1$ and $C_2$ s.t. $T_1(X_1) = C_1$ and $T_2(U_1) =C_2$ a.s., and at least one of $C_1$ and $C_2$ is non-zero. W.l.o.g. assume $C_1> 0$ which gives
$$ C_1=\mathbb{E}_{X_2} \big[ k(X_2, X_1) \big]  - \mathbb{E}_{U_2} \big[ k(X_1, F(U_2, \as)) \big]=:s_1(X_1)-s_2(X_1) .$$
This implies $s_1(x) - s_2(x) = C_1$ for almost every $x \in \mathrm{supp}(P)$. By definition of $T_2$, we have 
$C_2=T_2(U_1) = -s_1\big(F(U_1, \as)\big) + s_2\big(F(U_1, \alpha^*)\big) $ a.s.
Therefore, since $\text{supp}(P)\cap \text{supp}(P_\as)\not=\emptyset$, this yields $C_2=-C_1$ and, then, we have $\tilde{h}(X_1, U_1,\as)=0$ a.s., which is contradiction to $\mmd(P_\as,P)>0$. Thus $\sigma_{\as}>0$ when $P_\as\neq P$. The proof of $\sigma_{\as,\bs}>0$ whenever $P_\as\not=P$ or $Q_\bs\not=P $ is similar and is thus omitted. 
\end{proof}

 \subsection{Proof of Theorem~\ref{thm:nondegtestonemodel} (model specification)}
\label{proof_thm:nondegtestonemodel}


\textcolor{black}{We first state a general Lemma which is of interest per se.}

\begin{lem}
\label{lem:asympdistrmmd_simple}
Suppose that Assumptions~\ref{A_characteristic}-\ref{cond_residual_deriv} hold. 
\begin{enumerate}
    \item[(i)] If $\sqrt{n} \big[ \hmmd(P_\as,P)-\mmd^2(P_\as,P),\ \an-\as\big]$
weakly tends to a real-valued random vector $[Z_\as,V_\as]$ then 
$$\sqrt{n}\big\{ \hmmd(P_\an,P)-\mmd^2(P_\as,P)\big\}\stackrel{\text{law}}{\longrightarrow} Z_\as+\nabla_{\alpha^\intercal}\mmd^2(P_\alpha,P)\vert_{\alpha=\as} V_\as,$$
where $Z_\as \sim \Nc\lc 0,\sigma^2_\as\rc$. 
Moreover, $\textcolor{black}{\tilde{\sigma}}^2_\an \rightarrow \sigma^2_\as$ in probability.
    \item[(ii)] If \textcolor{black}{$\sqrt{n}(\an-\as)=O_{\Pb}(1)$ and }$P_\as=P$, \textcolor{black}{i.e. if $\Hc_{0,\Mc}$ is satisfied},  then $\sqrt{n} \hmmd(P_\an,P) =O_{\Pb}(n^{-1/2})$ and $\textcolor{black}{\tilde{\sigma}}^2_\an =O_{\Pb}(n^{-1})$.
    \item[(iii)] If $\sqrt{n} \big[ \hmmdq(P_\as,P)-\mmd^2(P_\as,P),\ \an-\as\big]$
weakly tends to a real-valued random vector $[Z_{q,\as},V_\as]$ then 
$$\sqrt{n}\big\{ \hmmdq(P_\an,P)-\mmd^2(P_\as,P)\big\}\stackrel{\text{law}}{\longrightarrow} Z_{q,\as} +\nabla_{\alpha^\intercal}\mmd^2(P_\alpha,P)\vert_{\alpha=\as} V_\as,$$ where $Z_{q,\as}\sim \Nc\lc 0,\sigma^2_{q,\as}\rc$.
    Moreover, $\textcolor{black}{\tilde{\sigma}}^2_{q,\an} \rightarrow \sigma^2_{q,\as}>0$ in probability.
\end{enumerate}
\end{lem}

In the latter lemma, the weak convergence of the two centered MMD estimators is guaranteed by standard $U$-statistics results (through H\'ajek projections). Here, the main point is their joint convergence with $\sqrt{n}(\an-\as)$, which is usually guaranteed when $\an-\as$ can be approximated by an i.i.d. expansion,
a typical situation with M-estimators \textcolor{black}{\citep[Section 5.3]{van2000asymptotic}}.
\textcolor{black}{For instance, when a log density is sufficiently regular (twice differentiable, in particular) w.r.t. its parameter, this is most often the case for maximum likelihood estimators. Such i.i.d. expansions directly appear when $\an$ are sample averages, differentiable functionals of them, or sample quantiles.}  
Note that the variance of $Z_\as$ is zero in the degenerate case $P_\as=P$, which will be the case under $\Hc_{0,\Mc}$.\\

\begin{proof}[of Lemma~\ref{lem:asympdistrmmd_simple}]
 \begin{enumerate}
     \item[\it (i):] A first order Taylor expansion yields
\begin{eqnarray}
   \hmmd(P_\an,P)&=&\frac{1}{n(n-1)} \sumijn   h\big([\Xbm, \Ubm]_{i,j};\as\big)   + O_{\Pb}(\Vert\an-\as\Vert^2)  \nonumber \\
&&+\, \frac{1}{n(n-1)} \sumijn  \nabla_{\alpha^\intercal} h\big([\Xbm, \Ubm]_{i,j};\as\big) \cdot (\an-\as) . \label{Taylor_F_differ}
\end{eqnarray}
Note that the remainder term $O_{\Pb}(\Vert\an-\as\Vert^2) $ is due to Assumption~\ref{cond_residual_deriv} and is $O_{\Pb}(n^{-1})$.
The weak convergence of $\sqrt{n}\big\{ \hmmd(P_\an,P)-\mmd^2(P_\as,P)\big\}$ is a direct consequence of weak convergence assumption, in addition to Leibniz's theorem applied to $\nabla_\alpha h$ at $\alpha=\as$ (second part of Assumption~\ref{cond_residual_deriv}). \textcolor{black}{Recalling~(\ref{hatsigma_an_V2})}, the convergence of the estimated variance $\hat\sigma^2_\an$ is easily obtained by a first order Taylor expansion of the maps $\alpha\mapsto g\big( [\xbm, \ubm]_{i,j,k};\alpha
\big)$
defined in~(\ref{def_g_alpha}) around $\as$, for every triplet of indices $(i,j,k)$, $i\neq j\neq k$. \textcolor{black}{Therefore, 
$\hat\sigma^2_\an=\Uc_n g(\cdot;\as)- 4\big\{\hmmd(P_\an,P)\big\}^2+o_{\Pb}(1)$ that tends to $\sigma_\as^2$ by the consistency of usual $U$-statistics. 
Note that we have again invoked 
Assumption~\ref{cond_residual_deriv} to manage the remainder term.
This immediately implies the consistency of $\tilde{\sigma}^2_{\an}$ by Lemma \ref{lemrelationvarest}}. 

\item[\it (ii):] With obvious notations, rewrite~(\ref{Taylor_F_differ}) as
$$  \hmmd(P_\an,P)= \Uc_n h(\cdot;\as) + \Uc_n \nabla_{\alpha^\intercal} h(\cdot;\as)  .(\an-\as)+O_{\Pb}(n^{-1}).$$
Since $\Uc_n h(\cdot;\as)$ is a degenerate $U$-statistic, it is $O_{\Pb}(n^{-1})$.
Moreover, the mean of the $U$-statistic $\Uc_n \nabla_{\alpha^\intercal} h(\cdot;\as)$ is zero because
$$ \e\big[ \nabla_\alpha h\big([\Xbm, \Ubm]_{i,j};\as\big)\big]=\nabla_\alpha \e\big[   h\big([\Xbm, \Ubm]_{i,j};\as\big)\big]=
 \nabla_\alpha\mmd^2(P_\alpha,P)\vert_{\alpha=\as}=0.
$$
Thus, the $U$-statistic $\Uc_n \nabla_{\alpha^\intercal} h(\cdot;\as)$ is $O_{\Pb}(n^{-1/2})$.
This implies $\sqrt{n}\,\hmmd(P_\an,P)=O_{\Pb}(n^{-1/2})$, the announced result.

Concerning the asymptotic variance, a Taylor expansion of $\alpha \mapsto g(\cdot;\alpha)$ around $\as$ and Assumption~\ref{cond_residual_deriv} yield
\begin{align*}
    \hat \sigma^2_\an&=\Uc_n g(\cdot;\an)-4\big\{\hmmd(P_\an,P\big\}^2\\
    &=\Uc_n g(\cdot;\as) + \Uc_n \nabla_{\alpha^\intercal} g(\cdot;\as) (\an-\as)+ O_\Pb\big(\|\an-\as\|^2\big)+ O_\Pb(n^{-2}).   
\end{align*}
Since it can be checked that $\Uc_n g(\cdot;\as)$ is a degenerate $U$-statistic, this term is $O_{\Pb}(n^{-1})$.

Moreover, $\Uc_n \nabla_\alpha g(\cdot;\as) $ tends in probability to
$\nabla_\alpha\e\big[  g\big([\Xbm, \Ubm]_{1:3};\as\big)\big]$ (Assumption~\ref{cond_residual_deriv}), that is zero because $\alpha \mapsto \e\big[   
g\big([\Xbm, \Ubm]_{1:3};\alpha\big)\big]$ is minimized at $\alpha=\as$.
This implies that the $U$-statistic $\Uc_n \nabla_\alpha g(\cdot;\as) $ is $O_{\Pb}(n^{-1/2})$.
Globally, we get $\hat \sigma^2_\an=O_{\Pb}(n^{-1})$\textcolor{black}{, which immediately implies $\tilde{\sigma}^2_{\an}=O_\Pb(n^{-1})$ by Lemma \ref{lemrelationvarest}}.
\item[\it (iii):]
        The positivity of $\sigma_{q,\as}$ had already been noted at the end of Section~\ref{sec:asymp_var_estim}.
        The rest can be proved exactly as for {\it (i)}.
 \end{enumerate}    
\end{proof}

The proof of Theorem~\ref{thm:nondegtestonemodel} follows from Lemma~\ref{lem:asympdistrmmd_simple}, with a few adjustments:
in the statement of Lemma~\ref{lem:asympdistrmmd_simple}, the term $\nabla_{\alpha^\intercal}\mmd^2(P_\alpha,P)\vert_{\alpha=\as} V_\as$ accounts for the influence of parameter estimation on the asymptotic distribution of $\hmmd(P_\an,P)$ and $\hmmdq(P_\an,P)$. \textcolor{black}{This will no longer be the case under $\Hc_{0,\Mc}$, 
because then $\nabla_{\alpha^\intercal}\mmd^2(P_\alpha,P)\vert_{\alpha=\as}=0$. This explains why only $\sqrt{n}(\an-\as)=O_{\Pb}(1)$ is required to prove Theorem~\ref{thm:nondegtestonemodel}. Moreover, since 
$\sqrt{n} \big\{ \hmmd(P_\as,P)-\mmd^2(P_\as,P)\big\}$ and  
$\sqrt{n} \big\{ \hmmdq(P_\as,P)-\mmd^2(P_\as,P)\big\}$ are usual $U$-statistics, they are jointly weakly convergent.}

\begin{enumerate}
\item Note that
$$ \sqrt{n}\frac{\hmmden(P_\an,P)}{\tilde{\sigma}_{\an}+\epsilon_n\tilde{\sigma}_{q,\an}}
=\frac{\sqrt{n}\,\hmmd(P_\an,P)}{\tilde{\sigma}_{\an}+\epsilon_n\tilde{\sigma}_{q,\an}}+\frac{\sqrt{n}\epsilon_n\;\sqrt{n}\,\hmmdq(P_\an,P)}{ \sqrt{n}\tilde{\sigma}_{\an}+\sqrt{n}\epsilon_n\tilde{\sigma}_{q,\an}}\cdot $$
Invoking Lemma \ref{lem:asympdistrmmd_simple} (ii), we obtain $$\frac{\sqrt{n}\,\hmmd(P_\an,P)}{\tilde{\sigma}_{\an}+\epsilon_n\tilde{\sigma}_{q,\an}}=\frac{O_{\Pb}(n^{-1/2})}{O_\Pb(n^{-1/2})+\epsilon_n\tilde{\sigma}_{q,\an}}=
\frac{O_{\Pb}(1)}{O_\Pb(1)+\sqrt{n}\epsilon_n \tilde{\sigma}_{q,\an}}\cdot $$
Since $\epsilon_n\sqrt{n}\to\infty$ in probability by assumption and $\tilde{\sigma}_{q,\an}\rightarrow \sigma_{q,\as}>0$ in probability, this yields
$$\frac{\sqrt{n}\,\hmmd(P_\an,P)}{\tilde{\sigma}_{\an}+\epsilon_n\tilde{\sigma}_{q,\an}}=o_{\Pb}(1).$$
Again, by Lemma \ref{lem:asympdistrmmd_simple} and the consistency of $\tilde{\sigma}_{q,\an}$, we have $\sqrt{n}\epsilon_n\tilde{\sigma}_{q,\an}\rightarrow \infty$ in probability, when
$\sqrt{n}\tilde{\sigma}_{\an} = O_{\Pb}(1)$.
This provides
$$ \frac{\sqrt{n}\epsilon_n \sqrt{n}\,\hmmdq(P_\an,P)}{\sqrt{n}\tilde{\sigma}_{\an} + \sqrt{n}\epsilon_n\tilde{\sigma}_{q,\an}}=
\frac{\sqrt{n}\,\hmmdq(P_\an,P)}{\tilde{\sigma}_{q,\an}}+o_{\Pb}(1)\stackrel{\text{law}}{\longrightarrow} \frac{\Nc(0,\sigma^2_{q,\as})}{\sigma_{q,\as}}\sim \Nc(0,1) ,$$
which proves the claim.
\item When $P_\as\not=P$, $\sqrt{n}\big\{\hmmd(P_\an,P)-\mmd^2(P_\as,P)\big\}\textcolor{black}{=Z_\as+O_{\Pb}(1)}$ 
where $Z_\as$ is a non-degenerate random variable and
$\tilde{\sigma}_\an\to\sigma_\as>0$ in probability, due to Lemmas \ref{lemrelationvarest} and \ref{lem:asympdistrmmd_simple} (i).
Therefore, the denominator of $\Tc_n(\Mc,P)$ tends to $\sigma_\as$ in probability and
$$ \frac{\sqrt{n}\,\hmmd(P_\an,P)}{\tilde{\sigma}_{\an}+\epsilon_n\tilde{\sigma}_{q,\an}} =
\sqrt{n}\frac{\mmd^2(P_\as,P)}{\sigma_\as} + O_{\Pb}(1).$$
Similarly, Lemma \ref{lem:asympdistrmmd_simple} (iii) \textcolor{black}{and $\sqrt{n}(\an-\as)=O_{\Pb}(1)$} yields
$$ \epsilon_n\frac{\sqrt{n}\,\hmmdq(P_\an,P)}{\tilde{\sigma}_{\an}+\epsilon_n\tilde{\sigma}_{q,\an}} =
\epsilon_n\sqrt{n}\frac{\mmd^2(P_\as,P)}{\sigma_\as} + o_{\Pb}(1).$$
This provides
$$\sqrt{n}\frac{\hmmden(P_\an,P)}{\tilde{\sigma}_{\an}+\epsilon_n\tilde{\sigma}_{q,\an}}= O_\Pb(1)+\sqrt{n}(1+\epsilon_n)\frac{\mmd^2(P_\as,P)}{\sigma_\as},$$
which implies the claim since $\mmd^2(P_\as,P)>0$.
\end{enumerate}

\subsection{Technical assumptions and proof of Theorem~\ref{thm:nondegtestonemodel_general} (model specification)}
\label{reg_assump_nondiff_generating_fct}

For a given $\delta >0$ \textcolor{black}{and in addition to~(\ref{def_Fc}) and~(\ref{def_Fcq})}, define the classes of symmetric functions
   $$   \Gc := \Big\{ [\xbm, \ubm]_{1:3} \mapsto  g\big([\xbm, \ubm]_{1:3};\alpha\big) \big\vert \ \alpha\in B_\delta (\as) \Big\},\; \text{ and}$$
    $$\Qc := \Big\{ [\xbm, \ubm]_{1:6} \mapsto   \xi \big([\xbm, \ubm]_{1:6};\alpha\big) \big\vert \ \alpha\in B_\delta (\as) \Big\} ,$$
where $g(\cdot,\alpha)$ and $\xi(\cdot,\alpha)$ were defined in (\ref{def_g_alpha}) and (\ref{def_xi_alpha}), respectively.

In this section, we will assume that certain centered empirical $U$-processes weakly converge to their appropriate limits in some function spaces: for some families $\Lc$ as above, we will assume that
\begin{align}
\label{deffunctionalconvuprocess}
\big( n^{r/2} (\Uc_n^{(m)}-\Pb)\ell\big)_{\ell\in\mathcal{L}}\overset{law}{\longrightarrow} \lc \Gb_r\ell\rc_{\ell\in\mathcal{L}} \text{ in }\lc L_\infty(\mathcal{L}),\Vert \cdot\Vert_\infty\rc,
\end{align}
where the non-negative integer $r-1$ denotes the degree of degeneracy of the class of functions $\mathcal{L}$ and $\Gb_r$ denotes a stochastic process on
$ \mathcal{L}$ with bounded and uniformly continuous sample paths w.r.t. the $L_2$-norm on the appropriate product space which is in accordance with the arguments of $\ell\in\mathcal{L}$. When~(\ref{deffunctionalconvuprocess}) is satisfied, we say that $n^{r/2} (\Uc_n^{(m)}-\Pb)$ indexed by ${\mathcal L}$ weakly converges. For example, when $r=1$, $\big(n^{1/2} (\Uc_n^{(2)}-\Pb)f\big)_{f\in\mathcal{F}_q} $ weakly converges to a Gaussian process $\big(\mathbb{G}_1 f\big)_{f\in\mathcal{F}_q}$ in $L_\infty(\Fc_q)$ which has uniformly continuous sample paths w.r.t.\ $\Vert\cdot \Vert_{L_2(\otimes_{i=1}^4 P \otimes P_U)}$.

Sufficient conditions ensuring the weak convergence in (\ref{deffunctionalconvuprocess}) are, among others, provided in \cite{arconesgine1993},~\cite{arcones1994bootstrap},~\cite{wellner2013weak}. In particular, there exist many sufficient conditions that do not rely on any differentiability property of the functions in $\mathcal{L}$, which makes (\ref{deffunctionalconvuprocess}) particularly useful when considering non-differentiable generating functions. Moreover, as another strength of the functional convergence (\ref{deffunctionalconvuprocess}), it also ensures that the centered empirical $U$-process satisfies some asymptotic equicontinuity property, as follows:
$$\lim_{\delta\to 0} \lim\sup_{n\rightarrow \infty} \Pb^* \bigg( \sup_{ f,g\in\mathcal{L},\Vert f-g\Vert_{L_2} <\delta} \vert n^{r/2}(\Uc_n^{(m)}-\Pb)(f-g)\vert>\epsilon \bigg) =0,$$
\textcolor{black}{denoting $\Pb^*$ the outer probability associated with $\Pb$ \citep[Section 1]{wellner2013weak}.}
To illustrate, when $r=1$, the latter asymptotic equicontinuity property will allow us to ``replace'' expressions of the form $ \sqrt{n}(\Uc_n-\Pb) h(\cdot; \an) $ by $\sqrt{n}(\Uc_n-\Pb) h(\cdot; \as)$, since $\sqrt{n}(\Uc_n-\Pb) \lc h(\cdot; \an)- h(\cdot; \as)\rc$ vanishes in probability due to asymptotic equicontinuity.

We impose the following assumption, which ensures that the centered empirical $U$-processes of interest are asymptotically equicontinuous.

\begin{assumpt}
\label{ass_equicont}
There exists some $\delta>0$ such that the empirical $U$-processes
$$
\big(\sqrt{n}(\Uc_n-\Pb)f\big)_{f\in\Fc} \  ,\ \big(\sqrt{n}(\Uc_n^{(2)}-\Pb)q\big)_{q\in\Fc_q}\ , \ \big(\sqrt{n}(\Uc_n-\Pb)g\big)_{g\in\Gc} \ ,\   \big(\sqrt{n}(\Uc_n^{(2)}-\Pb)\xi\big)_{f\in\Qc}
$$
weakly converge to their appropriate limits in the functional sense. Moreover,
$$\Vert h(\cdot; \an)-h(\cdot;\as)  \Vert_{L_2(\otimes_{i=1}^2 P\otimes P_U )}\ , \ \Vert q(\cdot; \an)-q(\cdot; \as) \Vert_{L_2(\otimes_{i=1}^4 P\otimes P_U )} \ ,\;\text{and}
$$
$$
\Vert g(\cdot;\an)-g(\cdot;\as)  \Vert_{L_2(\otimes_{i=1}^3 P\otimes P_U )} \ ,\text{and} \  \Vert \xi(\cdot;\an)-\xi(\cdot;\as)  \Vert_{L_2(\otimes_{i=1}^6 P\otimes P_U )}
$$
tend to zero with $n$ in probability.
\end{assumpt}

Instead of providing technical conditions that ensure the functional convergence of our centered empirical $U$-processes, we have opted to impose the latter rather ``high-level'' assumption to avoid an excessively technical discussion. Nonetheless, we provide explicit sufficient conditions for Assumption~\ref{ass_equicont} in Appendix~\ref{sec:appsuffcondempuprocess} and exemplarily verify the conditions in a ReLu-type generative neural network example in Appendix \ref{app:relu-type_example}.

To later determine the exact rates of convergence of some quantities related to our model specification test, we need to further introduce the auxiliary map
\begin{align}
\tilde{g}(x,y;\alpha):\ \alpha \mapsto \frac{4}{3}\tilde h^2 (x,y;\alpha)
   +\frac{8}{3}\e\Big[  h\big((X,F(U; \alpha)), (x,y)\big)   \tilde h \big(X,F(U; \alpha);\alpha\big)\Big].
\label{def_tilde_g}
\end{align}
Note that $\tilde g(\cdot;\as)=0$ when $P=P_\as$ since $\tilde h(\cdot;\as)=0$ in this case.
We impose the following regularity conditions on $\tilde{h}(x,y;\alpha)$ and $\tilde{g}(x,y;\alpha)$.

\begin{assumpt}
\label{regularity_tildeh}
The maps $\alpha\mapsto \tilde{h}(x,y;\alpha)$ and $\alpha\mapsto \tilde{g}(x,y;\alpha)$ are
twice continuously differentiable in a neighborhood of $\alpha=\as$, for every $(x,y)\in  \Sc^2$.
The maps $\iota:\alpha \mapsto \e \Big[ \nabla_\alpha \tilde{h} \big( X, F(U; \alpha);\as\big)\Big]$
and $\zeta:\alpha \mapsto \e \Big[ \nabla_\alpha \tilde{g} \big( X, F(U; \alpha);\as\big)\Big]$ are differentiable in a neighborhood of $\alpha=\as$.
Moreover, $\e\lk \nabla_\alpha \tilde{h} \big( X, F(U; \as);\as\big)\rk= \nabla_\alpha \e\lk\tilde{h} \big( X, F(U; \as);\as\big)\rk$ and $\e\lk \nabla_\alpha \tilde{g} \big( X, F(U; \as);\as\big)\rk= \nabla_\alpha \e\lk\tilde{g} \big( X, F(U; \as);\as\big)\rk$ and there exists a real constant $\delta>0$ such that
\begin{align}
\label{cond_nabla_halpha}
\begin{cases}
    \;\e\Big[ \sup_{\alpha_1; \alpha_2 \in B_\delta(\as)} \| \nabla^2_{\alpha,\alpha^\intercal} \tilde{h} \big(X, F(U; \alpha_2);\alpha_1\big)\|_2^2 \Big] <\infty,\; \text{and} \\
\;\Eb\Big[ \sup_{\alpha_1,\alpha_2 \in B_\delta(\as)} \| \nabla^2_{\alpha,\alpha^\intercal} \tilde{g} \big(X, F(U; \alpha_2);\alpha_1\big) \|_2^2 \Big] <\infty.
\end{cases}
 \end{align}
\end{assumpt}

For any $\alpha_0\in \Theta_1$, the map $\nabla_\alpha \tilde{h} \big(x,F(u; \alpha_0);\as\big)$ denotes hereafter the derivative of $\alpha\mapsto \tilde h \big(x,F(u; \alpha_0);\alpha\big)$ at $\alpha=\as$.
Thus, $\nabla_\alpha \tilde{h} \big(x,F(u; \as);\as\big)$ should not be confused with the derivative of the map $\alpha \mapsto \tilde{h}\big(x, F(u; \alpha);\alpha\big)$ at $\alpha=\as$,
that may not exist because $\alpha\mapsto F(u; \alpha)$ may not be differentiable.
Similarly, $\nabla_\alpha \tilde{g} \big(x,F(u; \alpha_0);\as\big)$ denotes the derivative of $\alpha\mapsto \tilde g \big(x,F(u; \alpha_0);\alpha\big)$ at $\alpha=\as$.

Even if the map $\alpha\mapsto F(u; \alpha)$ might not be differentiable, Assumption \ref{regularity_tildeh} imposes that $\tilde{h}(x,y;\alpha)$ and its derivatives are sufficiently smooth w.r.t.\ $\alpha$ for any $(x,y)$.  This is legitimate because $\e\lk h\big((x,y), (X,F(U; \alpha))\big)\rk$ might be interpreted as a smoothing of the (possibly) non-differentiable function $h\big( (x_1,y), (x_2,F(u; \alpha)) \big)$. A similar statement applies to $\tilde{g}(x,y;\alpha)$.
Moreover, we denote the classes of functions
\begin{eqnarray*}
\mathcal{F}_c:= \Big\{ [\xbm, \ubm]_{1:2}&\mapsto& h\big([\xbm, \ubm]_{1:2};\alpha\big)-\tilde{h}\big(x_1,F(u_1; \alpha);\alpha\big)- \tilde{h} \big(x_2,F(u_2; \alpha);\alpha\big)\\
& & +\, \mmd^2(P_\alpha,P) \,\big\vert \ \alpha\in B_\delta (\as) \Big\},\\
        \mathcal{G}_c:= \Big\{ [\xbm, \ubm]_{1:3}&\mapsto& g([\xbm, \ubm]_{1:3};\alpha)-\tilde{g}\big(x_1,F(u_1; \alpha);\alpha\big)-\tilde{g} \big(x_2,F(u_2; \alpha);\alpha\big) \\
       & &-\,\tilde{g}\big(x_3,  F(u_3; \alpha);\alpha\big) + 2\Pb g(\cdot;\alpha) \, \big\vert \ \alpha\in B_\delta (\as) \Big\},
\end{eqnarray*}
$$\tilde{\mathcal{F}}:= \Big\{ (x,u)\mapsto \nabla_{\alpha} \tilde{h} \big(x,F(u; \alpha);\as\big) \,\big\vert \ \alpha\in B_\delta (\as) \Big\},\; \text{and}$$
$$ \tilde{\mathcal{G}}:= \Big\{ (x,u)\mapsto \nabla_\alpha \tilde{g} \big(x, F(u; \alpha);\as \big) \,\big\vert \ \alpha\in B_\delta (\as) \Big\}.
$$

We assume that the centered empirical ($U$-)processes indexed by the classes of functions $\Fc_c$, $\Gc_c$, $\tilde{\Fc}$ and $\tilde{\Gc}$ converge to their appropriate functional limits.

\begin{assumpt}
\label{ass_degenerateuprocess}
    There exists some $\delta>0$ such that the centered empirical $U$-processes $\big( n(\Uc_n-\Pb)f\big)_{f\in\Fc_c}$ and $\big( n(\Uc_n-\Pb)g\big)_{g\in\Gc_c}$, and the empirical processes $\big(\sqrt{n}(\Pb_n-\Pb)f\big)_{f\in\tilde \Fc}$ and $\big(\sqrt{n}(\Pb_n-\Pb)g\big)_{g\in\tilde \Gc}$ weakly converge.
    Further, the map  $\alpha\mapsto \sigma_\alpha$ is continuously differentiable in a neighborhood of $\as$. Finally,
$$ \e_{X,U}\lk \Vert \nabla_\alpha\Tilde{h}\big(X,F(U; \an);\as\big)  -\nabla_\alpha \Tilde{h}\big(X,F(U; \as);\as\big) \Vert_2^2 \rk,\;\text{as well as} $$
$$ \e_{X,U}\lk \Vert \nabla_\alpha\Tilde{g}\big(X,F(U; \an);\as\big)  -\nabla_\alpha \Tilde{g}\big(X,F(U; \as);\as\big) \Vert_2^2\rk $$
tend to zero in probability.
\end{assumpt}

Again, we provide some sufficient conditions to ensure the functional convergence of the latter centered empirical $U$-processes and empirical processes in Appendix \ref{sec:appsuffcondempuprocess}.
Now, we can derive the asymptotic distribution of our estimators of the MMD with estimated parameters.

\begin{lem}
\label{lem:asympdistrmmdnonsmooth}
Under Assumptions~\ref{A_characteristic}-\ref{regular_model} and~\ref{ass_equicont}-\ref{ass_degenerateuprocess}, the statements of Lemma \ref{lem:asympdistrmmd_simple} are valid.
\end{lem}


\begin{proof}[of Lemma~\ref{lem:asympdistrmmdnonsmooth}]

We prove the statements (i)-(iii) of Lemma~\ref{lem:asympdistrmmd_simple} successively.
\begin{enumerate}
    \item[\it (i):] Note that
        \begin{eqnarray*}
        \lefteqn{ \sqrt{n}\big\{ \hmmd(P_\an,P)-\mmd^2(P_\as,P)\big\} }\\
        &=&\sqrt{n}\big\{ \hmmd(P_\an,P)-\mmd^2(P_\an,P)-\hmmd(P_\as,P)+\mmd^2(P_\as,P)\\
        &&+\,\hmmd(P_\as,P)-\mmd^2(P_\as,P)+\mmd^2(P_\an,P)-\mmd^2(P_\as,P)\big\}\\
        &=&\sqrt{n}(\Uc_n-\Pb) \big(h(\cdot; \an)-h(\cdot;\as)\big)+\sqrt{n}(\Uc_n-\Pb)h(\cdot; \as)\\
        &&+\,\sqrt{n}\nabla_{\alpha^\intercal}\mmd^2(P_\alpha,P)\vert_{\alpha=\as}(\an-\as) +\sqrt{n}\,o_{\Pb}(\Vert\an-\as\Vert),
        \end{eqnarray*}
        by a Taylor expansion of $\alpha\mapsto \mmd^2(P_\alpha,P)$.
    Since the process $\sqrt{n}(\Uc_n - \Pb)$, indexed by $\Fc$, is asymptotically equicontinuous by Assumption \ref{ass_equicont} and the convergence  $\Vert h(\cdot; \an)-h(\cdot; \as)  \Vert_{L_2(P\otimes P_U\otimes P\otimes P_U)}\to 0$ in probability holds, we get that
        $\sqrt{n}(\Uc_n-\Pb) \big(h(\cdot;\an)-h(\cdot; \as)\big)$ is $o_\Pb(1)$.
        Moreover, by \citet[Theorem 4.9]{arconesgine1993}, $\sqrt{n}(\Uc_n-\Pb)h(\cdot; \as)$ converges in law towards a normal random variable whose variance is
        $$\sigma_\as^2=\var \Big( 2\e_{X_2,U_2}\big[  h\big([\Xbm, \Ubm]_{1:2};\as\big) |\, X_1,U_1 \big] \Big).
$$
By assumption, $\sqrt{n}\nabla_{\alpha^\intercal}\mmd^2(P_\as,P)(\an-\as)$ converges to
$\nabla_{\alpha^\intercal}\mmd^2(P_\as,P) V_\as$
and the joint convergence follows from our assumptions.
The fourth term is clearly $o_{\Pb}(1)$, since $\Vert \an-\as\Vert=O_{\Pb}(n^{-1/2})$. Thus, we have
$$\sqrt{n}\big\{ \hmmd(P_\an,P)-\mmd^2(P_\as,P)\big\}\stackrel{\text{law}}{\longrightarrow} Z_\as+\nabla_{\alpha^\intercal}\mmd^2(P_\alpha,P)\vert_{\alpha=\as} V_\as.$$
Furthermore, $\hat{\sigma}^2_{\an}= \Uc_n g(\cdot;\an) - 4\Big\{ \hmmd(P_\an,P) \Big\}^2$.
We have just  deduced the convergence of  $\hmmd(P_\an,P)$ to $\mmd^2(P_\as,P)$ in probability.
Similarly, since $\sqrt{n}(\Uc_n -\Pb)$ indexed by $\Gc$ converges to a Gaussian limit and $ \Vert g(\cdot;\an)-g(\cdot;\as)  \Vert_{L_2(\otimes_{i=1}^3 P\otimes P_U )} \to 0$ in probability, we have
$$
\Uc_n g(\cdot;\an) = \Eb_{[\Xbm', \Ubm']_{1:3}}\big[ g(\cdot;\an) \big( [\Xbm', \Ubm']_{1:3}\big)  \big] + o_{\Pb}(1),\;\text{and}
$$
$$
 \Big\vert \Eb_{[\Xbm', \Ubm']_{1:3}} \lk g(\cdot;\an) \big([\Xbm', \Ubm']_{1:3}\big)
        - g \big([\Xbm', \Ubm']_{1:3};\as\big)  \rk \Big\vert\leq   \Vert g(\cdot;\an)-g(\cdot;\as)  \Vert_{L_2(\otimes_{i=1}^3 P\otimes P_U )} \stackrel{\Pb}{\rightarrow} 0,
$$
where $[\Xbm', \Ubm']_{1:3}$ is an independent copy of $[\Xbm, \Ubm]_{1:3}$.
Thus, this yields
$$\Uc_n g(\cdot;\an) = \Eb[ g \big([\Xbm, \Ubm]_{1:3};\as\big) ] + o_{\Pb}(1),
$$
proving that $\hat{\sigma}_\an^2\to\sigma^2_\as$ in probability. \textcolor{black}{Therefore, we also have $\tilde{\sigma}^2_{\an}\to\sigma^2_\as$ in probability by Lemma \ref{lemrelationvarest}.}

\item[\it (ii):]    Obviously, when $P_\as=P$, $\nabla_\alpha\mmd^2(P_\alpha,P)\vert_{\alpha=\as}=0$ since $\as$ belongs to the interior of $\Theta_1$ and is an argmin of $\alpha\mapsto \mmd^2(P_\alpha,P)$. Moreover, $\Uc_n h(\cdot;\as)=O_{\Pb}(n^{-1})$ by standard $U$-statistic arguments, because $h(\cdot;\as)$ is now a degenerate $U$-statistic kernel. Note that, by Assumption~\ref{regular_model}, $\sqrt{n}\,\mmd^2(P_\an,P)=O_{\Pb}(n^{-1/2})$ since
\begin{eqnarray*}
\mmd^2(P_\an,P)&=&(\an-\as)^\intercal\nabla^2_{\alpha,\alpha^\intercal}\mmd^2(P_\alpha,P)\vert_{\alpha=\as}(\an-\as)+ o_{\Pb}\lc\Vert \an-\as\Vert^2\rc\\
&=& O_{\Pb}\lc\Vert \an-\as\Vert^2\rc,
\end{eqnarray*}
which is $ O_{\Pb}(n^{-1})$. Let us decompose the quantity of interest as
        $$\sqrt{n}\, \hmmd(P_\an,P)  = \sqrt{n}\,\Uc_n h(\cdot;\an) = \sqrt{n}\,\Uc_n \big(h(\cdot; \an)-h(\cdot; \as)\big) + O_{\Pb}(n^{-1/2}).$$
Thus, it remains to prove that $\sqrt{n}\,\Uc_n\big( h(\cdot; \an)-h(\cdot; \as)\big)=O_{\Pb}(n^{-1/2})$. To this aim, we will use the asymptotic equicontinuity of degenerate $U$-statistic kernels.
Obviously, for any $\alpha\neq \as$, $h(\cdot;\alpha)-h(\cdot;\as)$ is generally not a degenerate $U$-statistic kernel. Nonetheless, we can rewrite
\begin{eqnarray}
 \sqrt{n} \,\Uc_n \big(h(\cdot;\an)-h(\cdot;\as)\big)
            &=&
\frac{\sqrt{n}}{n(n-1)}\sumijn \big\{ \psi\big([\Xbm, \Ubm]_{i,j};\an\big) - \psi\big([\Xbm, \Ubm]_{i,j};\as\big) \big\}  \nonumber \\
&&+\, \frac{2\sqrt{n}}{n}\sum_{i=1}^n  \tilde{h}\big(X_i,F(U_i; \an);\an\big) +O_{\Pb}(n^{-1/2}) ,\,  \label{degenerate_Term0}
\end{eqnarray}
\begin{equation*}
\psi\big([\xbm, \ubm]_{1:2};\alpha\big):=   h\big( [\xbm, \ubm]_{1:2};\alpha\big) -\tilde{h}\big(x_1,F(u_1; \alpha);\alpha\big)
        -\tilde{h}(x_2,F(u_2; \alpha);\alpha \big)
+\mmd^2(P_\alpha,P).
\end{equation*}
Note that $\psi\big([\xbm, \ubm]_{1:2};\as\big)= h\big([\xbm, \ubm]_{1:2};\as\big)$ because $\tilde{h}\big(x,y;\as)=0$ for every $(x,y)$, when $P=P_\as$.
Moreover, the $U$-statistic $\Uc_n \psi(\cdot;\alpha)$ is now degenerate, for every $\alpha \in \Theta_1$.
 For any $\epsilon>0$ and any positive constants $\delta_k$, $k\in \{1,2\}$, we get
\begin{eqnarray*}
\lefteqn{ \Pb\bigg(  \Big\vert \frac{\sqrt{n}}{n(n-1)}\sumijn \big\{ \psi\big([\Xbm, \Ubm]_{i,j};\an\big)- \psi\big([\Xbm, \Ubm]_{i,j};\as\big) \big\} \Big\vert > n^{-1/2}\epsilon \bigg) }\\
& \leq &
           \Pb\Big(  \sup_{f,g\in\Fc_c  \Vert f-g \Vert_{L_2(\otimes_{i=1}^2 P\otimes P_u)}<\delta_2 }  \big\vert n(\Uc_n-\Pb) (f-g)  \big\vert > \epsilon \Big)+ \Pb(\|\an-\as\|\geq  \delta_1) \\
           &&+\,
           \Pb\Big( \Vert \psi(\cdot;\an)-\psi(\cdot;\as)\Vert_{L_2(\otimes_{i=1}^2(P\otimes P_U))}\geq  \delta_2, \|\an-\as\|<  \delta_1 \Big) =:T_1+T_2+T_3.
        \end{eqnarray*}
By the asymptotic equicontinuity of $n(\Uc_n-\Pb)$ indexed by $\Fc_c$ (Assumption~\ref{ass_degenerateuprocess}), there exists $\delta_2$ s.t. $T_1$ is arbitrarily small for $n$ sufficiently large.
Since
$$\Vert \psi(\cdot;\an)-\psi(\cdot; \as) \Vert_{L_2(\otimes_{i=1}^2 (P\otimes P_U))}\leq  4\Vert h(\cdot; \an)-h(\cdot; \as) \Vert_{L_2(\otimes_{i=1}^2 (P\otimes P_U))} $$
that tends to zero in probability, $\delta_1$ can be chosen so that $T_2$ and $T_3$ are arbitrarily small for sufficiently large $n$. Globally, we have obtained that
        \begin{equation}
        \frac{\sqrt{n}}{n(n-1)}\sumijn \big\{\psi\big([\Xbm, \Ubm]_{i,j};\an\big) - \psi\big([\Xbm, \Ubm]_{i,j};\as\big)\big\}  =  o_{\Pb}(n^{-1/2}).
        \label{degenerate_Term1}
        \end{equation}
        Moreover, since $\tilde{h}(x,y;\alpha)$ is twice continuously differentiable w.r.t.\ $\alpha$ by Assumption~\ref{regularity_tildeh}, we obtain
        by a limited expansion
        \begin{eqnarray*}
        \lefteqn{ \frac{2\sqrt{n}}{n}\sum_{i=1}^n  \tilde{h} \big(X_i,F(U_i; \an);\an\big)
             =\frac{2\sqrt{n}}{n}\sumn \Big\{ \tilde{h} \big(X_i,F(U_i; \an);\as\big)   }\\
             &&+\,\nabla_{\alpha^\intercal}\tilde{h} \big(X_i,F(U_i; \an);\as\big) (\an-\as)+2^{-1}
             (\an-\as)^\intercal\nabla^2_{\alpha,\alpha^\intercal}\tilde{h} \big(X_i,F(U_i;\an);\tilde\alpha_n\big) (\an-\as) \Big\},
        \end{eqnarray*}
        for some random parameter $\tilde\alpha_n$ that satisfies $\|\tilde\alpha_n - \as\|< \|\an-\as\|$.
        Due to Assumption~\ref{regularity_tildeh} (Equation~(\ref{cond_nabla_halpha})) and since $\tilde h_\as=0$ under the null, this yields
        \begin{eqnarray*}
        \lefteqn{ \frac{2\sqrt{n}}{n}\sum_{i=1}^n  \tilde{h} \big(X_i,F(U_i; \an);\an\big)
        =\frac{2\sqrt{n}}{n}\sumn \nabla_{\alpha^\intercal}\tilde{h} \big(X_i,F(U_i;\an);\as\big) (\an-\as)+O_{\Pb}(n^{-1/2})  }\\
        &=&\sqrt{n} (\an-\as)^\intercal \frac{2}{n}\sumn \Big\{ \nabla_\alpha\tilde{h} \big(X_i,F(U_i;\an);\as\big)-\nabla_\alpha\tilde{h} \big(X_i,F(U_i;\as);\as\big)\Big\}\\
        &&+\,\sqrt{n} (\an-\as)^\intercal \frac{2}{n}\sumn \nabla_\alpha\tilde{h} \big(X_i,F(U_i;\as);\as\big) +O_{\Pb}(n^{-1/2}) \\
        &=& 2\sqrt{n} (\an-\as)^\intercal \Pb_n \Big\{\nabla_\alpha\tilde{h} \big( \cdot,F(\cdot;\an);\as\big) - \nabla_\alpha\tilde{h} \big( \cdot,F(\cdot;\as);\as\big) \Big\} \\
        &&+\,2\sqrt{n} (\an-\as)^\intercal \Pb_n \nabla_\alpha\tilde{h} \big(\cdot,F(\cdot;\as);\as\big) +O_{\Pb}(n^{-1/2}).
      \end{eqnarray*}
        Observe that $  \e \Big[ \nabla_\alpha \tilde h\big(X,F(U;\as);\as\big)\Big]$
        $= \nabla_\alpha \e \Big[  \tilde h\big(X,F(U;\as);\as\big)\Big]=\nabla_\alpha 0=0$ due to Assumption \ref{regularity_tildeh} and $P_\as=P$. 
        Moreover, since
        the map $\alpha \mapsto \iota(\alpha)=\e \Big[ \nabla_\alpha \tilde{h} \big( X, F(U;\alpha);\as\big)\Big]$ is assumed to be differentiable at $\alpha=\as$ (Assumption~\ref{regularity_tildeh}), a Taylor expansion provides
        $$ \Vert \iota(\an) \Vert_1= \Vert\iota(\as)+ (\an-\as)^\intercal \nabla_\alpha \iota(\as)\Vert_1+o_{\Pb}(\Vert \an-\as\Vert)=0+O_{\Pb}(n^{-1/2}).$$
        Deduce
        \begin{equation*}
        \Vert\Pb \Big\{\nabla_\alpha\tilde{h} \big( \cdot,F(\cdot;\an);\as\big) - \nabla_\alpha\tilde{h} \big( \cdot,F(\cdot;\as);\as \big) \Big\}\Vert_1=\Vert\iota(\an)-\iota(\as)\Vert_1=O_{\Pb}(n^{-1/2}).
        \label{relation_pi_deriv}
        \end{equation*}


        Therefore, we get
        \begin{eqnarray*}
        \lefteqn{ \frac{2\sqrt{n}}{n}\sum_{i=1}^n  \tilde{h} \big(X_i,F(U_i; \an);\an\big)
        = 2\sqrt{n} (\an-\as)^\intercal (\Pb_n - \Pb) \nabla_\alpha\tilde{h} \big(\cdot,F(\cdot;\as);\as\big)}\\
        &&+\, 2\sqrt{n} (\an-\as)^\intercal(\Pb_n - \Pb) \Big\{\nabla_\alpha\tilde{h} \big( \cdot,F(\cdot;\an);\as \big) - \nabla_\alpha\tilde{h} \big( \cdot;F(\cdot;\as);\as\big) \Big\}  + O_{\Pb}(n^{-1/2}).\hspace{2cm}
      \end{eqnarray*}

        Since $\e_{X,U}\Big[ \| \nabla_\alpha\tilde{h}\big(X,F(U;\an);\as\big)-\nabla_\alpha\tilde{h}\big(X,F(U;\as);\as \big)\|_2^2\Big]\stackrel{\Pb}{\rightarrow} 0 $ and  $\sqrt{n}\big (\Pb_n -\Pb)f \big)_{f\in\Tilde{\Fc}}$ converges to a tight Gaussian limit (Assumption~\ref{ass_degenerateuprocess}), the asymptotic equicontinuity of the latter process yields
        $$ 2\sqrt{n} (\an-\as)^\intercal(\Pb_n - \Pb) \Big\{\nabla_\alpha\tilde{h} \big( \cdot,F(\cdot;\an);\as\big) - \nabla_\alpha\tilde{h} \big( \cdot,F(\cdot;\as);\as\big) \Big\} =o_{\Pb}(n^{-1/2}).$$
        Moreover, the usual CLT yields
        $ \sqrt{n} (\an-\as)^\intercal (\Pb_n - \Pb) \nabla_\alpha\tilde{h} \big(\cdot,F(\cdot;\as);\as\big) =O_\Pb(n^{-1/2}).$
        In other words, we have obtained
        \begin{equation}
        \frac{2\sqrt{n}}{n}\sum_{i=1}^n  \tilde{h} \big(X_i,F(U_i;\an);\an\big) = O_\Pb(n^{-1/2}).
        \label{degenerate_Term2}
        \end{equation}
        Therefore, by~(\ref{degenerate_Term0}),~(\ref{degenerate_Term1}) and~(\ref{degenerate_Term2}), we get
         $\sqrt{n}\,\Uc_n \big(h(\cdot;\an)-h(\cdot;\as)\big)=O_\Pb(n^{-1/2})$ and $\hmmd(P_\an,P)=O_{\Pb}(n^{-1})$.

Furthermore, note that we have also just proven that
$$\hat{\sigma}^2_\an=\Uc_n g(\cdot;\an)-4\big(\hmmd(P_\an,P)\big)^2=\Uc_n g(\cdot;\an) +O_\Pb(n^{-2}).$$
Thus, it remains to show that $\Uc_n g(\cdot;\an) =O_{\Pb}(n^{-1})$ when $P_\as=P$. To this aim, observe that
\begin{align*}
         &3\e\lk g\big([\Xbm, \Ubm]_{1:3};\as\big) \mid X_1,U_1 \rk/4=\tilde{h}^2 \big(X_1,F(U_1;\as);\as\big)\\
         &\quad +\e\lk \e\lk h \big([\Xbm, \Ubm]_{2,1};\as\big)   h\big([\Xbm, \Ubm]_{2,3};\as\big)
         \mid [\Xbm, \Ubm]_{1,2}\rk\mid X_1,U_1 \rk\\
         &\quad +\e\lk \e\lk h\big([\Xbm, \Ubm]_{3,2};\as\big)   h\big([\Xbm, \Ubm]_{3,1};\as\big)
         \mid [\Xbm, \Ubm]_{1, 3}\rk\mid X_1,U_1 \rk\\
         &=\tilde{h}^2 \big(X_1,F(U_1;\as);\as\big)+2\e\lk h\big([\Xbm, \Ubm]_{2,1};\as\big)
         \tilde{h}\big( X_2,F(U_2;\as);\as \big) \mid X_1,U_1   \rk
         =0,
        \end{align*}
        since $\tilde{h}(\cdot,\cdot ;\as)=0$ when $P_\as=P$, i.e., $g(\cdot;\as)$ is a degenerate $U$-statistic kernel.
        Therefore, as a degenerate $U$-statistic, $\big(\Uc_n-\Pb\big)  g(\cdot;\as) $ is $O_{\Pb}(n^{-1})$.
        Since $\Pb g(\cdot;\an) = \sigma_\an^2+ O_{\Pb}(n^{-1})$, deduce
        $$  \Uc_n g(\cdot;\an)=(\Uc_n-\Pb)g(\cdot;\as)+(\Uc_n-\Pb)(g(\cdot;\an)-g(\cdot;\as)) +\sigma^2_\an+O_{\Pb}(n^{-1}).$$
        Since $\alpha\mapsto \sigma_\alpha$ is continuously differentiable in a neighborhood of $\as$, there exists $\bar\alpha\in \Theta_1$ s.t.
        $ \sigma_\alpha = ( \alpha - \as)^\intercal \nabla_\alpha \sigma_{\bar \alpha} $
        and $\|\bar\alpha-\as\|< \|\alpha-\as \|$. This yields $\sigma^2_\an=O_\Pb\big( \| \an-\as\|^2 \big)=O_\Pb(n^{-1})$.
        We have obtained
        $  \Uc_n g(\cdot;\an)=(\Uc_n-\Pb)(g(\cdot;\an)-g(\cdot;\as))+O_{\Pb}(n^{-1}).$
        By the same type of decomposition as above, we get
        \begin{eqnarray*}
           \lefteqn{  \sqrt{n} (\Uc_n -\Pb)\big(g(\cdot;\an)-g(\cdot;\as)\big)
            =  \frac{\sqrt{n}}{n(n-1)(n-2)}\sum_{\substack{i,j,k=1 \\ i\not=j\not=k}}^n \Big\{ g\big([\Xbm, \Ubm]_{i,j,k};\an\big)}\\
            &&-\,\tilde{g}\big(X_i,F(U_i;\an);\an\big) -\tilde{g}\big(X_j,F(U_j;\an);\an\big)-\tilde{g}\big(X_k,F(U_k;\an);\an\big)
            +2\Pb g(\cdot;\an) \\
            &&-\, g\big( X_i,F(U_i;\as), X_j,F(U_j;\as), X_k,F(U_k; \as)\big)  \Big\}
            -3\sqrt{n}\Pb g(\cdot;\an)  \\
            &&+\,\frac{3\sqrt{n}}{n}\sum_{i=1}^n  \tilde{g}\big(X_i,F(U_i;\an);\an\big).
        \end{eqnarray*}
        By the asymptotic equicontinuity of the degenerate process $n(\Uc_n-\Pb)$ indexed by $\Gc_c$, we obtain
        \begin{align*}
            \sqrt{n} (\Uc_n -\Pb)\big(g(\cdot;\an)-g(\cdot;\as)\big)
            &=  \frac{3\sqrt{n}}{n}\sum_{i=1}^n  \tilde{g}\big(X_i,F(U_i; \an);\an\big) +O_\Pb(n^{-1/2}).
        \end{align*}
        Now, since $\alpha\mapsto\Tilde{g}_\alpha(x,y)$ is twice continuously differentiable, a limited expansion yields
        \begin{eqnarray*}
            \lefteqn{ \frac{3\sqrt{n}}{n}\sum_{i=1}^n  \tilde{g}\big(X_i,F(U_i; \an);\an\big) =
            \sqrt{n}(\an-\as)\frac{3}{n}\sum_{i=1}^n  \nabla_\alpha \tilde{g}\big(X_i,F(U_i; \as);\as\big)   }\\
            &&+\,\frac{3(\an-\as)}{\sqrt{n}}\sum_{i=1}^n \Big\{ \nabla_\alpha \tilde{g}\big(X_i,F(U_i; \an);\as\big) 
            -\nabla_\alpha \tilde{g}\big(X_i,F(U_i; \as);\as\big)\Big\} +O_\Pb(n^{-1/2}).
        \end{eqnarray*}
        Note we have invoked Assumption~\ref{regularity_tildeh} to manage the remainder term.
        Finally, due to the asymptotic equicontinuity of the empirical $U$ process indexed by $\Tilde{\Gc}$ and the usual central limit theorem, we obtain
        $\sum_{i=1}^n  \tilde{g}\big(X_i,F(U_i;\an);\an\big)/\sqrt{n}=O_\Pb(n^{-1/2}). $
        This yields $\hat{\sigma}^2_\an=O_{\Pb}(n^{-1})$ \textcolor{black}{and therefore $\tilde{\sigma}^2_\an=O_{\Pb}(n^{-1})$ also by Lemma \ref{lemrelationvarest}}.
\item[\it (iii):] Follows exactly by the same arguments as the proof of {\it (i)}.
\end{enumerate}
\end{proof}

\textcolor{black}{Note that the proof of Theorem~\ref{thm:nondegtestonemodel} solely relied on Lemma \ref{lem:asympdistrmmd_simple}, which is why Theorem~\ref{thm:nondegtestonemodel_general} immediately follows from Lemma \ref{lem:asympdistrmmdnonsmooth}.}

\subsection{Proof of Theorem~\ref{thm:nondegtestmodelcomp_diff} (model comparison)}
\label{techn_proofs_model_comp}

If the maps $ \alpha \mapsto F(u;\alpha)$ and $\beta\mapsto G(v;\beta)$ are \textcolor{black}{twice} differentiable for every $u\in \Uc$ and $v\in\Vc$, the same arguments as in Section~\ref{case_differentiable_F} can be invoked to obtain the asymptotic behaviors of $\Tc_n(\Mc_1,\Mc_2,P)$'s numerator and denominator.
\textcolor{black}{As for Lemmas~\ref{lem:asympdistrmmd_simple} and~\ref{lem:asympdistrmmdnonsmooth}, 
we establish a more general result than we need. To this aim, we require the following joint convergence assumption.}

\begin{assumpt}
\label{weak_conv_model_joint}
When $n\rightarrow \infty$,
    $$\sqrt{n} \begin{pmatrix} &\hmmd(P_\as,P)-\mmd^2(P_\as,P) \\ &\hmmd(Q_\bs,P)-\mmd^2(Q_\bs,P)\\
    &\hmmdq(P_\as,P)-\mmd^2(P_\as,P)\\
    &\hmmdq(Q_\bs,P)-\mmd^2(Q_\bs,P)\\
    &\an-\as\\
    &\bn-\bs\end{pmatrix}   \stackrel{\text{law}}{\longrightarrow} \begin{pmatrix} & Z^{(1)}_\as\\ & Z^{(2)}_{\bs} \\ & Z^{(1)}_{q,\as} \\ & Z^{(2)}_{q,\bs}\\
    & V^{(1)}_\as \\ & V^{(2)}_\bs\end{pmatrix}.$$
When $\as$ (resp. $\bs$) minimizes the distance $\mmd(P_\alpha,P)$ over $\Theta_1$ (resp. $\mmd(Q_\beta,P)$ over $\Theta_2$),
the weak convergence of $\sqrt{n}(\an-\as)$ (resp. $\sqrt{n}(\bn-\bs)$) is no longer required and replaced by the weaker requirement $\sqrt{n}(\an-\as)=O_{\Pb}(1)$ (resp. $\sqrt{n}(\bn-\bs)=O_{\Pb}(1)$).
\end{assumpt}

\begin{lem}
\label{lem:asympdistmmdteststat_diff}
Assume that Assumptions~\ref{A_characteristic}-\ref{cond_resid_model_comp} and~\ref{weak_conv_model_joint} are satisfied by the competing models $\Mc_1$ and $\Mc_2$. Then the following is true:
\begin{enumerate}
    \item[(i)] 
    \begin{align*}
        &\sqrt{n}\big\{\hmmd(P_\an,P)-\mmd^2(P_\as,P)-\hmmd(Q_\bn,P)+\mmd^2(Q_\bs,P)\big\}\\
        &\stackrel{\text{law}}{\longrightarrow} W_{\as,\bs}
        +\nabla_{\alpha^\intercal}\mmd^2(P_\alpha,P)\vert_{\alpha=\as} V_\as^{(1)}-\nabla_{\beta^\intercal}\mmd^2(Q_\beta,P)\vert_{\beta=\bs} V_\bs^{(2)},
    \end{align*}
    where $ W_{\as,\bs}:= Z^{(1)}_\as - Z^{(2)}_\bs \sim \Nc\big( 0,\sigma^2_{\as,\bs}\big).$
    Moreover, $\textcolor{black}{\tilde{\sigma}}^2_{\an,\bn}\to \sigma^2_{\as,\bs}$ in probability.
    \item[(ii)] If $P_\as=Q_\bs=P$, then $\sigma_{\as,\bs}=0$.
    Moreover, $\textcolor{black}{\tilde{\sigma}}^2_{\an,\bn}=O_{\Pb}(n^{-1})$ and
    $$\sqrt{n}\big\{\hmmd(P_\an,P)-\hmmd(Q_\bn,P)\big\}=O_\Pb\big( n^{-1/2}\big).$$
    \item[(iii)] \begin{align*}
       & \sqrt{n}\big\{ \hmmdq(P_\an,P)-\mmd^2(P_\as,P)-\hmmdq(Q_\bn,P)+\mmd^2(Q_\bs,P)\big\}\\
       &\stackrel{\text{law}}{\longrightarrow} W_{q,\as,\bs}+\nabla_{\alpha^\intercal}\mmd^2(P_\alpha,P)\vert_{\alpha=\as} V_\as^{(1)}
       -\nabla_{\beta^\intercal}\mmd^2(Q_\beta,P)\vert_{\beta=\bs} V_\bs^{(2)},
         \end{align*}
         where $W_{q,\as,\bs} :=Z^{(1)}_{q,\as} - Z^{(2)}_{q,\bs} \sim \Nc\big( 0,\sigma^2_{q,\as,\bs}\big).$
    If the samples $(U_i)_{i\geq 1}$ and $(V_i)_{i\geq 1}$ are independent, then $\sigma^2_{q,\as,\bs}>0$. Finally, $\textcolor{black}{\tilde{\sigma}}^2_{q,\an,\bn}\to \sigma^2_{q,\as,\bs}$ in probability.
\end{enumerate}
\end{lem}

\begin{proof}[of Lemma~\ref{lem:asympdistmmdteststat_diff}]
As in Lemma~\ref{lem:asympdistrmmd_simple}, the proof is essentially based on two Taylor expansions. 
Recall that $\hmmd(P_\an,P)- \hmmd(Q_\bn,P)=\Uc_n h(\cdot;\an,\bn)$.
By a Taylor expansion of the $U$-kernel $h(\cdot;\an,\bn)$ around $(\as,\bs)$ and due to Assumptions~\ref{cond_diff_k}-\ref{cond_residual_deriv} for both models, we have
\begin{eqnarray}
  h(\cdot;\an,\bn)&=& h(\cdot;\as,\bs) +\nabla_{\alpha^\top} h(\cdot;\as,\bs).(\an-\as)+
  \nabla_{\beta^\top} h(\cdot;\as,\bs).(\bn-\bs) \nonumber\\
  &&+\,O_{\Pb}(\Vert \an-\as \Vert^2)+O_{\Pb}(\Vert \bn-\bs \Vert^2). 
\label{lim_expansion_hanbn}
\end{eqnarray} 
The convergence of $\sqrt{n}\big\{\hmmd(P_\an,P)-\mmd^2(P_\as,P)-\hmmd(Q_\bn,P)+\mmd^2(Q_\bs,P)\big\}$ immediately follows from Assumption~\ref{weak_conv_model_joint}. 
We prove the convergence of the estimated variance $\hat\sigma^2_{\an,\bn}$, recalling~(\ref{def_hat_sigma_alpha_beta}), by a first order Taylor expansion of the map $(\alpha,\beta)\mapsto g\big( [\xbm, \ubm,\vbm]_{i,j,k};\alpha,\beta\big)$
defined in~(\ref{def_g_alpha_modelselection}) around $(\as,\bs)$, for every triplet of indices $(i,j,k)$, $i\neq j\neq k$.
By Lemma~\ref{lemrelationvarest}, we get the convergence of $\tilde{\sigma}_{\an,\bn}$ to $\sigma_{\as,\bs}$. This yields (i).

In the case (ii), $P_\as=P=Q_\bs$ and $\as$ and $\bs$ are minimizers of $\alpha\mapsto \mmd(P_\alpha,P)$ and 
$\beta\mapsto \mmd(Q_\beta,P)$ respectively.
By direct calculation and recalling~(\ref{def_sigma_alpha_beta}), we easily obtain $\sigma_{\as,\bs}=0$.
Then, Equation~(\ref{lim_expansion_hanbn}) provides $\Uc_n h(\cdot;\an,\bn)=O_\Pb(n^{-1})$ because 
$$\e\big[ \nabla_\alpha h([X,U,V]_{1,2};\as,\bs)\Big]=\nabla_\alpha\e\Big[ h([X,U,V]_{1,2};\as,\bs)\Big]=\nabla_\alpha\mmd(P_\as,P)=0,$$
and similarly $\e\lk\nabla_\beta h([X,U,V]_{1,2};\as,\bs)\rk=0$. 
Moreover, a Taylor expansion of the $U$-kernel $ g(\cdot;\an,\bn) $ around $(\as,\bs)$ yields that $ \Uc_n g(\cdot;\an,\bn)=O_\Pb(n^{-1}) $ when $P_\as=P=Q_\bs$ since $\Uc_n g(\cdot;\as,\bs)$ is a degenerate $U$-statistic, as in the proof of Lemma \ref{lem:asympdistrmmd_simple} (ii).
This implies $\hat{\sigma}^2_{\an,\bn}=\Uc_n g(\cdot;\an,\bn)-4\big(\hmmd(P_\an,P)-\hmmd(Q_\bn,P)\big)^2=O_\Pb(n^{-1})$. 

The same reasonings apply to prove (iii), noting that $\hmmdq(P_\an,P)- \hmmdq(Q_\bn,P)= \Uc_n^{(2)}q(\cdot;\an,\bn)$.
In particular, mimick the same arguments as above for the $U$-kernels $q(\cdot;\an,\bn)$ and $\xi(\cdot;\an,\bn)$, recalling~(\ref{def_q_alpha_beta}) and~(\ref{def_xi_alpha_modelselection}). This yields $\sqrt{n}\big\{ \Uc_n^{(2)}q(\cdot;\an,\bn)-\big(\mmd(P_\as,P)-\mmd(Q_\bs,P)\big)\big\}\to \Nc(0,\sigma^2_{q,\as,\bs})$ and $\hat{\sigma}_{q,\an,\bn} \to\sigma_{q,\as,\bs}>0$.
\end{proof}


The proof of Theorem \ref{thm:nondegtestmodelcomp_diff} is a direct consequence of Lemma \ref{lem:asympdistmmdteststat_diff}, mimicking the proof of Theorem \ref{thm:nondegtestonemodel}. Note that we only need the joint weak convergence of the first four components of the random vector in Assumption~\ref{weak_conv_model_joint}, that is obtained by the Cramer-Wold device and usual H\'ajek projections of $U$-statistics based on the sample $(X_i,U_i,V_i)$.
Contrary to Theorem \ref{thm:nondegtestonemodel}, one must differentiate the case $P_\as=P=Q_\bs$ and the case $\mmd(P_\as,P)=\mmd(Q_\bs)$, but $P_\as\not =P\not =Q_\bs$ under the null. 
When $P_\as=P=Q_\bs$, the proof is identical to the proof of Theorem \ref{thm:nondegtestonemodel}. When $\mmd(P_\as,P)=\mmd(Q_\bs)$ but $P_\as\not =P\not =Q_\bs$ the proof structure is still identical to the latter case, but with the sole differences that $\sqrt{n}\,\Uc_n h(\cdot;\an,\bn)\to \Nc(0,\sigma^2_{\as,\bs})$, $\tilde{\sigma}_{\an,\bn}\to \sigma_{\as,\bs}>0$, $\epsilon_n\tilde{\sigma}_{q,\an,\bn}\to 0$, as well as $\sqrt{n}\epsilon_n\big\{ \hmmdq(P_\an,P)-\hmmdq(Q_\bn,P)\big\}\to 0$. The latter arguments yield (i) of Theorem \ref{thm:nondegtestmodelcomp_diff} and the result follows.

\begin{rem}
\label{rem_non_zero_var_modelcomp}
In Lemma \ref{lem:asympdistmmdteststat_diff}, it is possible that $\sigma^2_{q,\as,\bs}=0$ when the random variables $U$ and $V$ are ``perfectly dependent'' and $P_\as=Q_\bs$, i.e., when
$F(U_i; \as)=G(V_i; \bs)$ a.s.\ for every $i$. To be specific, by simple calculations, we always have
\begin{align*}
        \sigma^2_{q,\as,\bs}/8 :=&\var \Big( \e[ k\big( X,G(V_2; \bs)\big) | V_2 ] - \e [ k\big(X ,F(U_2; \as)\big) | U_2]  \\
        &+\, \e[ k\big(X_2, G(V; \bs) \big) | X_2]
        - \e [ k\big(X_2,F(U; \as)  \big)|X_2] \\
        & +\, \e [ k\big( F(U_1; \as), F(U; \as) \big) |U_1] - \e[k\big( G(V_1; \bs), G(V; \bs) \big) |V_1 ] \Big).
\end{align*}
When $F(U_i; \as)=G(V_i; \bs)$ a.s., which is for example the case when both generators of the optimal models are identical and $U_i$ and $V_i$ are generated by the same source of randomness, then it becomes clear that $\sigma^2_{q,\as,\bs}=0$. However, this undesired phenomenon is avoided when $U_i$ and $V_i$ are chosen to be independent, i.e., when the two competing models are generated by independent sources of randomness, which is why we have imposed this assumption.
\end{rem}

\subsection{Technical assumptions and proof of  Theorem~\ref{thm:nondegtestmodelcomp} (model comparison)}
\label{proofs_model_comp_nonsmooth}

Define
\textcolor{black}{
$    \tilde h(x,y,z;\alpha,\beta):=\tilde h^{(1)}(x,y;\alpha)- \tilde h^{(2)}(x,z;\beta),$
where  $\tilde h^{(1)}(x,y;\alpha)$ and  $\tilde h^{(2)}(x,z;\beta)$ are defined according to (\ref{def_h_tilde}).}
Moreover, define 
\begin{align*}
\tilde{g}(x,y,z;\alpha,\beta):=& \frac{8}{3}\e\big[ h\lc (X,F(U;\alpha),G(V;\beta)),(x,y,z) \rc \tilde{h}\lc X,F(U,\alpha),G(V;\beta);\alpha,\beta \rc\big]\\
& +\,\frac{4}{3}\tilde{h}^2(x,y,z;\alpha,\beta).
\end{align*}
Define some classes of functions as in Section~\ref{reg_assump_nondiff_generating_fct}, but now indexed by $(\alpha,\beta)$ instead of $\alpha$ only, as
$$   \Fc^{(\Mc)} := \Big\{  [\xbm, \ubm, \vbm]_{1:3} \mapsto  h \big( [\xbm, \ubm, \vbm]_{1:3};\alpha,\beta\big) \big\vert \ \alpha\in B_\delta (\as), \beta\in B_\delta (\bs) \Big\},$$
$$   \Gc^{(\Mc)} := \Big\{  [\xbm, \ubm, \vbm]_{1:3} \mapsto  g \big( [\xbm, \ubm, \vbm]_{1:3};\alpha,\beta\big) \big\vert \ \alpha\in B_\delta (\as), \beta\in B_\delta (\bs) \Big\},$$
\textcolor{black}{and, in the same spirit, the analogs $\Fc_q^{(\Mc)}$, $\Qc^{(\Mc)}$, $\Fc_c^{(\Mc)}$, $\Gc_c^{(\Mc)}$, $\tilde{\Fc}^{(\Mc)}$ and $\tilde{\mathcal{G}}^{(\Mc)}$ of
$\Fc_q$, $\Qc$, $\Fc_c$, $\Gc_c$, $\tilde{\Fc}$ and $\tilde{\mathcal{G}}$.
}
We need to adapt the regularity Assumptions \ref{regularity_tildeh} and \ref{ass_degenerateuprocess} to the new framework.


\begin{assumpt}
\label{regularity_tildeh_modelcomp}
The maps $(\alpha,\beta)\mapsto \tilde{h}(x,y,z;\alpha,\beta)$ and $(\alpha,\beta)\mapsto \tilde{g}(x,y,z;\alpha,\beta)$ are twice continuously differentiable in a neighborhood of $(\as,\bs)$ at every $(x,y,z)\in \Sc\times\Sc\times \Sc$. Additionally, the maps $$(\alpha,\beta)\mapsto \e \Big[ \nabla_{(\alpha,\beta)} \tilde{h} \big( X, F(U; \alpha),G(V; \beta);\as,\bs\big)\Big],\; \text{and}$$
$$(\alpha,\beta)\mapsto \e \Big[ \nabla_{(\alpha,\beta)} \tilde{g} \big( X, F(U; \alpha),G(V; \beta);\as,\bs\big)\Big]$$
 are differentiable in a neighborhood of $(\as,\bs)$.  
 We assume 
 $$\e\lk \nabla_{(\alpha,\beta)} \tilde{h} \big( X, F(U; \as);\as,\bs\big)\rk= \nabla_{(\alpha,\beta)} \e\lk\tilde{h} \big( X, F(U; \as),G(V;\bs);\as,\bs\big)\rk,\; \text{and}$$ 
 $$\e\lk \nabla_{(\alpha,\beta)} \tilde{g} \big( X, F(U; \as),G(V;\bs);\as,\bs\big)\rk= \nabla_{(\alpha,\beta)} \e\lk\tilde{g} \big( X, F(U; \as),G(V;\bs);\as,\bs\big)\rk.$$
Moreover,
$$\Vert h(\cdot; \an,\bn)-h(\cdot;\as,\bs)  \Vert_{L_2(\otimes_{i=1}^2 P\otimes P_U \otimes P_V )}\ , \ \Vert q(\cdot; \an,\bn)-q(\cdot; \as,\bs) \Vert_{L_2(\otimes_{i=1}^4 P\otimes P_U \otimes P_V)} \ ,
$$
$$
\Vert g(\cdot;\an,\bn)-g(\cdot;\as,\bs)  \Vert_{L_2(\otimes_{i=1}^3 P\otimes P_U \otimes P_V)} \ , \text{and}\  \Vert \xi(\cdot;\an,\bn)-\xi(\cdot;\as,\bs)  \Vert_{L_2(\otimes_{i=1}^6 P\otimes P_U \otimes P_V)}
$$
tend to zero with $n$ in probability.
Finally, for some real constant $\delta>0$, we have
\begin{align*}
 &\Eb\bigg[ \sup_{\substack{(\alpha_1,\beta_1) \in B_\delta(\as)\times B_\delta(\bs),\\ (\alpha_2,\beta_2) \in B_\delta(\as)\times B_\delta(\bs)} }  \| \nabla^2_{(\alpha,\beta),(\alpha,\beta)^\intercal} \tilde{h} \big(X, F(U; \alpha_2),G(V; \beta_2);\alpha,\beta\big)\vert_{(\alpha,\beta)=(\alpha_1,\beta_1)} \|_2^2 \bigg]    \\
  & + \Eb\bigg[ \sup_{\substack{(\alpha_1,\beta_1) \in B_\delta(\as)\times B_\delta(\bs),\\ (\alpha_2,\beta_2) \in B_\delta(\as)\times B_\delta(\bs)} }  \| \nabla^2_{(\alpha,\beta),(\alpha,\beta)^\intercal} \tilde{g} \big(X, F(U; \alpha_2),G(V; \beta_2);\alpha,\beta\big)\vert_{(\alpha,\beta)=(\alpha_1,\beta_1)} \|_2^2 \bigg] <\infty.
 \end{align*}
\end{assumpt}
\begin{assumpt}
\label{ass_degenerateuprocess_modelcompp}
    There exists some $\delta>0$ such that the centered empirical $U$-processes \textcolor{black}{   $\lc\sqrt{n}(\Uc_n-\Pb)f\rc_{f\in\Fc^{(\Mc)}}$, $\lc\sqrt{n}(\Uc_n-\Pb)f\rc_{f\in\Fc_q^{(\Mc)}}$, $\lc n(\Uc_n-\Pb)f\rc_{f\in\Fc_c^{(\Mc)}}$,  } $\lc\sqrt{n}(\Uc_n^{(2)}-\Pb)f\rc_{f\in\Qc^{(\Mc)}}$, $\lc\sqrt{n}(\Uc_n-\Pb)g\rc_{g\in\Gc^{(\Mc)}}$ and $\lc n(\Uc_n-\Pb)g\rc_{g\in\Gc^{(\Mc)}_c}$ and the empirical processes $\lc\sqrt{n}(\Pb_n-\Pb)f\rc_{f\in\tilde{\Fc}^{(\Mc)}}$ and $\lc\sqrt{n}(\Pb_n-\Pb)g\rc_{g\in\tilde{\Gc}^{(\Mc)}}$ weakly converge. Finally, the map  $(\alpha,\beta)\mapsto \sigma_{\alpha,\beta}$ is continuously differentiable in a neighborhood of $(\as,\bs)$ and
$$ \e_{X,U,V}\lk \Vert \nabla_{(\alpha,\beta)}\Tilde{h}\big(X,F(U; \an),G(V;\bn);\as,\bs\big)  -\nabla_{(\alpha,\beta)} \Tilde{h}\big(X,F(U; \as),G(V;\bs);\as,\bs\big) \Vert_2^2 \rk,\; $$
as well as
$$
\e_{X,U,V}\lk \Vert \nabla_{(\alpha,\beta)}\Tilde{g}\big(X,F(U; \an),G(V; \bn);\as,\bs\big)  -\nabla_{(\alpha,\beta)} \Tilde{g}\big(X,F(U; \as),G(V; \bs);\as,\bs\big) \Vert_2^2\rk
$$
converge to 0 in probability.

\end{assumpt}

Again, sufficient conditions to ensure the functional convergence of the considered empirical $U$-processes can be found in Appendix \ref{sec:appsuffcondempuprocess}. 


\begin{lem}
\label{lem:modelcompteststat}
Assume the models $\Mc_1$ and $\Mc_2$ satisfy Assumptions~\ref{A_characteristic}-\ref{regular_model} and \ref{weak_conv_model_joint}-\ref{ass_degenerateuprocess_modelcompp}. Then the conclusions of Lemma \ref{lem:asympdistmmdteststat_diff} apply.
\end{lem}
Essentially, the proof of Lemma~\ref{lem:modelcompteststat} goes along the same lines as the proof of Lemma \ref{lem:asympdistrmmdnonsmooth}, replacing each quantity defined for model specification with the corresponding quantity for model selection. Therefore, it has been omitted.
Then, assuming Lemma \ref{lem:modelcompteststat} is satisfied, the proof of Theorem \ref{thm:nondegtestmodelcomp} follows identically to the proof of Theorem \ref{thm:nondegtestmodelcomp_diff}.

\section{Sufficient condition for functional convergence of centered empirical $U$-processes}
\label{sec:appsuffcondempuprocess}

In Section \ref{general_specif_test} and Section \ref{general_comparison_test}, we require that some centered empirical $U$-processes indexed by certain classes of functions converge to their appropriate limits in the functional sense (\ref{deffunctionalconvuprocess}). In this section, we provide some sufficient conditions that ensure such a functional weak convergences.

Consider a generic class $\mathcal{L}$ of symmetric real-valued measurable functions on some product probability space $\lc \Zc^q,\otimes_{i=1}^q P_Z\rc$.
A class $\Lc$ is degenerate of order $r-1$, $r\geq 1$, if
$$ \e_{(Z_i)_{1\leq i\leq q-r+1}\sim\otimes_{i=1}^{q-r+1} P_Z}\big[  \ell(Z_1,\ldots,Z_q)\big] =\text{const} \;\;\; \otimes_{q-r+2}^q P_Z \text{-a.s. },\;\text{and}$$
$$\var\lc \e_{(Z_i)_{1\leq i\leq q-r}\sim\otimes_{i=1}^{q-r} P_Z}\big[  \ell(Z_1,\ldots,Z_q)\big] \rc>0   $$
for every  $\ell\in\Lc$.
When $r=1$, this simply means $\e\lk \ell(Z_1,\ldots,Z_q)\rk=0$ and  $\e[ \ell(Z_1,\ldots,Z_q)\mid $ $ Z_1]$ is not a constant; thus, the class $\Lc$ is also called non-degenerate in this case. 

Let us recall the usual definition of covering numbers: for a given norm $\Vert\cdot\Vert$ on $\mathcal{L}$, define the $\epsilon$-covering number of $(\mathcal{L},\Vert\cdot\Vert)$ as
$$ N(\epsilon,\mathcal{L},\Vert\cdot\Vert):=\min \Big\{n: \exists f_1,\ldots,f_n\in \mathcal{L}  \;\text{s.t.}\; \sup_{f\in \mathcal{L}}\min_{i\leq n} \Vert f-f_i\Vert \leq \epsilon    \Big\}.$$
Essentially, the covering number measures the size of $\mathcal{L}$ w.r.t.\ $\Vert \cdot\Vert$ and can be interpreted as a measure of complexity of $\mathcal{L}$ w.r.t.\ $\Vert\cdot\Vert$. Hereafter, we define some regularity condition on a generic class of functions $\mathcal{L}$ that is based on the covering numbers of $\mathcal{L}$. This condition will ensure the weak convergence of the centered empirical $U$-process $n^{r/2}\big((\Uc_n-\Pb)\ell\big)_{\ell\in\mathcal{L}}$ and it is mainly based on (simplified) conditions provided by \cite{arconesgine1993}.

\begin{defn}
\label{CS_weak_conv_Uprocess}
Let $\mathcal{L}$ denote a class of symmetric measurable functions on some product probability space $\big( \Zc^q,\otimes_{i=1}^q P_Z\big)$ and let $r$ be a positive integer. Then, $\mathcal{L}$ is called $r$-regular if the following is satisfied:
\begin{enumerate}
    \item $ S(z_1,\ldots,z_q):=\sup_{\ell\in\mathcal{L}}\vert \ell(z_1,\ldots,z_q)\vert<\infty$ for all $(z_1,\ldots,z_q)\in\mathcal{Z}^q$;
    \item $\e_{(Z_i)_{1\leq i\leq q}\sim\otimes_{i=1}^q P_Z}\lk  S(Z_1,\ldots,Z_q)^2\rk<\infty$;
    \item $\lim_{t\to\infty} t P_Z \Big( \e_{(Z_i)_{1\leq i\leq q-1}\sim\otimes_{i=1}^{q-1} P_Z}\big[  S(Z_1,\ldots,Z_{q-1},Z)^2\big]  >t  \Big)\to 0$;
    \item $\mathcal{L}$ is image admissible Suslin;
    \item for all probability measures $Q$ such that $\e_{(Z_i)_{1\leq i\leq q}\sim Q}\lk S(Z_1,\ldots,Z_q)^2\rk<\infty$, we have
    \begin{align}
        &\int_0^\infty \bigg\{ \sup_{Q} \log N\Big(\epsilon \sqrt{\e_{(Z_i)_{1\leq i\leq q}\sim Q}\big[  S(Z_1,\ldots,Z_q)^2\big] } ,\mathcal{L},\|\cdot \|_{L_2(Q)}\Big)\bigg\}^{r/2} \rmd \epsilon<\infty,\, \label{finiteitegcovnmb} \\
        &\text{ and }\int_0^\infty \bigg\{  \log N\Big(\epsilon ,\mathcal{L},\|\cdot \|_{L_2(\otimes_{i=1}^q P_Z)}\Big)\bigg\}^{r/2} \rmd \epsilon<\infty.    \nonumber
    \end{align}

\end{enumerate}
\end{defn}

Any centered empirical $U$-process $n^{r/2}\big((\Uc_n-\Pb)\ell\big)_{\ell\in\mathcal{L}}$ that is
degenerate of degree $r-1$ and $r$-regular
weakly converges to a limit process in $L_\infty(\mathcal{L})$, which is asymptotically uniformly equicontinuous w.r.t.\ the norm $\Vert \cdot\Vert_{L_2(\otimes_{i=1}^q P_Z)}$: see~\cite{arconesgine1993}, p.\,1535.

In Definition~\ref{CS_weak_conv_Uprocess}, the concept of being image admissible Suslin is a measurability property that may not be familiar to many readers. 
This is why we briefly discuss some sufficient conditions to ensure the image admissible Suslin property of a considered class of function $\mathcal{L}$.
Assume that $\mathcal{L}$ is parameterized by a vector-valued parameter $\alpha$ (\textcolor{black}{resp.} $\beta$) which belongs to a neighborhood of $\as$ (\textcolor{black}{resp.} $\bs$).
In other words, we assume any class of functions $\mathcal{L}$ may be rewritten as
$\mathcal{L}=\{\ell_\theta: \Ec \mapsto \R \,\vert \, \theta \in \Theta_0\}$, for some non-empty compact subset $\Theta_0$ of a finite dimensional Euclidean space and $\Ec$ is the Cartesian product of $\Sc$, $\Uc$ (\textcolor{black}{resp.} $\Vc$) spaces. From~\cite{dudley1984course}, p.\,101,
this implies that  $\mathcal{L}$ is image admissible Suslin if
\begin{itemize}
    \item $\Ec$ endowed with its Borel $\sigma-$algebra is a Polish space (i.e., is metrizable to become a separable and complete metric space), and
    \item the map $(x,\theta) \mapsto f_\theta(x)$ from $\Ec\times \Theta_0$ to $\R$ is jointly measurable.
\end{itemize}
These conditions are often met in practice and ensure the image admissible Suslin property of $\mathcal{L}$.
Moreover, for certain classes of functions, it is well known that (\ref{finiteitegcovnmb}) is satisfied, such as for VC-subgraph classes of functions or classes of functions that satisfy a Lipschitz-property, see \cite[Section 2.6 and 2.7]{wellner2013weak}. In particular, these conditions may be directly checked provided the kernel $k$, $\big( F(\cdot;\alpha)\big)_{\alpha\in\Theta_1}$ \textcolor{black}{and} $\big( G(\cdot;\beta)\big)_{\beta\in\Theta_2}$.

We are ready to provide sufficient conditions for the weak convergence statements in Assumptions \ref{ass_equicont}, \ref{ass_degenerateuprocess} and \ref{ass_degenerateuprocess_modelcompp} to hold. A sufficient condition to ensure that there exists some $\delta>0$ such that the empirical $U$-processes
$\big(\sqrt{n}(\Uc_n-\Pb)f\big)_{f\in\Fc}$, $\big(\sqrt{n}(\Uc_n^{(2)}-\Pb)q\big)_{q\in\Fc_q}$, $\big(\sqrt{n}(\Uc_n-\Pb)g\big)_{g\in\Gc}$
and $\big(\sqrt{n}(\Uc_n^{(2)}-\Pb)\xi\big)_{f\in\Qc}$ weakly converge to their appropriate limits in the functional sense as claimed in Assumption \ref{ass_equicont} is as follows:
 \textit{there exists some $\delta>0$ such that the classes of functions $\Fc$, $\Fc_q$,  $\Gc$ and $\Qc$ are $1$-regular.}

Moreover, a sufficient condition to ensure that there exists $\delta>0$ such that the centered empirical $U$-processes
$\big( n(\Uc_n-\Pb)f\big)_{f\in\Fc_c}$ and $\big( n(\Uc_n-\Pb)g\big)_{f\in\Gc_c}$ and the empirical processes $\big(\sqrt{n}(\Pb_n-\Pb)f\big)_{f\in\tilde \Fc}$ and $\big(\sqrt{n}(\Pb_n-\Pb)g\big)_{g\in\tilde \Gc}$ converge to their appropriate functional limits as claimed in Assumption \ref{ass_degenerateuprocess} is:
\textit{there exists some $\delta>0$ such that the classes of functions $\tilde \Fc$ and  $\tilde \Gc$ are $1$-regular and that the classes of functions $\Fc_c$ and $\Gc_c$ are $2$-regular}.

A sufficient condition to ensure that there exists some $\delta>0$ such that the centered empirical $U$-processes $\lc\sqrt{n}(\Uc_n-\Pb)f\rc_{f\in\Fc^{(\Mc)}}$, $\lc\sqrt{n}(\Uc_n-\Pb)f\rc_{f\in\Fc_q^{(\Mc)}}$, $\lc n(\Uc_n-\Pb)f\rc_{f\in\Fc_c^{(\Mc)}}$, $\lc\sqrt{n}(\Uc_n^{(2)}-\Pb)f\rc_{f\in\Qc^{(\Mc)}}$, $\lc\sqrt{n}(\Uc_n-\Pb)g\rc_{g\in\Gc^{(\Mc)}}$ and $\lc n(\Uc_n-\Pb)g\rc_{g\in\Gc^{(\Mc)}_c}$ and the empirical processes $\lc\sqrt{n}(\Pb_n-\Pb)f\rc_{f\in\tilde{\Fc}^{(\Mc)}}$ and $\lc\sqrt{n}(\Pb_n-\Pb)g\rc_{g\in\tilde{\Gc}^{(\Mc)}}$ weakly converge to their appropriate functional limits as claimed in Assumption \ref{ass_degenerateuprocess_modelcompp} is:
\textit{there exists some $\delta>0$ such that the classes of functions   $\Fc^{(\Mc)}$, $\Fc_q^{(\Mc)}$, $\tilde{\Fc}^{(\Mc)}$,  $\Qc^{(\Mc)}$, $\Gc^{(\Mc)}$ and $\tilde{\Gc}^{(\Mc)}$ are 
$1$-regular and that the class of functions $\Fc_c^{(\Mc)}$ and $\Gc^{(\Mc)}_c$ is $2$-regular}.
    
Note that the $1$-regularity of the classes of functions $\tilde \Fc$, $\tilde \Gc$,  $\tilde{\Fc}^{(\Mc)}$ and  $\tilde{\Gc}^{(\Mc)}$ is simply a sufficient condition to ensure that they are Donsker classes \citep[Section 2.5]{wellner2013weak}.
Finally, it should again be emphasized that the concept of $r$-regularity does not rely on any differentiability property of the considered functions. Therefore, it is a suitable concept to ensure the asymptotic convergence of centered empirical $U$-processes indexed by classes of non-differentiable functions, such as in Section \ref{general_specif_test} and Section \ref{general_comparison_test}. See Appendix~\ref{app:relu-type_example} for an example.

\section{A ReLu-type neural network model}
\label{app:relu-type_example}

The goal of this section is to provide a short example which illustrates that the conditions for an application of Theorem \ref{thm:nondegtestonemodel_general} are verifiable when the generating function $\alpha\mapsto F(\cdot,\alpha)$ is non-smooth. Similar calculations yield that the assumptions of Theorem \ref{thm:nondegtestmodelcomp} are also satisfied for this example, but due to space limitations we leave the detailed calculations to the reader.
The generative model with which we conduct our analysis is a special type of a one-layer ReLu neural network given by
\begin{align}
F(u;\alpha): u\mapsto \sum_{k=1}^m a_k \max(u - b_k,0) + c, \label{defrelumodel}
\end{align}
a map that is parametrized by $\alpha:=(a_1,\ldots,a_m,b_2,\ldots,b_m,c)$, since we impose $b_1=0$.
We try to model $P$ in terms of a generative model $(P_\alpha)_{\alpha\in\Theta_1}$ where the law of any $P_\alpha$ is the law of $F(U;\alpha)$, $U\sim \text{Unif}([0,1])$, and $\Theta_1$ denotes some subset of $\R^{2m}$ with non-empty interior. 
This model is clearly not differentiable w.r.t.\ (some of) the parameters $b_k$. Thus, one cannot rely on the results of Theorem \ref{thm:nondegtestonemodel}. 
To make the exposition easier we assume that $\as=\argmin_{\alpha} \mmd(P_\alpha,P)$ is unique. This latter property is satisfied when all parameters $a_k$ are positive and $b_1=0<b_1<\ldots< b_m<1$ (Lemma~\ref{lemidentifytoymodel} below), an identifiability condition that is assumed from now on. Further, we assume our kernel is bounded by one and is twice continuously differentiable with bounded first and second derivatives, which is e.g.\ satisfied by the Gaussian kernel. This implies that it is globally Lipschitz continuous and we denote the corresponding Lipschitz constant as $L_k$.

We will frequently use Theorem 2.10.20 in \cite{wellner2013weak}. It implies that sums of Donsker classes are again Donsker. Moreover, products of uniformly bounded Donsker classes are Donsker. Let $\mathcal{L}$ denote a generic class of uniformly bounded functions with an upper bound $C$. We will show that each $\mathcal{L}\in \{ \mathcal{F},\mathcal{F}_q,\mathcal{Q},\mathcal{G},\mathcal{F}_c,\mathcal{G}_c,\tilde{\mathcal{F}},\tilde{\mathcal{G}}\}$ has polynomially bounded covering number w.r.t.\ $\Vert\cdot\Vert_{L_2(Q)}$, where $Q$ denotes an arbitrary discrete probability measure. This then implies that the entropy integral in (\ref{finiteitegcovnmb}) is finite, since \citep[footnote 1 p.84]{wellner2013weak} implies that the entropy integral in (\ref{finiteitegcovnmb}) only has to be considered for discrete probability measures. From this it immediately follows that $\mathcal{L}$ is $r$-regular for $r\in \{1,2\}$.

First, recall that, for every norm $\Vert\cdot\Vert$ on $\Lc$, the $\epsilon$ covering number $N(\epsilon,\mathcal{L},\Vert\cdot\Vert)$ is bounded by the corresponding $2\epsilon$ bracketing number, denoted $N_{[]}(2\epsilon,\mathcal{L},\Vert\cdot\Vert)$ \citep[p. 84]{wellner2013weak}. Further, every ReLu neural network with a bounded parameter set and bounded inputs is globally Lipschitz w.r.t.\ its parameter $\alpha$, denoting the corresponding Lipschitz constant as $L$. Then, we have 
$$ \vert k\big(F(U_1;\alpha_1),F(U_2;\alpha_2)\big)-k\big(F(U_1;\alpha_3),F(U_2;\alpha_4)\big) \vert\leq L_k  L \Vert (\alpha_1,\alpha_2)-(\alpha_3,\alpha_4)\Vert_1,\; \text{a.e.} $$
By \citet[Theorem 2.7.11]{wellner2013weak}, the $2L_k  L\epsilon$ bracketing number of the classes $\mathcal{K}_1:=\{ k\big(F(\cdot;\alpha_1),F(\cdot;\alpha_2)\big)\mid \alpha_1,\alpha_2\in \Theta_1\}$ and $\mathcal{K}_2:=\{ k(\cdot,F(\cdot;\alpha_2)\big)\mid \alpha\in \Theta_1\}$ w.r.t.\ an arbitrary norm $\Vert \cdot\Vert$ are bounded by the $C\epsilon$ covering number of $\Theta_1\times \Theta_1$ and $\Theta_1$  w.r.t. $\Vert\cdot\Vert_1$, respectively, where $C$ is a suitable constant. These covering numbers are polynomial in $\epsilon$ as $\Theta_1\times \Theta_1$, resp. $\Theta_1$, is contained in a ball of radius $R>0$ and such covering numbers are bounded by an expression of the form $C_1 \epsilon^{-2(2m+1)}$, $C_1>0$ \citep[Corollary 4.2.13]{vershynin2018}. 
Now, consider a class $\mathcal{L}\in\{\mathcal{F},\mathcal{F}_q,\mathcal{Q},\mathcal{G},\mathcal{F}_c,\mathcal{G}_c \}$. Any element in $\Hc$ is written as sums and/or products of uniformly bounded functions from $\mathcal{K}_{1}$ and $\mathcal{K}_{2}$. Thus, the proof of \citet[Theorem 2.10.20]{wellner2013weak} - in their notation, $\alpha_i=1$ - implies the existence of $C_2>0$ s.t.
$$ \sup_{Q} N(\epsilon,\mathcal{L},L_2(Q))\leq \sup_{Q}  N(C_2\epsilon,\mathcal{K}_1,\Vert\cdot\Vert_{L_2(Q)})^{j_1}N(C_2\epsilon,\mathcal{K}_2,\Vert\cdot\Vert_{L_2(Q)})^{j_2}\ \ i=1,2,$$
where the powers $j_1,j_2\in\N$ depend on the respective function class $\mathcal{L}$, and $Q$ ranges over all discrete probability measures. Due to the derivations above, 
$$N\big(C_2\epsilon,\mathcal{K}_i,L_2(Q)\big)^{j_i}\leq N_{[]}(2C_2\epsilon,\mathcal{K}_i,L_2(Q))^{j_i}\leq C_3\epsilon^{-2j_i(2m+1)}$$
for a suitable $C_3>0$. Thus, $\sup_{Q} N(\epsilon,\mathcal{L},L_2(Q))$ is bounded by an expression of the form $C_4\epsilon^{-2(j_1+j_2)(2m+1)}$, $C_4>0$, implying that the entropy integrals in (\ref{finiteitegcovnmb}) are finite for every positive integer $r$. Thus, $ \{\mathcal{F},\mathcal{F}_q,\mathcal{Q},\mathcal{G},\mathcal{F}_c,\mathcal{G}_c \}$ are $r$-regular for every $r\in\N$ and for every ReLu neural network with bounded inputs and a bounded parameter set. 

Let us continue by verifying the existence of moments for the derivatives of $\tilde{h}(x,\alpha)$ and $\tilde{g}(x,y;\alpha)$ w.r.t.\ $\alpha$ (recall~(\ref{def_h_tilde}) and~(\ref{def_tilde_g})). It is easy to see that the differentiability of $\alpha\mapsto\tilde{h}(x,y;\alpha)$ and $\alpha\mapsto \tilde{g}(x,y;\alpha)$ follows from the differentiability of terms of the form $\e\lk k\big(x,F(U;\alpha)\big)\rk$, since they are sums, squares and products of these terms.
In our model (\ref{defrelumodel}) and setting $b_{m+1}=1$, these terms can be written as
$$ \e\lk k\big(x,F(U;\alpha)\big)\rk = \sum_{k=1}^m \int_{b_k}^{b_{k+1}} k\big(x, c+ \sum_{j=1}^k a_j(u-b_j) \big)\rmd u.$$
Thus, $\e\lk k(x,F(U;\alpha)\rk$ is twice continuously differentiable w.r.t.\ $\alpha$ in a neighborhood of $\as$ for every fixed $x$, with bounded derivatives. This implies the existence of $\nabla_\as \tilde{h}(x,y;\as)$ and $\nabla^2_\as \tilde{h}(x,y;\as)$ and its respective moment conditions. Similar arguments show that the same is true for $\nabla_\as \tilde{g}(x,y;\as)$ and $\nabla^2_\as \tilde{g}(x,y;\as)$. Therefore, Assumption \ref{regularity_tildeh} is satisfied. Note that this also implies that $\alpha\mapsto \sigma^2_\alpha$ and $\alpha\mapsto \sigma^2_{q,\alpha}$ are differentiable in a neighborhood of $\as$.

As $\nabla_\as \tilde{h}(x,y;\as)$ and $\nabla_\as \tilde{g}(x,y;\as)$ are continuously differentiable in a neighborhood of $\as$ and $k$ has bounded derivatives, we also have that $\tilde{\mathcal{F}}$ and $\tilde{\mathcal{G}}$ are classes of Lipschitz continuous bounded functions indexed by $\alpha\in B_\delta(\as)$, choosing a $\delta>0$ small enough. Therefore, mimicking the arguments above we can show that $\{\tilde{\mathcal{F}},\tilde{\mathcal{G}}\}$ have polynomially bounded covering numbers and are $r$-regular for every $r\in\N$.
Finally, all conditions mentioned in Assumption \ref{ass_equicont}-\ref{ass_degenerateuprocess} of the form $\Vert l(x,\an)-l(x,\as)\Vert_{L_2}\to 0$ are satisfied for all $l\in \mathcal{L}$. 
Indeed, $l(x,\an)-l(x,\as)\to 0$ in probability for every consistent estimator $\an$ of $\as$, because all considered functions are bounded and Lipschitz w.r.t. $\alpha$.

It only remains to spell out conditions under which an estimator $\an$ for $\as$ is asymptotically normal. To this purpose, define 
$$\an=\argmin_{\alpha\in\Theta_1}n^{-1} \sum_{i=1}^{n/2} h\Big( \big(X_{2i},F(U_{2i};\alpha)\big),\big(X_{2i-1},F(U_{2i-1};\alpha)\big)\Big).$$
Note that $\an$ is an M-estimator of $\as:=\argmin_{\alpha\in\Theta_1}\mmd(P_\alpha,P)$. By the same reasoning as above, 
the map $\alpha \mapsto  h\Big(\big(x_1,F(u_1;\alpha)\big),\big(x_2,F(u_2;\alpha)\big)\Big)$ is differentiable in a neighborhood of $\as$ for $P\otimes P_U\otimes P\otimes P_U$ almost every $(x_1,u_1,x_2,u_2)$, with Lipschitz continuous derivative w.r.t.\ $\alpha$. Additionally, the corresponding Lipschitz constant is independent of $(x_1,u_1,x_2,u_2)$. Since $\alpha\mapsto\mmd(P_\alpha,P)$ is twice continuously differentiable in a neighborhood of $\as$ we obtain that $\an$ is asymptotically normal \citep[Theorem 5.23]{van2000asymptotic}.

Note that the most difficult task in this example is to ensure the uniqueness of $\as$, which is necessary to apply standard results to obtain the asymptotic normality of $\an$. All other statements above do not require the existence of a unique $\as$ and are also valid when there are multiple argmins. We were able to show that, 
under the assumptions of Lemma \ref{lemidentifytoymodel} below, that our model (\ref{defrelumodel}) always has a unique argmin. Thus, under the identifiability conditions from Lemma \ref{lemidentifytoymodel} (a result that is of interest per se), Assumptions~\ref{A_characteristic}-\ref{cond_residual_deriv} and Assumptions \ref{ass_equicont}-\ref{ass_degenerateuprocess} are satisfied and Theorem \ref{thm:nondegtestonemodel_general} applies.

\begin{lem}
\label{lemidentifytoymodel}
    If $a_k$, $k\in\{1,\ldots,m\}$ are strictly positive and $b_1=0<b_2<\ldots <b_{m-1} < b_m < 1$, then model~(\ref{defrelumodel}) is identifiable.
\end{lem}

\begin{proof}
Denote by $\theta:=(a_1,\ldots,a_m,b_2,\ldots,b_{m},c)$ the vector of unknown parameters.
Let $F_\theta$ be the cdf of the random variable $\sum_{k=1}^m a_k \max(U - b_k,0) + c$.
For notational convenience, set $s_k:=a_1+\ldots +a_k$ and $v_k:=a_1b_1+a_2b_2+\ldots+ a_kb_k$, $k\in \{1,\ldots, m\}$.
Note that $s_k\geq v_k$ for any $k$. 
The support of our law is then $\Dc_\theta:=(c,c+s_m-v_m)$. Set $b_{m+1}=1$. For any $t$ in $\Dc_\theta$, write 
\begin{eqnarray*}
F_\theta(t)  
&=& \sum_{k=1}^{m} \Pb\bigg( \sum_{j=1}^m a_j \max(U - b_j,0) + c \leq t; b_k\leq U \leq b_{k+1} \bigg) \\
&=& \sum_{k=1}^{m} \Pb\big( s_k U - v_k + c \leq t; b_k\leq U \leq b_{k+1} \big) \\
&=& \sum_{k=1}^{m} \big\{ \1(s_k b_k < t-c+v_k < s_k b_{k+1}) \big(\frac{t-c+v_k}{s_k} - b_k \big) 
+ \1( t-c+v_k> s_k b_{k+1}) (b_{k+1}-b_k) \big\}.
\end{eqnarray*}
Any cdf $F_{\theta}$ is piecewise linear, with successive positive linear slopes $1/s_1$, $1/s_2$,...,$1/s_{m}$.
Indeed, the interior of the intervals $I_k:=[s_k b_k+c-v_k, s_k b_{k+1}+c-v_k)$ are never empty 
since $b_{k+1}>b_k$. Moreover, their intersections with $\Dc_\theta$ are never empty since 
$s_k b_k - v_k\geq 0$ and 
$$ s_k b_{k+1} - v_k \leq s_k - v_k  \leq s_m-v_m .$$
Note that any $I_k$ starts at $t_k=s_kb_k +c-v_k$, ends at $t'_k=s_kb_{k+1} +c-v_k$, and check that 
$t'_k=t_{k+1}$, $k\in \{1,\ldots,m-1\}$. Since $t_1=c$ and $t'_m=c+s_m-v_m$, the intervals $I_1,\ldots,I_{m-1}$ are disjoint and yield a partition of $\Dc_\theta$. 

Now consider two model parameters $\theta^{(1)}$ and $\theta^{(2)}$ s.t. $F_{\theta^{(1)}}(t)=F_{\theta^{(2)}}(t)$ for any real number $t$. In particular, assume their supports are the same. With obvious notations, this implies $c^{(1)}=c^{(2)}=:c$.
Moreover, their sequences of slopes have to be the same, implying $s_k^{(1)}=s_k^{(2)}$ for every $k=1,\ldots,m$.
This implies $a_k^{(1)}=a_k^{(2)}$ for every $k=1,\ldots,m$, now denoted $a_k$ simply and similarly for their sums $s_k$.
Considering the starting points of the upward sloping segments, we have to satisfy $s_k b_k^{(1)}-v_k^{(1)}=s_k b_k^{(2)}-v_k^{(2)}$, $k\in\{2,\ldots,m\}$. In particular, 
$s_2 b_2^{(1)}-v_2^{(1)}=s_2 b_2^{(2)}-v_2^{(2)}$, or $a_1 b_2^{(1)}=a_1 b_2^{(2)} $ equivalently. Since $a_1>0$, we get
$b_2^{(1)}=b_2^{(2)}$. 
Recursively, it can be proved that $ b_k^{(1)}= b_k^{(2)}$ for any $k\in \{2,\ldots,m\}$. Indeed, 
assume $b_j^{(1)}=b_j^{(2)}$ for $j\in\{1,\ldots,k-1\}$.
Then $s_k b_k^{(1)}-v_k^{(1)}=s_k b_k^{(2)}-v_k^{(2)}$ implies 
$ \big( s_j - a_j\big) b_k^{(1)}=\big( s_j - a_j\big) b_k^{(2)}$ and then $ b_k^{(1)}= b_k^{(2)}$, proving the result.
\end{proof}

\vskip 0.2in
\bibliography{bibliography}

\end{document}